\documentclass[twocolumn,twocolappendix,times]{aastex63}

\usepackage{graphicx}
\usepackage{amsmath}
\usepackage{amssymb}
\usepackage{booktabs}
\usepackage{url}
\usepackage{xcolor}
\usepackage{xspace}

\newcommand{\lya}{\ensuremath{{\rm Ly}\alpha}\xspace}

\newcommand{\secpoint}{\mbox{$''\mskip-7.6mu.\,$}}

\newcommand{\msun}{\ensuremath{\rm M_\odot}}
\newcommand{\mstar}{\ensuremath{M_*}}

\newcommand{\nhi}{\ensuremath{N_{\rm HI}}}

\newcommand{\kms}{\rm km~s\ensuremath{^{-1}\,}}
\newcommand{\dv}{\ensuremath{\Delta v}\xspace}
\newcommand{\dvla}{\ensuremath{\Delta v(\lya)}}
\newcommand{\dvciv}{\ensuremath{\Delta v(\text{\civ})}\xspace}

\newcommand{\ztwo}{\ensuremath{z\sim2}}

\newcommand{\ztwothree}{\ensuremath{z\sim2-3}\xspace}
\def\ltsima{$\; \buildrel < \over \sim \;$}
\def\simlt{\lower.5ex\hbox{\ltsima}}
\def\gtsima{$\; \buildrel > \over \sim \;$}
\def\simgt{\lower.5ex\hbox{\gtsima}}
\def\arcs{$''~$}

\def\ltsima{$\; \buildrel < \over \sim \;$}
\def\simlt{\lower.5ex\hbox{\ltsima}}
\def\gtsima{$\; \buildrel > \over \sim \;$}
\def\simgt{\lower.5ex\hbox{\gtsima}}

\def\kms{km s$^{-1}$}
\def\ltsima{$\; \buildrel < \over \sim \;$}
\def\simlt{\lower.5ex\hbox{\ltsima}}
\def\gtsima{$\; \buildrel > \over \sim \;$}
\def\simgt{\lower.5ex\hbox{\gtsima}}
\def\arcs{$''~$}

\def\sun{\ifmmode_{\mathord\odot}\else$_{\mathord\odot}$\fi}

\newcommand{\ionl}[3]{\ion{#1}{#2} $\lambda\, #3$}

\newcommand{\ew}{$W_\lambda$\xspace}

\newcommand{\hone}{\ion{H}{1}\xspace}

\newcommand{\siiii}{\ion{Si}{3}\xspace}
\newcommand{\siiiiw}{\ion{Si}{3} $\lambda1206$\xspace}

\newcommand{\siii}{\ion{Si}{2}\xspace}
\newcommand{\siiiwl}{\ion{Si}{2} $\lambda1260$\xspace}
\newcommand{\siiiwm}{\ion{Si}{2} $\lambda1304.4$\xspace}
\newcommand{\siiiwh}{\ion{Si}{2} $\lambda1526$\xspace}

\newcommand{\cii}{\ion{C}{2}\xspace}

\newcommand{\alii}{\ion{Al}{2}\xspace}

\newcommand{\oi}{\ion{O}{1}\xspace}
\newcommand{\oiw}{\ion{O}{1} $\lambda1302.2$\xspace}

\newcommand{\siiv}{\ion{Si}{4}\xspace}

\newcommand{\civ}{\ion{C}{4}\xspace}
\newcommand{\civd}{\ion{C}{4} $\lambda\lambda1548, 1550$\xspace}
\newcommand{\civl}{\ion{C}{4} $\lambda1548$\xspace}
\newcommand{\civh}{\ion{C}{4} $\lambda1550$\xspace}

\newcommand{\feiiw}{\ion{Fe}{2} $\lambda1608$\xspace}

\newcommand{\vlos}{\ensuremath{v_{\text{LOS}}}\xspace}
\newcommand{\zfg}{\ensuremath{z_\mathrm{fg}}\xspace}
\newcommand{\zbg}{\ensuremath{z_\mathrm{bg}}\xspace}
\newcommand{\dtran}{\ensuremath{D_{\rm tran}}\xspace}

\shorttitle{Mapping the $z\sim 2$ CGM with KBSS Pairs}
\shortauthors{Prusinski, Steidel, \& Chen}

\turnoffedit

\begin{document}

\title{\replaced{Tomography of}{Mapping} the \boldmath$z\sim 2$\unboldmath\ Circumgalactic Medium with KBSS Galaxy Pairs\footnote{Based on data obtained at W.\ M.\ Keck Observatory, which is operated as a scientific partnership among the California Institute of Technology, the University of California, and the National Aeronautics and Space Administration. The Observatory was made possible by the generous financial support of the W.\ M.\ Keck Foundation.}}

\author[0000-0001-5847-7934]{Nikolaus Z.\ Prusinski}
\affiliation{Cahill Center for Astronomy and Astrophysics, California Institute of Technology, MC 249-17, Pasadena, CA 91125, USA; \url{nik@astro.caltech.edu}, \url{ccs@astro.caltech.edu}} 

\author[0000-0002-4834-7260]{Charles C.\ Steidel}
\affiliation{Cahill Center for Astronomy and Astrophysics, California Institute of Technology, MC 249-17, Pasadena, CA 91125, USA; \url{nik@astro.caltech.edu}, \url{ccs@astro.caltech.edu}}

\author[0000-0003-4520-5395]{Yuguang Chen}
\affiliation{Department of Physics and Astronomy, University of California Davis, 1 Shields Avenue, Davis, CA 95616, USA}
\affiliation{Department of Physics, The Chinese University of Hong Kong, Shatin, N.T., Hong Kong SAR, China; \url{yuguangchen@cuhk.edu.hk}}

\received{2025 March 24}
\accepted{2025 September 15}
\submitjournal{ApJ}

\begin{abstract}
    
We present new results on the spatial structure and kinematics of the circumgalactic medium (CGM) of $z\sim2$ star-forming galaxies drawn from the Keck Baryonic Structure Survey, using \lya\ and metallic ion absorption recorded in spectra of background galaxies whose sightlines are projected within $\theta \sim 30$\arcs\ (physical distances $\dtran \lesssim 250$ kpc) of the foreground galaxy. The sample of 1033 foreground galaxies ($\langle\zfg\rangle=2.03 \pm 0.36$; $8 \lesssim {\rm log(\mstar/M_{\odot})} \lesssim 11$) is probed by the spectra of 736 background galaxies 
obtained using Keck/KCWI ($R\sim 1800$) 
and Keck/LRIS ($R \sim 1200$).
Using the galaxy pair ensemble, we present measurements of absorption strength (\ew) and line-of-sight velocity dispersion ($\sigma$) as a function of $\dtran$, which together with ``down-the-barrel'' spectra of the foreground galaxies provide unprecedented maps of neutral hydrogen and ionized metals throughout the CGM of star-forming galaxies at ``cosmic noon''.
A 2D map of \lya absorption depth (vs.\ line-of-sight velocity and \dtran) for the full ensemble shows a distinct transition near $\dtran \sim 80$ kpc (close to the virial radius of the average foreground galaxy) where the line-of-sight velocity dispersion reaches a minimum and beyond which the \lya absorption depth flattens. Splitting the sample into subsets based on properties of the foreground galaxies, we find a strong and monotonic correlation of both $W_{\lambda}$(\ion{C}{4}) and $\sigma$(\ion{C}{4}) with stellar mass (\mstar) at all \dtran including DTB (i.e., $\dtran \sim 0-250$ kpc). Taken together, our results suggest that the kinematics, covering fraction, and spatial extent of CGM gas traced by \lya\ and (especially) \ion{C}{4} are modulated primarily by halo mass, 
and that a significant fraction of the CGM gas may be unbound.

\end{abstract}

\section{Introduction}
\label{sec:intro}

Galaxies evolve via the cycling of baryons between gas reservoirs in the intergalactic medium (IGM) and their own interstellar media (ISM). The circumgalactic medium (CGM) represents the sphere of influence for a galaxy and acts as a boundary layer where one can directly observe the exchange of baryons to and from the IGM. Despite its importance in galaxy formation, the CGM and the underlying processes driving its evolution with time remain largely unconstrained \citep{tumlinson_circumgalactic_2017}.

Observational and theoretical work has established that the CGM of star-forming galaxies comprises metal-enriched gas \citep[e.g.,][]{peeples_budget_2014,peroux_cosmic_2020,dutta_muse_2020,dutta_metal-enriched_2021} in a clumpy morphology \citep[e.g.,][]{rubin_galaxies_2018-1,rudie_column_2019, erb_circumgalactic_2023,li_alpaca_2024} across a range of temperatures, densities, and ionization states \citep[e.g.,][]{byrohl_cosmic_2023,faucher-giguere_key_2023}. The multiphase nature of the CGM is the result of a combination of baryonic feedback processes i.a., supernova ejecta \citep{leitherer_starburst99_1999,veilleux_galactic_2005}, active galactic nuclei \citep{faucher-giguere_physics_2012}, and radiation pressure \citep{murray_radiation_2011} from massive stars all acting in concert to shape the gas distribution and physical conditions. Numerous studies across a range of redshifts have observed outflows which inject energy and momentum into the ISM and CGM of their host galaxy \citep[e.g.,][]{shapley_rest-frame_2003,weiner_ubiquitous_2009,steidel_structure_2010,martin_demographics_2012,erb_galactic_2012,prusinski_connecting_2021}; signatures of inflowing material are much less common \citep[e.g.,][]{keres_how_2005,rubin_direct_2012,bouche_signatures_2013}, but the connections between such baryonic flows, the underlying CGM gas distribution, and the host galaxy properties remain uncertain.

A natural epoch in which to study the CGM around galaxies is at \ztwothree where the global star formation rate, baryonic accretion rate, and supermassive black hole growth all reached their peak \citep[e.g.,][]{richards_sloan_2006,faucher-giguere_baryonic_2011, behroozi_average_2013, madau_cosmic_2014}. At these redshifts, many of the rest-frame far-UV and optical diagnostics fall into wavelength windows with high atmospheric transmission and can therefore be observed using ground-based telescopes \citep[see e.g.,][]{steidel_strong_2014}. Indeed, galaxies in this redshift range show prominent outflows, extended \hone halos in both emission and absorption, and substantial metal absorption in their CGM \citep{bouche_dynamical_2007,steidel_structure_2010,steidel_diffuse_2011,rudie_gaseous_2012,leclercq_muse_2020,chen_kbsskcwi_2021,weldon_mosdef-lris_2022}.

Prior to the early 1990s, the connection between diffuse, metal-enriched gas and galaxies was based largely on the statistical incidence of various classes of intervening absorption systems in spectroscopic surveys of QSOs -- the characteristic ``size'' of extended gas in galaxy halos was derived from a combination of the observed incidence per unit redshift path length of absorption systems (e.g.\ \ion{Mg}{2}, \ion{C}{4}, damped \lya)  and the space density of galaxies with which the gas was assumed to be bound, yielding ``predicted'' halo sizes of $R_{\rm gal} \simeq 20-100$ pkpc, depending on the sensitivity of the QSO absorption system samples, the tracer ion, and the assumed scaling relation between galaxy luminosity and galaxy gaseous cross-section. More direct association of QSO absorption systems and intervening galaxies began at low redshift:\ deep CCD images of QSO sightlines with known $z \simeq 0.2-0.5 $ \ion{Mg}{2} absorption systems were used to identify candidate absorbing galaxies within several arcsec of the QSO, followed by spectroscopic observations of the relatively faint ($R \gtrsim 20$) galaxies to confirm their redshifts \citep{bergeron_sample_1991}. The initial success in identifying $\simeq L^{\ast}$ galaxies within $\lesssim 50$ kpc of the QSO sightline appeared to support the general expectations that had been based on simple statistical arguments.

However, the small samples of identified absorbing galaxies, in addition to being observationally expensive, were subject to potential observational selection biases -- e.g., one cannot be certain that the galaxy identified is the one responsible for the absorption system, or simply the most easily-observed galaxy at the same redshift near the QSO sightline. A less biased means of establishing the connection between galaxies and absorbers would be to first observe galaxies near QSO sightlines, where each would have known luminosity, stellar mass, SFR, etc., as well as separation from the QSO sightline \dtran, and subsequently look for absorption signatures (or sensitive limits) in the QSO spectrum for each \added{\citep[e.g.][]{adelberger_galaxies_2003,crighton11,rudie_gaseous_2012,werk+2012,turner_metal-line_2014,chen+2020_CUBS,lofthouse_muse_2020}}. Such a ``galaxy-centered'' approach -- where galaxy samples are observed in fields deliberately chosen to have one or more bright background sources (usually QSOs) that can provide high-quality measurements of diffuse gas over the relevant range of redshift -- has become more practical on large telescopes equipped with efficient multi-object spectrographs (MOS) or integral field units (IFUs). 

\begin{figure*}[htbp!]
    \epsscale{1.18}
    \plotone{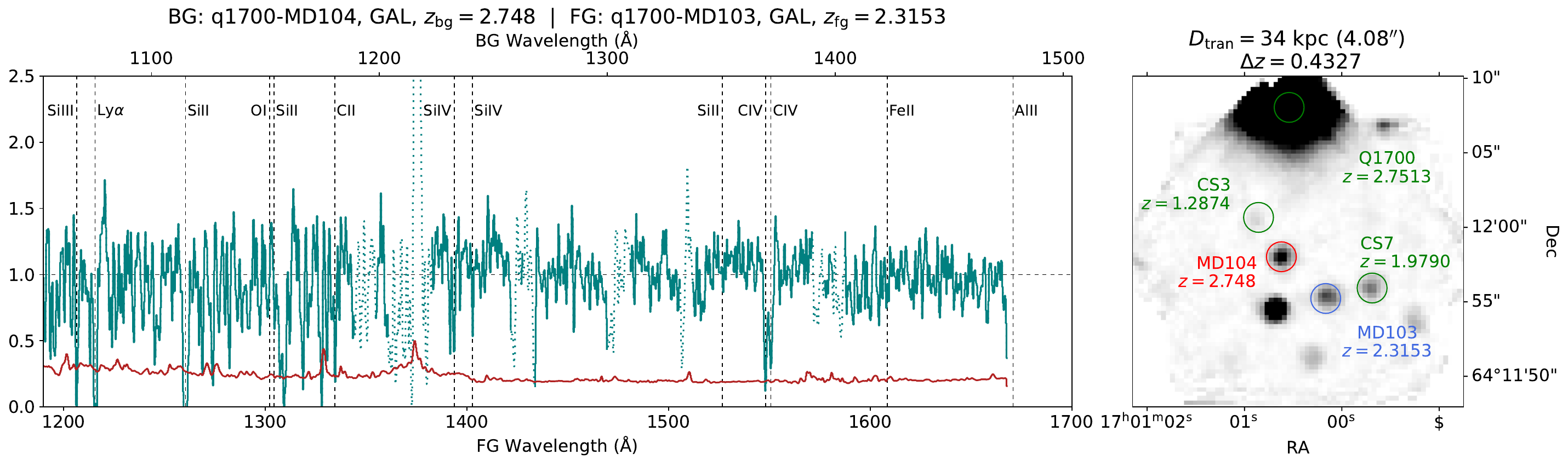}
    \plotone{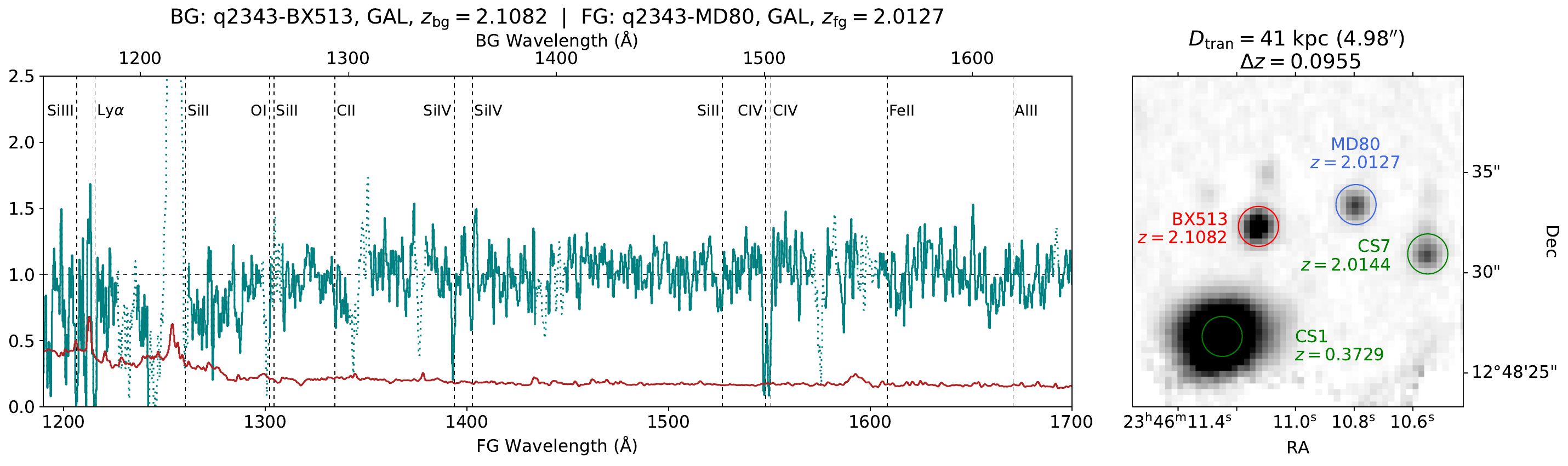}
    \plotone{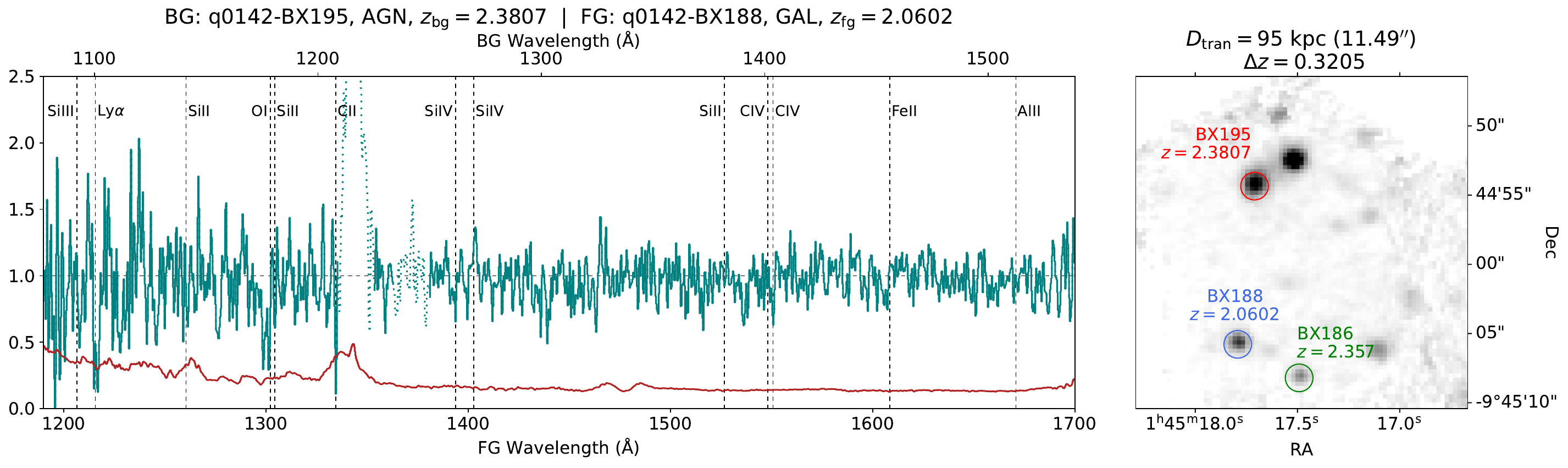}
    
    \caption{Example KCWI background galaxy spectra, normalized and shifted to the rest frame of the foreground galaxy of the pair, from the KBSS-KCWI sample. The righthand images show the white light (3530 -- 5530 \AA) image from the relevant KCWI pointing, where the background galaxy is labeled in red and the foreground galaxy in blue; green labels indicate other galaxies with spectroscopic identifications within the field of view.  Spectral features at the redshift of the FG in each pair are indicated with vertical dashed lines and labels. The location of the same set of features at the BG redshift have been highlighted with a dotted line style. The three cases have $D_{\rm tran}\simeq 34$, 41, and 95 kpc, respectively. 
    }
    \label{fig:indobjs}
\end{figure*}

An alternative to using background QSOs is to use unrelated foreground-background galaxy pairs, which are 50--100 times more numerous within the Keck Baryonic Structure Survey (KBSS) than QSO-galaxy sightlines (e.g.\ \citealt{adelberger_connection_2005,steidel_structure_2010,chen_keck_2020}; see also \citealt{bordoloi_radial_2011,rubin_galaxies_2018,rubin_galaxies_2018-1,mendez-hernandez_metal_2022}). The CGM of a particular foreground galaxy (at $z_\mathrm{fg}$) can be probed using spectroscopy of background galaxies ($z_\mathrm{bg} > z_\mathrm{fg}$), each of which forms a sightline through the foreground CGM. 
\added{The use of faint galaxies as background sources, while individually of much lower $S/N$ and typically of significantly lower spectral resolution compared to a QSO spectrum, has additional advantages other than the larger number of foregound-background pairs that complement QSO-based measurements. As emphasized by \citet{steidel_structure_2010}, using background galaxies is equivalent to averaging over a beam with a diameter of several kpc passing through the CGM of the foreground galaxy \citep[e.g.][]{rubin_galaxies_2018-1,Afruni+2023}, whereas the projected beam of a background QSO sightline is $\sim 1$ pc; this means that, in principle, spectra of an ensemble of resolved background sources converge more quickly to a spatially averaged absorption strength and representative kinematics of gas associated with a foreground galaxy, in a statistical sense -- albeit without resolving the individual velocity components necessary for measuring accurate column densities within a foreground galaxy's CGM. An ensemble of foreground/background galaxy pairs is thus highly complementary to surveys of galaxies near the lines of sight to bright QSOs.}
Three representative examples of foreground/background galaxy pairs are shown in Figure \ref{fig:indobjs}, in order to illustrate the technique. 
Note that the projected distance (\dtran) for the Q0142-BX195-BX188 pair is $\simeq 2-3$ times larger that of the pairs in the top two panels and has comparatively weaker absorption features at the redshift of the foreground galaxy. An ensemble of such pairs can be used to characterize the CGM absorption as a function of both \dtran and with respect to the properties of the foreground galaxies.

Galaxy foreground-background close angular pairs drawn from what would become the KBSS survey were used by both \citet{adelberger_connection_2005} and \citet{steidel_structure_2010}; the latter study constructed a sample of 512 galaxy pairs with angular separations $\theta \le 15$\arcsec\ from a sample of \ztwothree star forming galaxies. The galaxy pairs were subdivided into 
three composite spectra ($\langle \dtran\rangle\in \left[31,63,103\right]$ kpc) to make a low resolution map of absorption as a function of \dtran (their Figure 21). They used the measurements of absorption line equivalent widths ($W_{\lambda}$) versus $D_{\rm tran}$, combined with the stack of ``down the barrel'' (DTB; $\dtran = 0$ kpc) spectra of the foreground galaxies in each angular pair, to model the velocity field and galactocentric radial distribution of absorbing gas in the ``mean'' CGM surrounding galaxies at $\langle z \rangle = 2.2$. They found that all of the observed metallic species were observable to $\dtran \simeq 70-100$ kpc, at which point the equivalent widths fell below the detection limit of $W_\lambda \simeq 0.1$ \AA. Excess \lya\ absorption remained detectable to at least $D_{\rm tran} \sim 250$ kpc, as expected given the earlier results using background QSOs to probe similar galaxies in the same fields \citep{adelberger_galaxies_2003,adelberger_connection_2005}.

\citet{chen_keck_2020} extended the galaxy pair technique by incorporating a larger sample of galaxy spectra ($N\simeq 3000$ with $1.7 \lesssim z \lesssim 3.4$), with an increased range of pair separations, drawn from the KBSS survey regions. The authors assembled more than 200,000 foreground/background galaxy pairs with angular separations $\Delta\theta\simeq3-500\arcsec$ ($20\lesssim \dtran/\mathrm{kpc} \lesssim 4000$) and produced maps of \lya absorption strength as a function of line of sight velocity (with respect to the foreground galaxy systemic redshifts) \vlos and \dtran. They found that, if expressed as a single power law, $W_\lambda(\lya)\propto \dtran^{-0.4}$ over the full range of impact parameters, in agreement with \citet{steidel_structure_2010}. The number of galaxy pairs sampling separations $D_{\rm tran} \le 100$ kpc -- i.e., the range over which metal lines had been detected by \citet{steidel_structure_2010} -- was not significantly larger, so only the distribution and kinematics of \lya\ were analyzed. %

In the present work, our goal has been to construct an ensemble of foreground/background galaxy pairs that provides improved sampling of gas within $\dtran \leq 250$ kpc, the approximate scale of the CGM surrounding individual $L_{\rm uv}^{\ast}$ galaxies at $z \simeq 2-3$ established by previous work using KBSS QSO sightlines. We are also interested in improving the constraints on CGM gas-phase kinematics (both \ion{H}{1} and metals) and its relation to the properties of the foreground galaxies; to this end, from \citet{chen_keck_2020} we include only the subset ($\sim 30$\%) of background galaxy spectra observed with the higher spectral resolving power configuration of LRIS-B ($\langle R \rangle \sim 1300$; see Appendix B of \citealt{chen_keck_2020}). The most significant improvement -- the one that has motivated our revisiting the use of close angular pairs of galaxies -- comes from incorporating deep IFU pointings using the Keck Cosmic Web Imager (KCWI; \citealt{morrissey_keck_2018}). The KCWI data  set (KBSS-KCWI; see \citealt{chen_kbsskcwi_2021}) consists of IFU observations of many contiguous $\sim 20$\arcsec\ diameter ($\simeq 170$ kpc at $z \sim 2$) regions within the KBSS survey fields. The KBSS-KCWI subsample provides deep ($\gtrsim 5$ hour total integrations per pointing) spectra with resolving power $\langle R \rangle \simeq 1800$ of a large number of galaxy angular pairs with projected separations $\le 20''$, including previously-known KBSS galaxies as well as fainter galaxies newly-identified from their extracted KCWI spectra. The KCWI pointings are embedded within larger KBSS survey fields, each of which has extensive photometric and spectroscopic ancillary data. All of the galaxy spectra included in the combined LRIS+KCWI sample have resolving power $R\simeq 1500\pm 300$, such that the \civd doublet is resolved, as are \oiw and \siiiwm.

This paper is organized as follows:\ in Section \ref{sec:obs}, we discuss the observations and data reduction associated with each instrument subsample; Sections \ref{sec:paircons} and \ref{sec:stacking} cover the construction of the galaxy pair sample and stacking of galaxies in bins of \dtran; Section \ref{sec:metalabs} features results from \lya and metal absorption features in 1-D and 2-D; Section \ref{sec:model} presents a semianalytic model that can be used to fit the 1-D composite spectra and the 2-D map of \lya absorption; Section \ref{sec:bin} has results from composite spectra binned by inferred foreground galaxy properties; these results are placed in a broader context in Section \ref{sec:disc}; and finally, we summarize our conclusions in Section \ref{sec:conc}. Appendix \ref{app:wavetrim} justifies why an explicit redshift cut is not required for our sample; Appendix \ref{app:hires} uses pairs generated from HIRES spectra of background QSOs to elucidate the profiles we observe at low resolution; Appendix \ref{app:stackmethod} shows how different stacking methods affect the composite spectra; and Appendix \ref{app:filling} details how, for galaxy pairs with close angular separations, \lya emission from the foreground galaxy can modulate the observed background \lya absorption signal.  

Throughout this paper, we adopt the \textit{Planck} 2018 cosmology \citep{planck_collaboration_planck_2020}:\ a flat $\Lambda$CDM model with $H_0=67.7$ km s$^{-1}$ Mpc$^{-1}$, $\Omega_{m,0}=0.31$, and $\Omega_{\Lambda,0}=0.69$. In this cosmology, at the median redshift of the sample ($z_\mathrm{med} = 2.3$), 1\arcsec\ corresponds to 8.4 (physical) kpc. We use \dtran to represent the projected physical separation between a foreground galaxy and the line of sight to a background object, evaluated at the redshift of the foreground galaxy ($z_{\rm fg}$); we direct the reader to to Figure 23 of \citet{steidel_structure_2010} for a schematic.

\section{Observations}
\label{sec:obs}

\subsection{KBSS-KCWI}
\label{sec:kbss-kcwi}

The KBSS-KCWI spectra \citep[][2025]{chen_kbsskcwi_2021} used in this paper comprise 193 galaxies observed in 46 different KCWI pointings, selected to include objects from KBSS. KBSS-KCWI pointings maximize the number of previously identified galaxies in each field, in particular those with data from LRIS-B \citep{steidel_survey_2004}, MOSFIRE \citep{mosfire}, and \textit{HST} WFC3-IR imaging. 
Of the 193 targets (excluding QSOs), 7 are optically faint AGN, while the remainder are star-forming galaxies.

KCWI observations were conducted between 2017 September and 2022 June, after which KCWI went offline for red-channel integration. The integral field spectra used in this study were obtained using KCWI's low resolution grating (BL; central wavelength $= 4500$ \AA) and Medium slice resulting in data cubes providing a contiguous 20\farcs4 $\times$ 16\farcs5 FoV and spectral resolving power $\langle R\rangle \approx 1800$ over the $3530-5530$ \AA\ bandpass. Each KBSS-KCWI field was observed for a typical total integration time of five hours, with dithers and/or position angle (PA) rotations between each 20-minute exposure. The final mosaicked data cube for each pointing maps a region with a 20\arcsec\ diameter (170 kpc at $\langle z \rangle=2.3$) with approximately uniform exposure and uniform spatial sampling. 

Individual exposures were reduced using a customized version of the IDL KCWI Data Reduction Pipeline (DRP; \citealt{neill_kderp_2023})\footnote{\url{https://github.com/Keck-DataReductionPipelines/KcwiDRP}} described in detail in \citet{chen_kbsskcwi_2021}. Briefly, each twenty minute science frame was converted into a 3D flux and wavelength calibrated, rectified data cube, with the original spatial sampling (0\secpoint3 $\times$0\secpoint68 and spectral sampling of 1.0~\AA/pixel. All exposures for a given pointing were then combined into a stacked, mosaicked data cube using the \texttt{KCWIKit} package \citep{chen_kbsskcwi_2021,prusinski_kcwikit_2024}. \texttt{KCWIKit} reads in a list of cubes to be stacked; a parameter file enumerating e.g.\ the final stack pixel scale, dimensions, and orientation; and if necessary, a list of pre-alignment cube world coordinate system (WCS) shifts if the header WCS is incorrect or significantly offset from the rest of the stack. The code runs natively in a \texttt{jupyter} \citep{loizides_jupyter_2016} environment and begins by aligning each cube with respect to a fiducial frame (the first in the list of cubes by default). The code de-rotates and cross-correlates each frame to determine sub-pixel relative offsets, then drizzles them onto a common grid with rectilinear ($0\farcs3 \times 0\farcs3$) spatial sampling, weighted by exposure time using the \texttt{Montage}\footnote{\url{http://montage.ipac.caltech.edu/}} \citep{jacob_montage_2010} package. Since all data were taken in the KCWI ``Medium BL'' configuration, each cube has the same common wavelength range (3530-5530~\AA) and sampling (1 \AA\ pix$^{-1}$). The WCS of the stacked cube is then corrected using an external observation (typically \textit{HST} F160W images) to match the astrometry of existing KBSS ancillary data products. The final output is a WCS-registered $\sim 20$\arcsec\ (170 kpc) diameter pointing.

For each stacked cube, we also constructed a ``white light'' (3530 -- 5530 \AA) image and used \texttt{Source Extractor} \citep{bertin_sextractor_1996} to detect all significant continuum sources; all detected continuum sources were extracted to 1D from the data cubes by weighting each spatial element within the detected segmentation region by the $(S/N)^2$ of the white light signal. Some of the continuum sources detected in each KCWI pointing were not KBSS catalog objects (normally limited to ${\cal R} \le 25.5$), but yielded spectroscopic redshifts placing them in the targeted range for the present purposes, $1.4 \le z \le 3.5$. Very deep photometric catalogs exist in all of the KBSS survey fields (in addition to the KCWI white light images), generally for at least $UG{\cal R}JHK_s$ bands as well as one or more deep pointings of HST-WFC3-IR F140W and/or F160W, so that photometry sufficient for constraining the stellar masses of the newly identified galaxies was available. These sources (of which there are 51 in all) have been labeled ``CS'' (continuum serendipitous) along with their segmentation IDs (e.g., CS3 -- see Figure \ref{fig:indobjs}). Of the 51 galaxies, 
22 are fainter than the nominal KBSS catalog continuum limit ($\mathcal{R}\geq25.5$), with a limiting $\mathcal{R}$-band magnitude of 27.33.\footnote{A small number of additional serendipitous objects were spectroscopically identified but lacked detections in any of the available near-IR photometric bands necessary to constrain their stellar masses; they have not been included in the analyses below.}

\subsection{LRIS-B}
\label{sec:lris}

LRIS-B spectra comprise the bulk of the KBSS rest-frame far-UV spectroscopic sample, having been accumulated over $\sim$two decades, and as such we direct the reader to \citet{steidel_lyman_2003, steidel_survey_2004} and \citet{adelberger_optical_2004} for color selection criteria and data reduction strategies of the QSO field observations used in this study. LRIS observations are used to supplement KCWI observations and include a portion of the KBSS galaxy pair sample (KGPS) compiled by \citet{chen_keck_2020}, which at the time comprised a broad superset of the galaxies used previously in numerous projects (e.g.\ \citealt{steidel_lyman_2003,steidel_survey_2004,adelberger_connection_2005,reddy_spectroscopic_2006,steidel_structure_2010}). Notably, in the current work we include only galaxy pairs with angular separations $\Delta \theta \leq 30\arcsec$ ($\sim 250$ kpc) and background galaxy spectra obtained with the higher resolution 600 lines mm$^{-1}$ LRIS-B grism. The latter condition translates to LRIS-B spectra with $R\sim 1300$ \citep{chen_keck_2020}, somewhat lower resolution than KCWI at the same wavelengths. 

All of the LRIS-B data used in this sample were obtained between 2002 and 2016. Of the 2862 spectra used in the KGPS sample, 872 were observed with the 600 lines mm$^{-1}$ grism blazed at 4000 \AA, and of those, 400/872 were observed on more than one slitmask. Each slitmask was observed for a minimum of 1.5 hours and $30\%$ of the 600 lines mm$^{-1}$ grism slitmasks have exposure times of $>3$ hours. Only the blue channel LRIS data have been used in the present work; the spectral range covered is similar to that 
of the KCWI data described above, with a red cutoff near 5600 \AA\, extending down to 3400-3500 \AA\ in the UV.

Of the 872 galaxies that make up the LRIS sample ($1.4 \leq z \leq 3.6$), 135 were observed with particularly long total integrations ($\sim 10-12$ hours). This ``KBSS-LM1'' sample was compiled between 2014 and 2016 to be a deep collection of LRIS rest-UV spectra focused on the redshift range $2.10\leq z_\mathrm{gal}\leq 2.58$, corresponding to the highest quality existing KBSS-MOSFIRE spectra. Sample selection criteria are described by \citet{steidel_reconciling_2016} and \citet{theios_dust_2019}, but in short, the galaxies were chosen to maximize the overlap with the existing KBSS-MOSFIRE sample, and especially pertinent for this work, the KBSS fields were prioritized so that the largest number of galaxies could be observed on a single slitmask.

\section{Galaxy Pair Sample}
\label{sec:paircons}

\begin{figure*}[htbp!]
    \plottwo{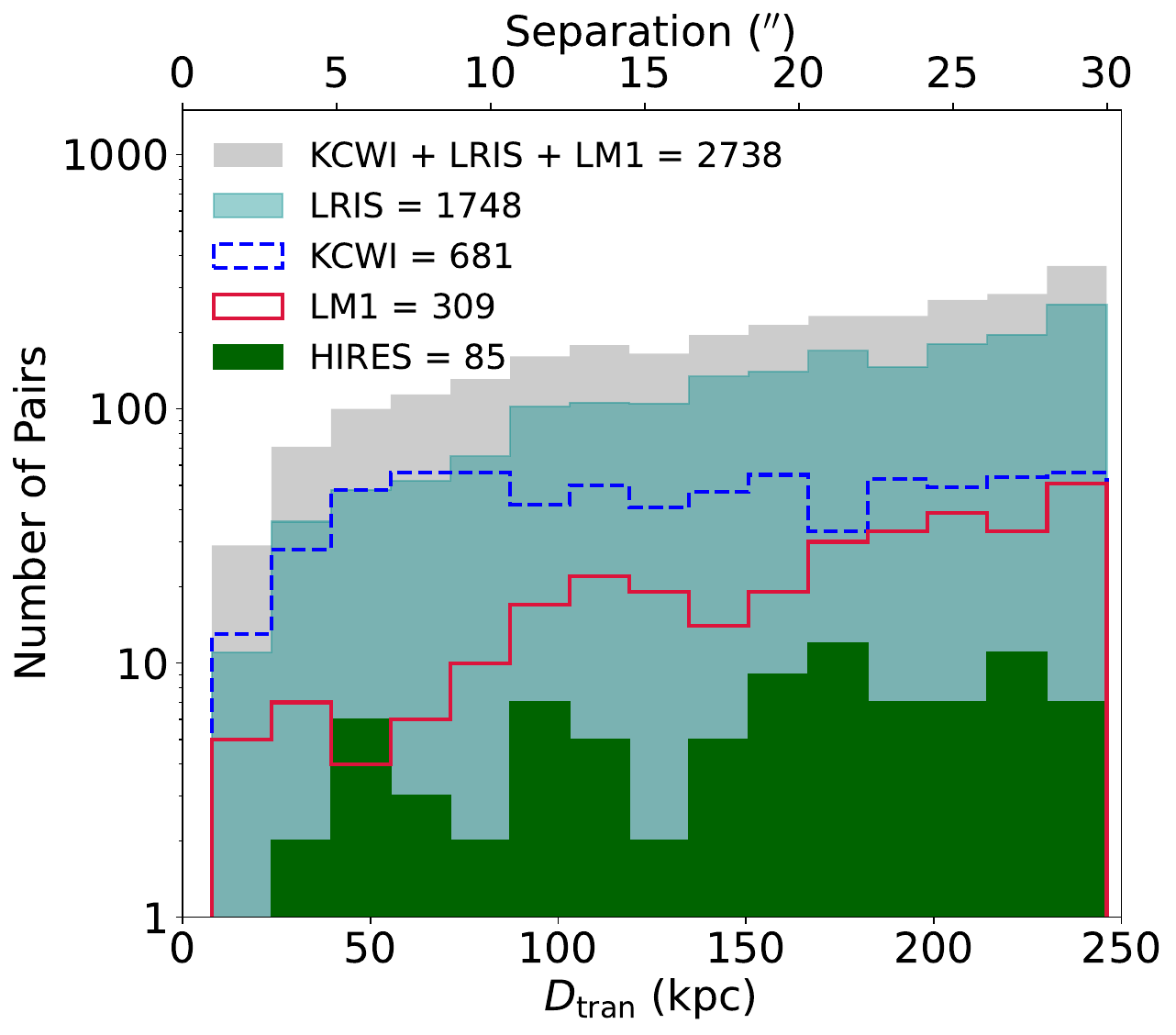}{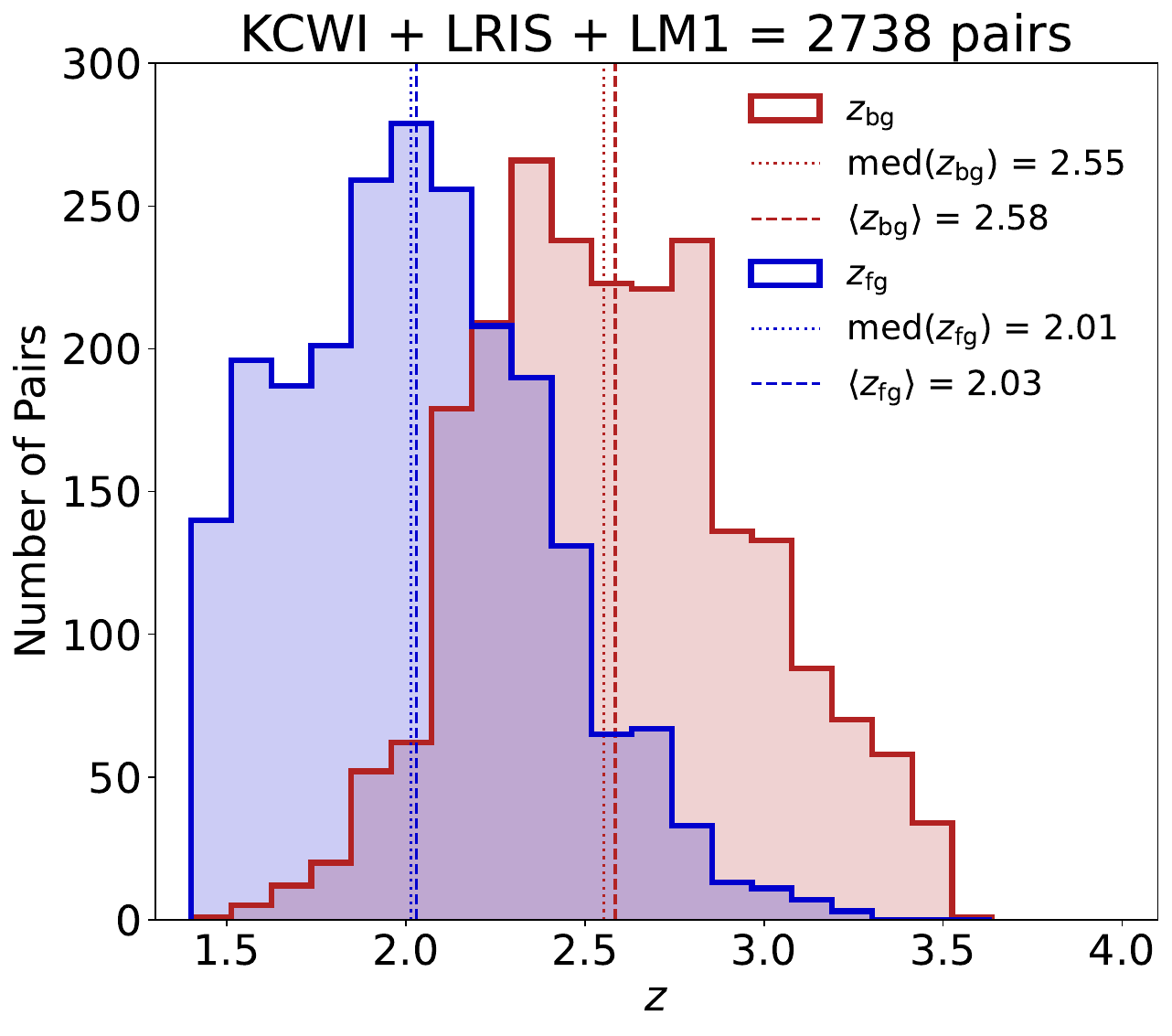}
    \caption{\textit{Left:} Histogram of \dtran for the galaxy pair sample. The full distribution of projected distances is shown in gray and KCWI pairs are shown in the dashed blue histogram. KCWI background spectra are supplemented by LRIS-B (cyan histogram) and LM1 (red outline) observations, the former of which begins to dominate at $\dtran\gtrsim 100$ kpc. There are 2738 pairs in total and respective totals are enumerated in the plot legend. The HIRES pair distribution constructed from background QSOs is shown for comparison in green. \textit{Right:} The redshift distribution of foreground (blue) and background (red) galaxies in the galaxy pair sample along with vertical dotted and dashed lines denoting the median and mean redshifts respectively.}
    \label{fig:hist}
\end{figure*}

\begin{figure}[htbp!]
    \epsscale{1.2}
    \plotone{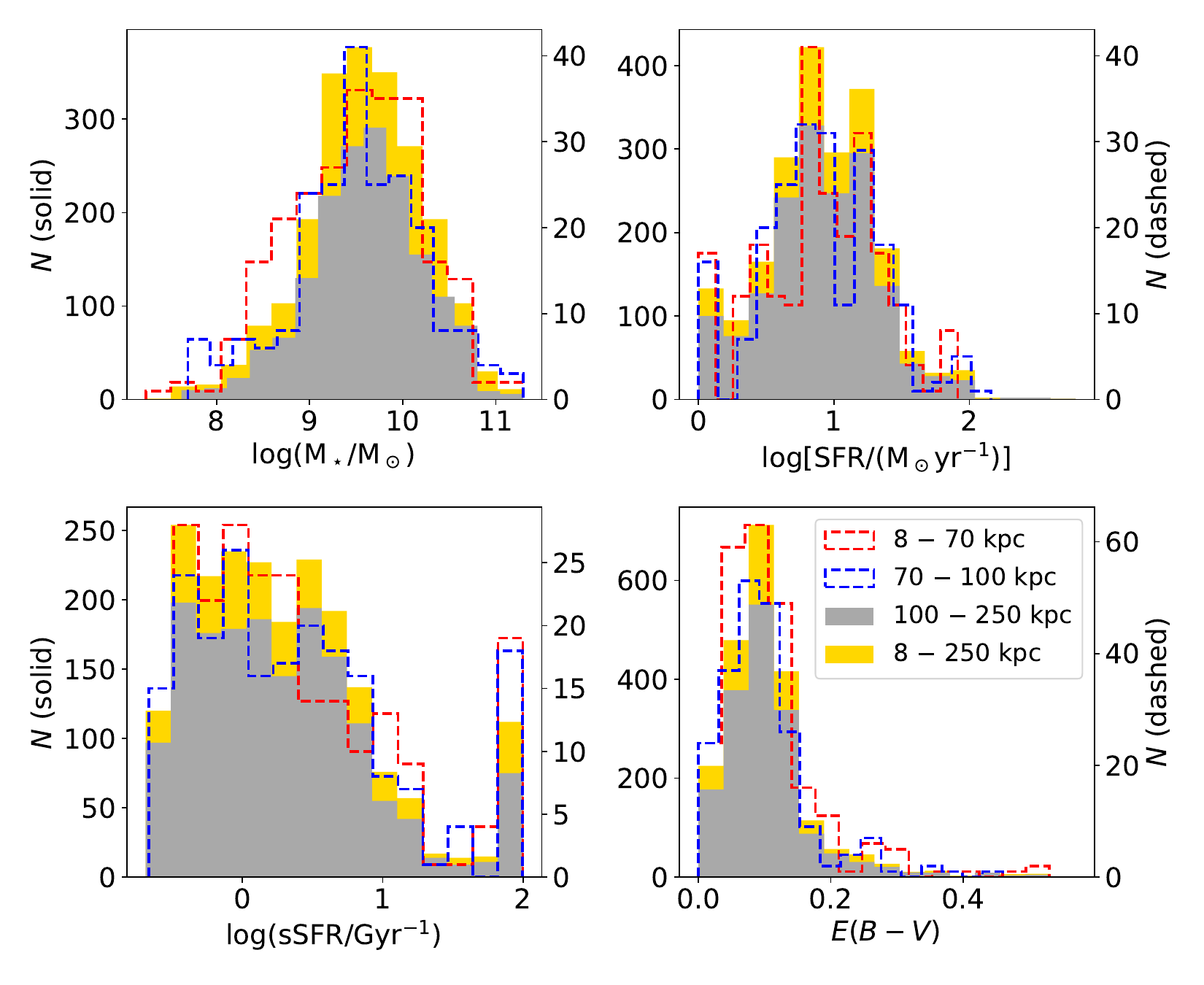}
    \caption{Foreground galaxy property distributions for \mstar\ (top left), SFR (top right), sSFR (bottom left), and $E(B-V)$ (bottom right). The yellow histogram shows the properties over the full range of \dtran probed in this sample, while the red outline, blue outline, and solid gray histogram correspond to samples drawn from $\dtran =8-70$ kpc, $70-100$ kpc, and $100-250$ kpc, respectively. 
    The number of pairs in the solid histograms is shown on the left vertical axis, while the numbers in the dashed histograms (with considerably fewer pairs but a similar overall distribution) are shown on the right vertical axis.}

    \label{fig:sedprops}
\end{figure}

To understand the bulk properties of gas in galaxies' circumgalactic media, we assembled a sample of galaxy pairs from the KCWI and LRIS subsamples. A galaxy pair consists of a foreground and background galaxy (at redshifts \zfg and \zbg respectively) with a projected separation on the sky satisfying the following requirements:
\begin{itemize}

    \item $\Delta \theta < 30\arcsec$ so that the background galaxy probes the foreground galaxy's CGM within 250 kpc. %

    \item $\Delta v = \frac{c\Delta z}{1+\zbg} \gtrsim$ 3,000 \kms, where $\Delta z = z_\mathrm{bg} - z_\mathrm{fg}$ to ensure the foreground and background galaxies are not part of the same large-scale structure (i.e., they are causally disconnected from each other). For the average foreground redshift $\langle \zfg\rangle=2.3$, $\Delta z \gtrsim 0.033$ \added{or 12.9 physical Mpc along the line of sight assuming pure Hubble flow}. We do not impose an upper bound on $\Delta v$ (see Appendix \ref{app:wavetrim}).

    \item $\zfg > 1.4$, such that at least \civ falls at $\lambda > 3700$ \AA, where both LRIS-B and KCWI spectra have relatively low noise. 

    \item The background galaxy has been observed with KCWI or LRIS-B with the 600 line mm$^{-1}$ grism. 
    
    \item The background object in a pair is not a QSO. The rationale behind this decision is discussed in Section \ref{sec:intro} and Appendix \ref{app:hires}: in short, we are using only spatially-resolved background sources because they average over (foreground galaxy) CGM sub-structure on projected scales of $2-4$ kpc, and so are physically distinct from QSO measurements, which sample a region several thousand times smaller (see \citealt{steidel_structure_2010} \added{and \citealt{rubin_galaxies_2018,rubin_galaxies_2018-1} for discussions} of this point).

    \item The foreground object in a foreground-background pair is not classified as an AGN (type 1 or type 2). Although such measurements would be interesting, here we focus solely on star-forming galaxies without evidence for current AGN activity. 

    \item For the purpose of binning the sample by galaxy properties (Section \ref{sec:bin}), we require the foreground galaxy to have an SED fit that allows for a constraint on its stellar mass.

\end{itemize}

Since we are interested in using background galaxy spectra to measure foreground absorption, a well-determined foreground galaxy redshift is essential to ensure accurate velocity measurements in the shifted spectra. To that end, where possible, nebular redshifts from KBSS-MOSFIRE \citep{steidel_strong_2014,strom_nebular_2017,strom_measuring_2018} were used to determine galaxy systemic velocities. The KBSS-MOSFIRE nebular redshifts have systemic redshift uncertainties $\Delta v_\mathrm{MOSFIRE}\simeq 18$ \kms\ \citep{steidel_strong_2014}, considerably higher velocity accuracy than either LRIS ($\sigma_\mathrm{LRIS}\simeq 107$ \kms) or KCWI ($\sigma_\mathrm{KCWI}\simeq 71$ \kms). When nebular emission line measurements were not available, rest-frame UV emission (i.e.\ \lya) and low ionization interstellar (LIS) absorption features (e.g.\ \siii, \cii) were used to calculate the systemic redshift, using the redshift calibration rules outlined by \citet{chen_keck_2020}; in such cases the typical redshift accuracy is $\sigma_\mathrm{UV}\simeq 100$ \kms\ \citep{steidel_keck_2018}.

To assemble the sample, we started by constructing galaxy pairs for all known foreground galaxies in KBSS where the corresponding background galaxy had been observed with KCWI; 681 pairs with a KCWI-observed background galaxy satisfied the criteria above, 341 ($51\%$) of which had a foreground galaxy with a nebular redshift. The number of KCWI pairs reaches a peak count at $\dtran \simeq 50$ kpc and plateaus for $\dtran > 100$ kpc (see the blue dashed histogram outline in the left panel of Figure \ref{fig:hist}). The lack of pairs with $\dtran > 50$ kpc is an artifact of the KCWI observing strategy:\ 252/681 (37\%) of the KCWI galaxy pairs have their foreground and background galaxies on the same KCWI pointing (which are $\approx 170$ kpc in diameter). The remaining 429 (63\%) pairs involve a KCWI background galaxy and a foreground galaxy that falls off the KCWI footprint but still probes gas (at \zfg) in the KCWI background galaxy spectrum.

To supplement the number of pairs sampling larger \dtran, we then formed pairs where the background galaxy was one observed with LRIS-B (the blue channel of LRIS). This added 1748 galaxy foreground/background pairs from the general LRIS sample ($t_\mathrm{int}\simeq 1.5-4.5$ hours; cyan solid histogram in the left panel of Figure \ref{fig:hist}) and 309 pairs from LM1 masks ($t_\mathrm{int}\simeq 10-12$ hours; red histogram outline in Figure \ref{fig:hist}) to the initial KCWI sample, mostly sampling $100\lesssim \dtran/\mathrm{kpc} \lesssim 250$; 879 of the 1748 (50\%) LRIS pairs and 186/309 (60\%) of the LM1 pairs had foreground galaxies with nebular redshifts. In the end, 2738 pairs (1406 [51\%] with foreground nebular redshifts) sampling $8\leq \dtran/\mathrm{kpc} \leq 250$ ($\Delta\theta\sim1\arcsec - 30\arcsec$) met all of our selection criteria. The final sample of galaxy pairs is summarized in Figure \ref{fig:hist}. The foreground galaxy stellar mass (\mstar), star formation rate (SFR), specific star formation rate ($\operatorname{sSFR}\equiv \operatorname{SFR}/\mstar$), and the inferred reddening $E(B-V)$ distributions are shown in Figure \ref{fig:sedprops}. A thorough description of the SED fitting underlying these measurements can be found in \citet{steidel_reconciling_2016,steidel_keck_2018} and \citet{theios_dust_2019}. We will return to the stellar population parameters of the foreground galaxies in Section \ref{sec:bin}.

\section{Stacking}
\label{sec:stacking}

\added{Although some background galaxy spectra show clear evidence of intervening absorption at the redshift of the foreground galaxy (see Figure \ref{fig:indobjs} for some examples), most of the background spectra do not have sufficient $S/N$ to detect weak absorption ines against the faint continuum. Stacking (in bins of \dtran) increases the effective $S/N$ and facilitates the detection of weaker (metallic) absorption features. In doing so, however, each constituent background spectrum contributes absorption at \zbg, \zfg, and possibly additional ``contaminating'' redshifts (especially in the \lya forest) to the stack. 
Averaging over a sufficiently large number of different (\zfg, \zbg) pairs  (see Appendix \ref{app:stackmethod}) can mitigate the effects of contamination because the locations of the discrete contaminating absorption lines from redshifts other than \zfg will be ``randomized'' with respect to the features of interest at \zfg.  It is useful to keep in mind that what one is trying to measure from the stacked background spectrum (redshifted to $\zfg$) is the {\it excess} absorption signal from a particular transition at a particular redshift (\zfg) compared to the mean absorption from all lines at other redshifts that happen by chance to coincide in wavelength with the feature of interest.  In the limit of a very large number of (\zfg, \zbg) pairs, the contamination will eventually manifest as a local continuum that is lowered relative to the original by an amount that depends on both rest wavelength and the particular ensemble of (\zfg, \zbg) pairs being combined (see, e.g., Appendix~\ref{app:hires}). For stacks based on the combination of a finite number of (\zfg, \zbg) pairs, a significant source of noise in the measurements of interest arises from fluctuations in the local (mean) continuum when the absorption line ``background'' has not converged to a smooth distribution.
These effects are exacerbated by the fact that there is a wide range of S/N among the spectra contributing to a given wavelength pixel in the stacked spectrum, and the number of individual spectra contributing to each wavelength pixel varies according to the distribution of \zfg, \zbg, and $\Delta z$ (the difference between them). }

Before stacking, each background galaxy spectrum was trimmed in wavelength. For KCWI pairs, background spectra were restricted to observed wavelengths where all 24 slices were sensitive i.e., $3530-5530$ \AA\ for the KCWI configuration we used. For the LRIS spectra, similar wavelength cuts were imposed both in the short and long wavelength ends of the blue channel data. The \lya forest is primarily responsible for noise at the short wavelength end, while the LRIS dichroic cutoff at $\sim 5600$ \AA\ and the sky subtraction residuals around [\ion{O}{1}] $\lambda5577$ occasionally introduce undulations in the red end of the blue channel. As for the KCWI spectra, $3500-5555$ \AA\ turned out to be a good fiducial wavelength to mitigate both effects. In addition to the wavelength endpoints, spectral regions near \lya in the background galaxy spectrum were masked to avoid contamination in the final stack.\footnote{We experimented with masking other far-UV absorption features (e.g.\ \civ, \siii) but found that additional wavelength masks added more discontinuities to the constituent spectra and resulted in overall noisier stacks.}

In order to give each background spectrum an approximately equal weight in the composite, each spectrum was normalized with a cubic spline. The spline knots used to fit the background spectrum were chosen via a sparsely sampled median, ignoring wavelength regions near resonance lines. In the \lya forest, we found points 1$\sigma$ above the local ($\Delta \lambda\approx 50$ \AA) median reliably traced the suppressed continuum. The KCWI and LM1 observations have associated error spectra; for the LRIS galaxies where error spectra do not always exist, we use the local standard deviation of the spectrum to construct a coarse error spectrum. \added{We impose a $S/N \gtrsim 1$ cut on all constituent background spectra, i.e.\ only spectral regions where the cubic spline fit is greater than the error spectrum contribute to the stack.} Background spectra where less than 15\% of the bandpass had continuum $S/N \gtrsim 1$ were excluded entirely to avoid adding patches of sporadic noise to the stack.

In stacking, each normalized background galaxy spectrum was shifted to the rest frame of its corresponding foreground galaxy 
and resampled onto a common wavelength grid with 0.2 \AA\ sampling using linear interpolation. For a given range of \dtran, the associated background spectra were combined to make a final stack using a trimmed mean where spectral pixels above (below) the 95th (5th) percentile were rejected. For the stacks presented below, this resulted in 4 -- 6 objects being excluded, comparable to a $\sim 2\sigma$ clipped average, where $\sigma$ is defined based on the variance of the input spectra at a given wavelength. \added{We experimented with various methods for combining the spectra and several measures of central tendency before deciding on the trimmed mean}; further details on the stacking method \added{and the results of these tests} are described in Appendix \ref{app:stackmethod}. Even with stacks of coarsely normalized spectra, the fact that each background spectrum has been shifted to an unrelated foreground redshift necessitates a manual
adjustment to the continuum fit of the composite in order to measure accurate absorption profiles (discussed further in Appendix \ref{app:hires}).

As highlighted above, error spectra for KCWI and LM1 masks exist and can be propagated through to the stacks; however, we adopt a bootstrap approach since it includes sample variance and reveals how much individual spectra contribute to the absorption observed in the combined spectrum. For each background stack, we ran 100 bootstrap realizations with randomly selected galaxy pairs (with replacement) from that bin's range of \dtran. The error spectrum is then the standard deviation of the 100 bootstrap stacks. As expected, the error spectra qualitatively resemble those propagated from each spectrum's respective noise estimated by the DRP, but there tends to be a spike in variability around strong absorption features due to the variance in foreground absorption between galaxies. 

Foreground galaxy spectra within each sub-sample (\dtran bin) were used to create a composite ``down the barrel'' average foreground galaxy spectrum for each. The individual spectra were combined in a similar fashion to their background galaxy counterparts, with a few notable exceptions. To ensure every foreground galaxy had a roughly equal contribution to the stack, each spectrum was approximately normalized based on its median continuum ($1410-1510$ \AA) flux density before being combined. Each spectrum with a continuum $S/N\gtrsim 1$ was then shifted to its rest frame and galaxies within the 5th -- 95th percentiles (as a function of wavelength) were averaged. 
The final stack was continuum-normalized by fitting a BPASSv2.2 \citep{stanway_re-evaluating_2018} constant star formation rate stellar population synthesis model with age $10^8$ years, metallicity $Z=0.14\,Z_\odot$, and including the effect of binary evolution of massive stars; an SMC extinction curve\footnote{$A_V=2.74\, E(B-V)$ for the SMC extinction curve.} was assumed. More details on the choice of model and the fitting procedure can be found in \citet{steidel_reconciling_2016,steidel_keck_2018} and \citet{theios_dust_2019}.

\begin{figure*}[htbp!]
    \plotone{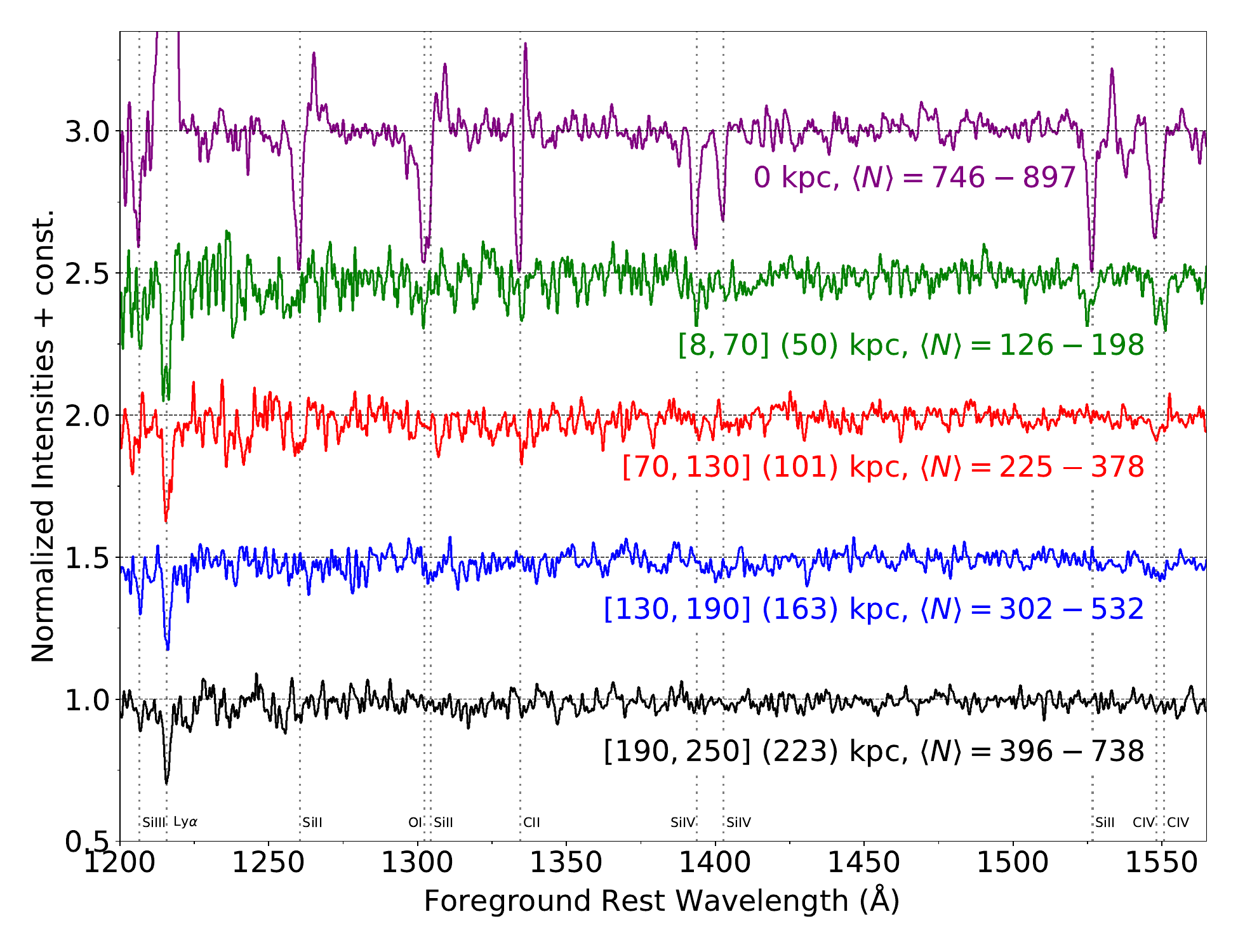}
    \caption{Normalized composite spectra of the foreground galaxies in this sample when viewed down the barrel (top, purple) plus four \dtran bins ranging from 8 -- 250 kpc utilizing background galaxies as a continuum source. The spectra are offset by 0.5 from one another for ease of visualization. Each spectrum's minimum, maximum, and median (shown in parentheses) sampled \dtran is labeled in the plot along with the number of pairs contributing to each stack. Note that the number of pairs varies as a function of wavelength with fewer pairs contributing at shorter wavelengths. The two counts specify the number of pairs contributing at \lya and \civ respectively. Typical far-UV resonant transitions are marked with vertical dotted lines and labeled at the bottom of the plot.}
    \label{fig:fullstack}
\end{figure*}

Stacks from several \dtran bins are shown in Figure \ref{fig:fullstack}, offset from one another for clarity. 
We observe prominent \lya and \civ over the widest range of impact parameters, plus \siiii, \siiv, \siiiwl, and \siiiwh primarily in the small impact parameter bin ($\langle\dtran\rangle=50$ kpc; green spectrum). The number of objects contributing to the final stack varies by a factor of $\sim 2$, largely due to the underlying redshift distribution ($1.4 \lesssim \zfg \lesssim 3.2$; see the right panel of Figure \ref{fig:hist}) and the short wavelength UV cutoff of the spectra ($\sim 3500$ \AA). As one moves to higher \zfg, correspondingly higher \zbg objects are required to measure shorter rest-wavelength transitions (e.g.\ \siiiiw) which invariably fall in the \lya forest of the background spectrum. The noise level is also somewhat higher at the short wavelength end of the spectrum as compared with a region around e.g.\ 1500 \AA\ because many of the spectra cut off at shorter wavelengths and fewer pairs end up being averaged. There is also some small scale variability in the number of spectra contributing at a particular wavelength that arises from trimming and/or masking of particular spectral features at the redshift of the background galaxies. 

\section{Absorption Trends from Composite Spectra}
\label{sec:metalabs}
With a method for stacking galaxy pairs in hand, we proceeded to construct a suite of composite spectra for \dtran between 8 kpc (the minimum pair separation) and 250 kpc, with the goal of sampling the changing absorption profiles over increasing galactocentric radius. Bins sampling projected distances $\dtran \lesssim 100$ kpc are typically 30 kpc wide and are spaced in 10 kpc intervals (i.e. 8 -- 40 kpc\footnote{All but one of the (\zfg,\zbg) pairs have $\dtran > 10$ kpc.}, 20 -- 50 kpc, \ldots, 100 -- 130 kpc). The number of pairs contributing to a particular stack ranges from $\sim 50$ at $\dtran \simeq 30$ kpc to $\sim 200$ at $\dtran \simeq 100$ kpc which ensures any supposed absorption is real and easily discernible (Appendix \ref{app:stackmethod}). As an illustration, \lya and \civ absorption in several such stacks is shown in Figure \ref{fig:lyaciv}; we will revisit these spectra in Section \ref{sec:lyaciv}. These bins are not unique however; choosing different bin endpoints and/or spacing does not significantly alter the results presented below because the underlying distribution being sampled is the same.

Since the number of pairs increases with \dtran (Figure \ref{fig:hist}), it becomes possible to sample smaller impact parameter intervals while maintaining a comparable continuum $S/N$. On the other hand, the absorption features become weaker and less variable outside of the virial radius ($R_\mathrm{vir}\simeq 90$ kpc for a typical halo in our sample). As a compromise, for $100 \lesssim \dtran/\mathrm{kpc} < 250$ we maintain 30 kpc bins but increase the spacing to 15 kpc (i.e.\ 115 -- 145 kpc, 130 -- 160 kpc, \ldots, 220 -- 250 kpc). The 10--15 kpc spacing between the bins means that every other or every third bin contains a unique set of background/foreground pairs or, depending on \dtran, at least $\sim 30-50\%$ of the pairs differ between adjacent bins. The benefit to this ``moving average" type sampling is a reduced susceptibility to outlier spectra and an increased resolution at small \dtran. We use these bins for the majority of the analysis with the full pair sample and specify cases for which the binning differs.

\subsection{Quantitative measurements of absorption features}

\added{Absorption features in the stacked composite spectra represent averages over many lines of sight through the CGM of foreground galaxies, each of which is already a spatial average over several kpc as discussed above, so that we do not attempt to measure column densities or Doppler parameters as one might using high resolution QSO spectra. The kinematics of the absorption line profiles convey information about the average velocity distribution (relative to \zfg) of gas giving rise to a given transition, while the relative depth of the absorption as a function of velocity is proportional to the fraction of the area of the beam footprint with significant optical depth at a particular LOS velocity in the transition; both are affected by the effective resolving power of the composite spectra.} 

\begin{figure*}[thbp!]
\epsscale{0.9}
          \plotone{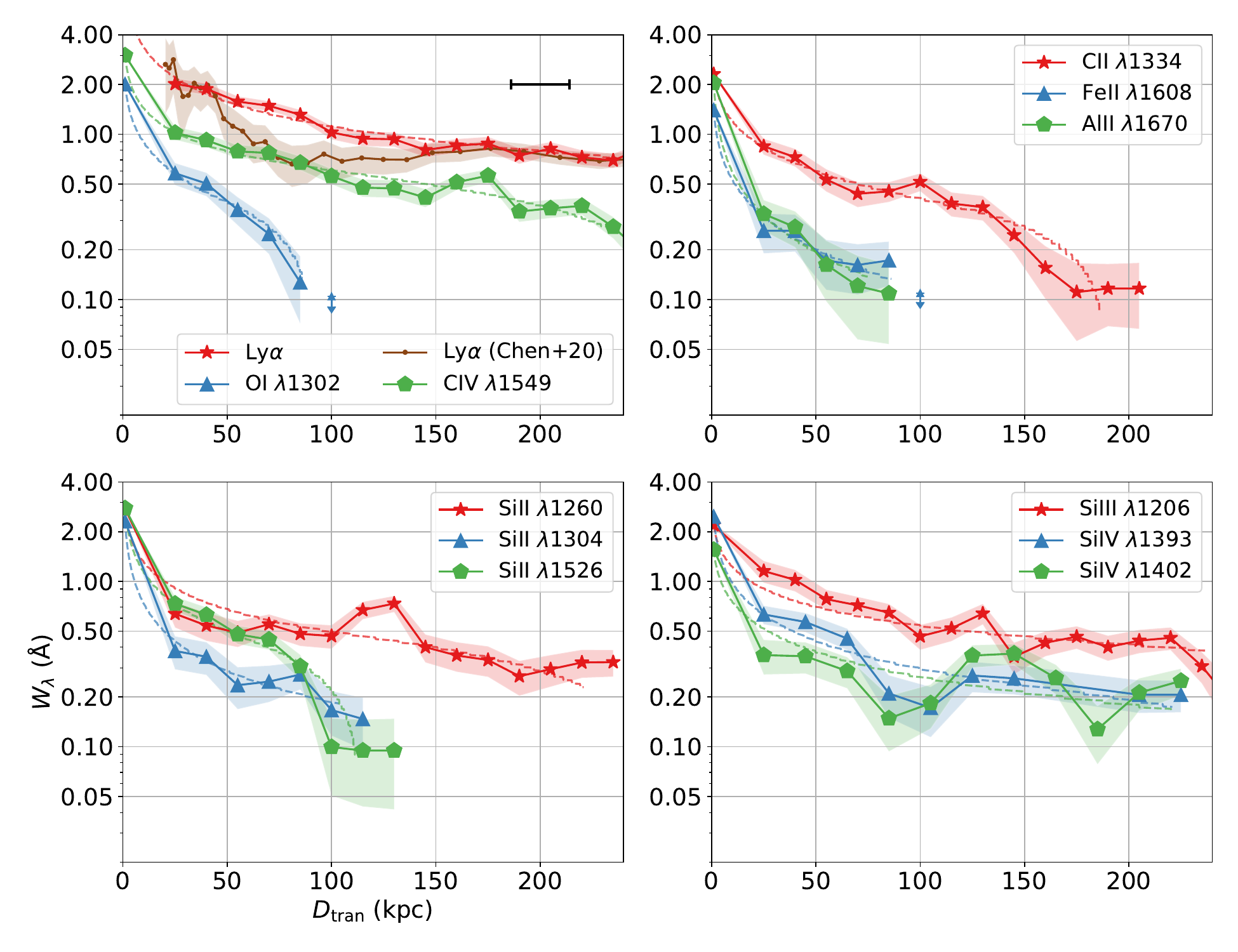}
        \caption{Absorption line \ew vs.\ \dtran for several transitions (labeled in the plot legends). The green \ionl{C}{4}{1549} points show the total equivalent width integrating over the entire feature. Color coded bands show \ew uncertainties and \added{dashed lines are predicted \ew from the model described in Section~\ref{sec:model} (best fit parameters are listed in Table \ref{tab:ewfits})}. In the top left plot, the horizontal error bar shows typical bin sizes for all the transitions shown. The \citet{chen_keck_2020} $W_\lambda(\lya)$ measurements (with uncertainties) are shown in brown.}
    \label{fig:ew_b}
\end{figure*}

To quantify the absorption features in a model-independent way, for each transition in the composite spectra we measured the rest-frame equivalent width $W_\lambda$, the velocity centroid  (first moment) of the absorption feature relative to the systemic velocity of the foreground galaxy  (\dv), and the second moment ($\sigma$) as a proxy for the velocity dispersion. For the DTB composite, each quantity was computed via direct integration over the region where the flux was below the continuum.\footnote{At this spectral resolution, the DTB $\sigma(\text{\cii})$ measurement is contaminated by the presence of the associated nonresonant \ion{C}{2}* $\lambda 1335.62$ emission line which ``fills in'' the red side of the absorption feature. To compensate, we fit a double Gaussian to both features and define $\sigma(\text{\cii})$ as the standard deviation of the Gaussian absorption component.} The $\dtran \ge 8$ kpc composites have comparatively lower continuum $S/N$ so each metric was instead computed over the velocity range $|v| \leq 700$ \kms.\footnote{Given their close proximity, the integration region for \oiw and \siiiwm was truncated at $\lambda=1303.3$ \AA, the midpoint between the two features. Note that measuring the $\dtran > 0$ kpc composites in the same manner as the DTB spectra increases the overall noise but does not significantly alter the results presented below.} Since the background galaxy spectra ($\dtran \ge 8$ kpc) measure absorption throughout the foreground galaxy halo and DTB spectra probe only gas ``in front" of the galaxy along our line of sight, we corrected the DTB \ew and $\sigma$ by multiplying the observed DTB quantities by a factor of two (Table \ref{tab:ewfits}) to approximate the total absorption and velocity spread that would be observed along a line of sight at $\dtran=0$ kpc. Similar to the background error spectrum calculation, we ran 2000 Monte Carlo resamples of the stacked spectrum perturbed by a Gaussian whose width was equal to the pixel noise level at each wavelength. Uncertainties on \ew, \dv, and $\sigma$ were determined from the standard deviation of the resulting parameter distributions. Absorption line \ew, $\sigma$, all as functions of \dtran, are shown in Figures~\ref{fig:ew_b} and \ref{fig:sigma_b}
, respectively. These figures comprise the principal results of this study, and we will discuss them further in Sections \ref{sec:lyaciv} and \ref{sec:lowions}. 

We note that a few of the low ionization lines are detected in only a subset of the \dtran bins, leading to some non-uniform sampling. In general, if low ions or \lya are detected in a stack, then \civ tends to be as well simply because there is a factor of $\simeq 2$ increase in the number of pairs contributing to a stack between \lya and \civ (recall the $S/N$ is roughly proportional to the number of pairs being averaged). 
The net result is that \lya and \civ have the densest sampling in \dtran, especially at $\dtran\lesssim 100$ kpc since bins where the low ionization features are detected can almost always be reused to measure \lya or \civ.

\subsection{Instrumental contributions to observed kinematics}
\label{sec:sigma_inst}

\edit1{In order to evaluate the significance of the observed kinematics of absorption features in the stacked spectra, we first evaluate the contributions of the instrumental resolving power and redshift uncertainties to observed kinematic profiles.} As discussed in Section \ref{sec:paircons}, $1406/2738$ ($f_\mathrm{neb} = 51\%$) of the foreground galaxies among the foreground/background galaxy pairs have nebular redshift measurements from KBSS-MOSFIRE, which have very small redshift uncertainties  $\sigma_\mathrm{neb} \simeq 18$ \kms, or $\Delta z \lesssim 0.0002$ at $z \sim 2$\ \citep[see][]{steidel_strong_2014}; the remaining ($f_\mathrm{UV} = 49\%$) have uncertainties $\sigma_\mathrm{UV}\simeq 100$ \kms\ \citep[see][]{steidel_keck_2018,chen_keck_2020}. Combining the two velocity errors in quadrature weighted by their fractional contribution i.e.\ 
\begin{equation}
\label{eqn:veff}
    \delta v_\mathrm{sys} = \sqrt{f_\mathrm{neb}\sigma_\mathrm{neb}^2+f_\mathrm{UV}\sigma_\mathrm{UV}^2}
\end{equation}
implies the full galaxy pair sample has a systemic velocity uncertainty of $\delta v_\mathrm{sys}\approx 71$ \kms. 

The spectral resolving power for individual background galaxy spectra depends on the instrument used to obtain them: for spectra obtained with LRIS-B using the 600 line mm$^{-1}$ grism, we adopt $\sigma_\mathrm{LRIS}=107$ \kms\ ($\langle R\rangle\simeq 1200$; \citealt{chen_keck_2020}), and for KCWI we use $\sigma_\mathrm{KCWI}=71$ \kms, consistent with a spectral resolution $\langle R \rangle \simeq 1800$.

In a similar fashion to Equation \ref{eqn:veff},
\begin{equation}
\label{eqn:Reff}
    \sigma_\mathrm{comb} = \sqrt{f_\mathrm{LRIS}\sigma_\mathrm{LRIS}^2+f_\mathrm{KCWI}\sigma_\mathrm{KCWI}^2}
\end{equation}
where $f_\mathrm{LRIS}$ and $f_\mathrm{KCWI}$ are the fractional contributions of background spectra from each instrument to the total stack and $\sigma_\mathrm{inst}$ are the respective spectral resolving power.
For the full $\dtran < 250$ kpc sample, $\sigma_\mathrm{comb} \approx 99$ \kms, and for $\dtran<100$ kpc, $\sigma_\mathrm{comb} \approx 90$ \kms.

Combining in quadrature the two sources of line broadening affecting the composite (stacked) spectra -- i.e, foreground galaxy systemic redshift uncertainties and spectral resolution of the background galaxy spectra -- we find \mbox{$\sigma_\mathrm{tot} = \sqrt{\delta v_\mathrm{sys}^2+ \sigma_\mathrm{comb}^2} \approx 122$ \kms}\ for the full sample and $\sigma_\mathrm{tot}\approx 121$ \kms\ at $\dtran < 100$ kpc. 

\begin{figure*}[htbp!]
\epsscale{0.9}
\plotone{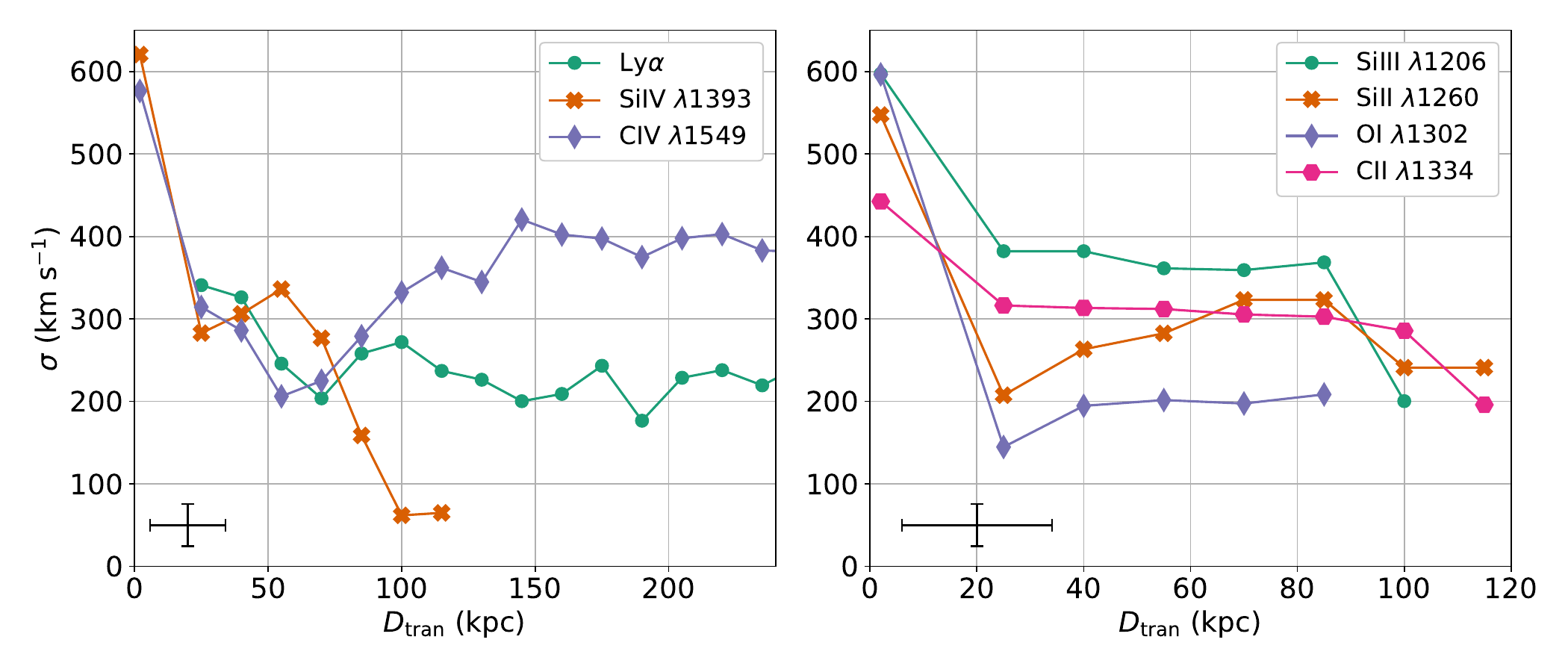}
        \caption{Second moment of the normalized absorption profiles ($\sigma$) vs.\ projected distance (\dtran) for transitions color coded in the legends of each panel. As for Figure \ref{fig:ew_b}, the \ionl{C}{4}{1549} (thin purple diamonds) measurements are obtained by integrating over the entire doublet. The values of $\sigma$ have been corrected for the instrumental contribution as discussed in Section~\ref{sec:sigma_inst}. The black error bar in the lower lefthand corner of each panel shows typical bin sizes (horizontal) and average bootstrap uncertainties on the second moment (vertical) for each transition.}
    \label{fig:sigma_b}
\end{figure*}
We adopt $\sigma_{\rm tot}=122$ \kms\ as the effective instrumental resolution for all composite spectra, meaning that the velocity width of any measured (single line) spectral feature must be corrected by subtracting this value in quadrature, and that real spectral features should have second moments larger than $\sigma_{\rm tot}$. In the case of the \ion{C}{4} doublet, which is usually blended in composite spectra, the effective instrumental contribution to a measured second moment for the full absorption feature (assuming each component is broadened by $\sigma_{\rm tot} = 122$ \kms) is $\sigma_{\rm tot,CIV} \sim 260$ \kms. 

We have subtracted (in quadrature) the appropriate instrumental contribution $\sigma_{\rm tot}$ (or $\sigma_{\rm tot,CIV}$) from all second moment measurements presented (Figure~\ref{fig:sigma_b}) to facilitate comparisons between velocity dispersions from different transitions.

\subsection{Ly$\alpha$ and \civ}
\label{sec:lyaciv}

\begin{figure*}[htbp!]
\epsscale{0.8}
\plottwo{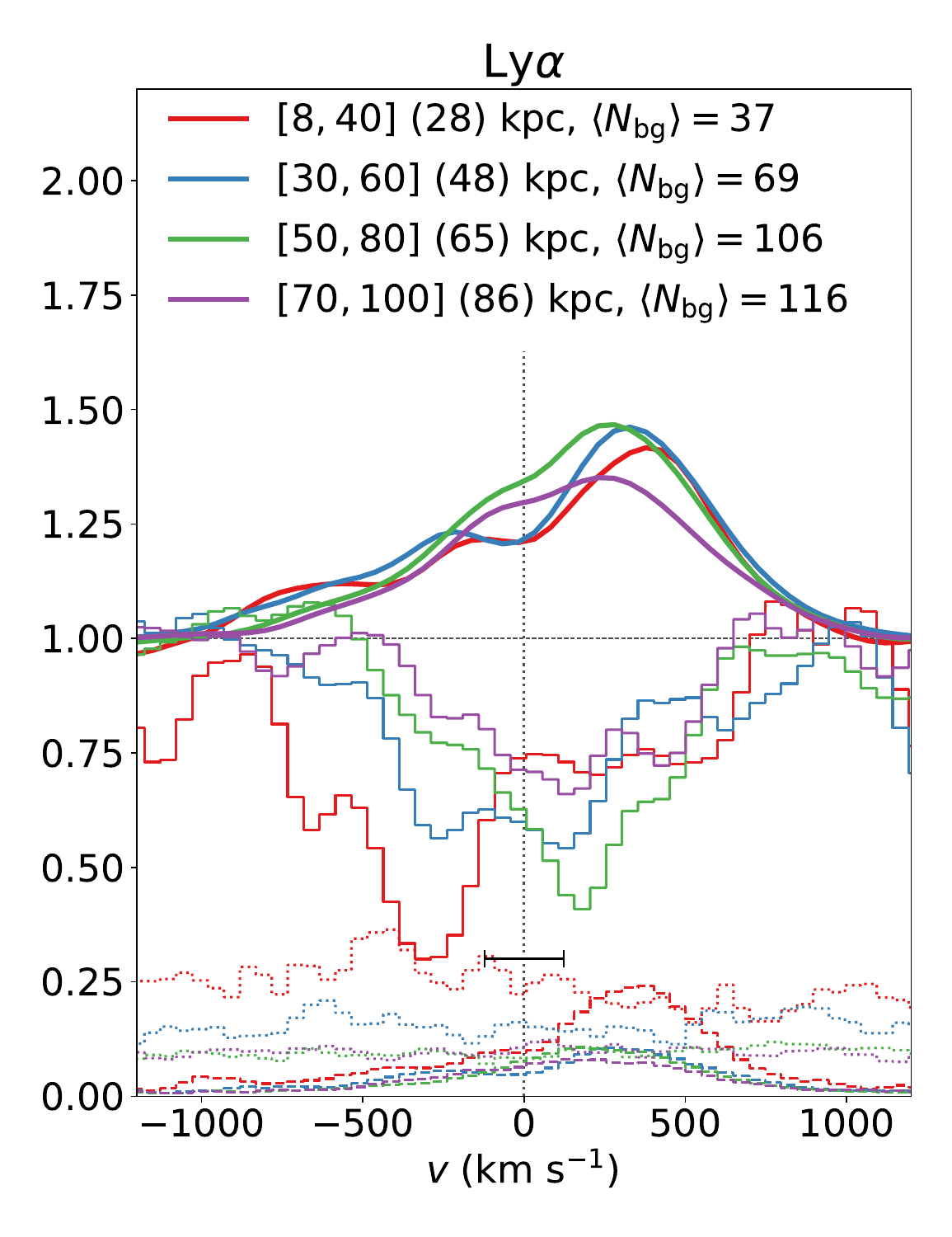}{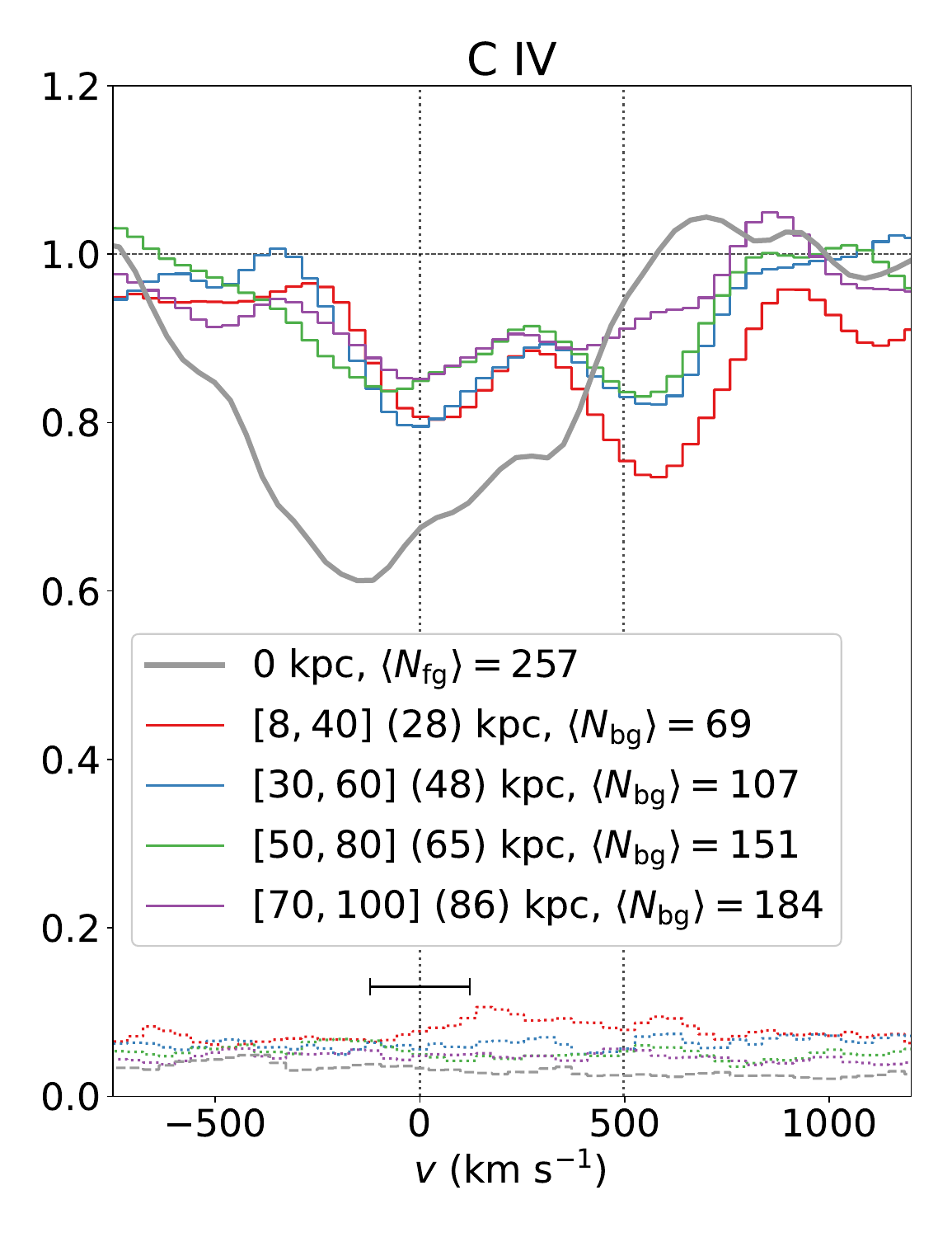}
        \caption{\textit{Left:} Normalized and scaled \lya emission (thick line style; $\dtran = 0$ kpc) and absorption (thin histogram line style; $\dtran > 0$ kpc) profiles plotted for several impact parameter bins. For a given impact parameter range, DTB foreground galaxy spectra were averaged to produce an emission spectrum while background spectra were similarly averaged to measure the absorption induced by the foreground galaxy's CGM. 
        The range of \dtran and the median are shown for each bin. Note that the \lya emission is scaled down by a factor of 15 and translated vertically by $+1$ for display purposes. \textit{Right:} \civ absorption profiles for the same bins in \dtran (with the same color scheme) as the left panel. The DTB (thick gray) spectrum includes all foreground galaxies in pairs with $\dtran < 100$ kpc. In both panels, the number of pairs contributing to each stack at the wavelength of interest is shown in the plot legends; ``fg" subscripts correspond to DTB spectra, while ``bg" are measurements utilizing background spectra. The vertical dashed lines show $v_\mathrm{sys}=0$ \kms\ for each transition and the black error bar shows the effective resolution element of the stack. Note that \civh is shown at $v=497$ \kms\ on the right hand panel. Normalized background (DTB) error spectra from bootstrap resampling are plotted in a dotted (dashed) line style and are color coded to match the stacked spectra.}
        \label{fig:lyaciv}
\end{figure*}

\lya and \civ (in that order) are by far the strongest absorption features detected in the composite spectra of background galaxies. We therefore present these results before moving onto the low ions. Starting in one dimension, some representative stacks are presented in Figure \ref{fig:lyaciv} that show how \lya (left) and \civ (right) change with \dtran. The same \dtran bins are shown for ease of comparison, although as can be seen in, e.g., Figure \ref{fig:fullstack}, excess \lya absorption extends well beyond $\dtran = 100$ kpc. For \lya and \civ, absorption starts out both stronger and more extended in velocity close to the center of the galaxy and the profiles become shallower and narrower with increasing \dtran. Between $\dtran \simeq 30$ kpc and $\dtran \simeq 100$ kpc, $W_\lambda(\text{\lya})$ drops by a factor of $\simeq 2$ from $2$ \AA\ to $1$ \AA\ (Figure \ref{fig:ew_b}) and $\sigma(\text{\lya})$ decreases from $340$ \kms\ to $270$ \kms; a $21\%$ reduction (Figure \ref{fig:sigma_b}). In the same range of projected distances, $W_\lambda(\text{\civ})$ drops by a factor of $\simeq 2$ from $1.0$ \AA\ to $0.5$ \AA\ and $\sigma(\text{\civ})$ broadly follows $\sigma(\lya)$ with a velocity spread $\simeq 300$ \kms.
In this case, $W_\lambda(\text{\civ})$ and $\sigma(\text{\civ})$ were measured via direct integration over the full doublet, without distinguishing between the individual transitions, and $\sigma(\text{\civ})$ were corrected as described in Section~\ref{sec:sigma_inst}.
\begin{figure}[htbp!]
    \epsscale{1}
    \plotone{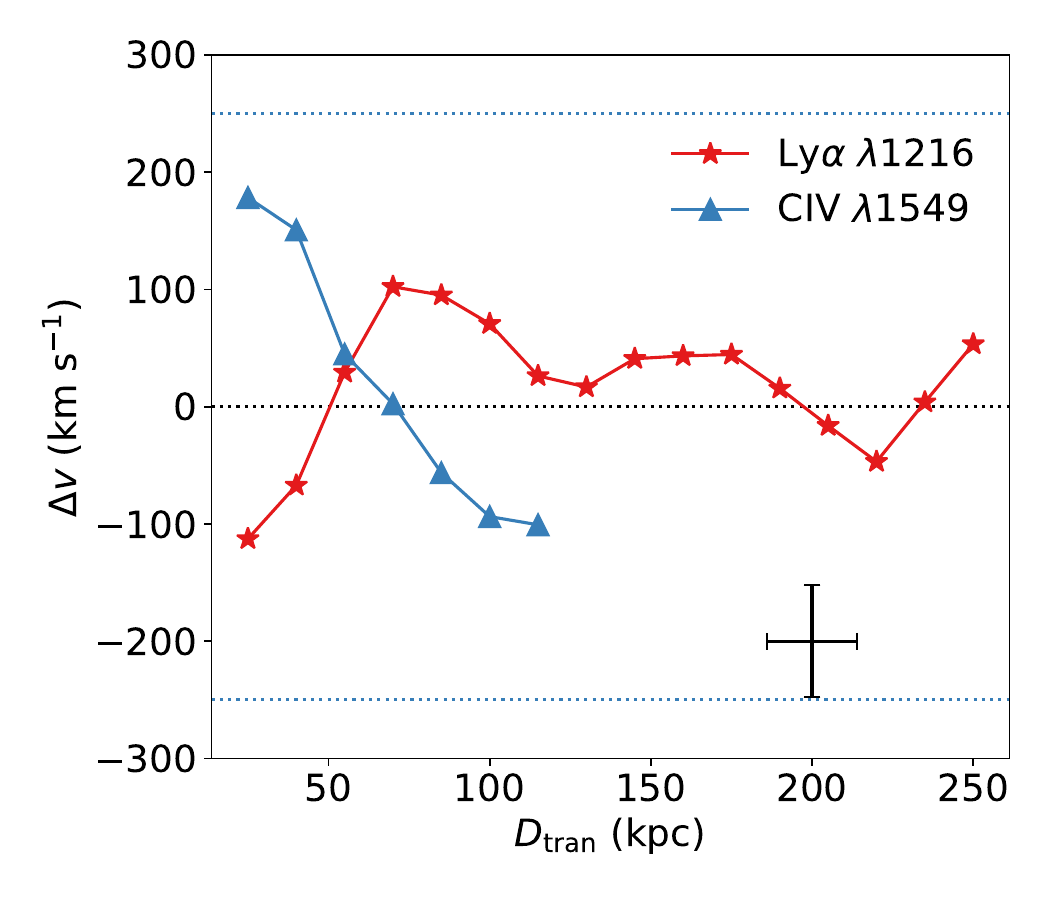}
    \caption{\lya (red) and \civ (blue) velocity centroids ($\dv$) as a function of \dtran. The \civ centroid was computed by integrating over the entire doublet and $\dvciv=0$ \kms\ corresponds to the midpoint between the two \civ transitions ($\lambda \approx 1549.480$ \AA). The  dotted blue horizontal lines at $\Delta v = -250,\, +250$ \kms\ correspond to \civl and \civh relative the fiducial velocity. The black error bar shows the average bin width and bootstrap error on the velocity centroid.}
    \label{fig:dv}
\end{figure}

For \lya in the left panel of Figure \ref{fig:lyaciv} (see also Figure \ref{fig:dv}), we observe a shifting velocity centroid with \dtran: at $\dtran \lesssim 50$ kpc, $\dvla<0$ \kms; between $\dtran \sim 50-125$ kpc, 
$\dvla \simeq 100$ \kms, after which point \dvla\ is consistent with zero. We note that the relative scarcity of galaxy pairs within $\simeq 100$ kpc results in noisier stacks compared with those at $\langle \dtran\rangle \gtrsim 100$ kpc where more lines of sight are probed. This can be seen by comparing the dotted red and dotted purple error spectra; the former has fewer (37) pairs contributing than the latter (116) and the corresponding bootstrap uncertainty is increased by a factor of $\simeq 2$. Despite the larger uncertainty in absorption strength, the shifted trough persists across multiple independent bins. We discuss the net blueshifted velocity centroid at $\dtran \lesssim 50$ kpc further in Appendix \ref{app:filling}.

While both \lya and \civ present a similar trend in decreasing \ew and $\sigma$ with increasing \dtran, the shape of the \civ absorption troughs changes more significantly. In the $\dtran = 8-40$ kpc bin (right panel of Figure \ref{fig:lyaciv}), both \civl and \civh present strong absorption near the systemic velocity with an extended trough in velocity space out to $\vlos\simeq -300$ \kms\ in \civl, and $\vlos\simeq +300$ \kms\ in \civh. We note that $W_\lambda(1550)$ is at least a factor of two larger than $W_\lambda(1548)$ which would of course be unphysical if one were measuring a single absorbing complex. In Appendix \ref{app:hires}, we use stacked HIRES spectra of background QSOs to show how such foreground absorption profiles can arise. 
\added{At lower $S/N$, small numbers of galaxy-galaxy pairs with a variety of absorption line strengths and kinematic profiles (e.g.\ those with $\sigma(\text{\civ}) \gtrsim 500$ \kms) can result in net $W_\lambda(1548)/W_\lambda(1550) < 1$. At the same time, averages over small numbers of pairs are also more sensitive to contamination by unrelated absorption features from other redshifts.} 
Moving to larger \dtran, the subsequent bins in Figure \ref{fig:lyaciv} show a contracting velocity spread centered near zero velocity and an increasing $W_\lambda(1548)/W_\lambda(1550)$ ratio, approaching the DTB presentation. As an example, between $\dtran \simeq 28$ kpc and $\dtran \simeq 48$ kpc, $W_\lambda(1550)$ diminishes by $\sim 25\%$ and the velocity dispersion reduces slightly from 300 \kms\ to 230 \kms.

We used the $W_\lambda(1548)/W_\lambda(1550)$ ratio and \dvciv to estimate the change in $\tau(\text{\civ})$ with \dtran. 
Between $\dtran \simeq 30$ kpc and $\dtran \simeq 90$ kpc, the average $W_\lambda(1548)/W_\lambda(1550)$ ratio transitions from approximately one to two (Figure \ref{fig:lyaciv}). When $\sigma(\text{\civ})$ is large, a more robust approach to measuring the gas opacity involves the overall velocity centroid (\dvciv) integrating over the full doublet. Measuring the velocity offset from $\lambda_\mathrm{cent}=1549.480$ \AA\ (the average of the two \civ rest wavelengths), we would theoretically expect a velocity centroid of $\dv\simeq -90$ \kms\ for optically thin \civ absorption with $W_\lambda(1548)/W_\lambda(1550)=2$. Naturally, the velocity centroid would be centered at $\dv=0$ \kms\ for the optically thick doublet ($W_\lambda(1548)/W_\lambda(1550)=1$). 

Between $\dtran\simeq 30$ kpc and $\dtran\simeq 100$ kpc, \dvciv blueshifts by $\simeq 300$ \kms\ from $\dvciv \simeq +200$ \kms\ to $\dvciv \simeq -100$ \kms\ (Figure \ref{fig:dv}). The majority of the \civ bins show $\dv\lesssim -90$ \kms\ beyond $\dtran\simeq 60$ kpc and the average $W_\lambda(1548)/W_\lambda(1550)$ changes from a 1:1 to a 2:1 ratio, consistent with a transition from optically thick to optically thin gas with increasing \dtran. The gradient in line strength is largest at $\dtran\simeq 50$ kpc, but the noise in both the \dvciv and $W_\lambda(1548)/W_\lambda(1550)$ measurements suggests that some optically thick \civ may persist to $\dtran \simeq 90$ kpc.

Beyond $\dtran \simeq 100$ kpc, the \civ absorption signal becomes weaker and the marginal $S/N$ prevents an accurate determination of $W_\lambda(\text{\civ})$ ratios. Physically, the equivalent width ratio likely remains at or around two but the number of pairs in our sample is insufficient to distinguish \civh from the continuum. Indeed, the consistently negative \dvciv values at $\dtran \gtrsim 80$ kpc indicate that optically thin \civ is still present out to at least $\dtran \simeq 125$ kpc; beyond this projected distance, the $S/N$ is not sufficient to make a reliable velocity centroid measurement.

\subsection{Low Ions}
\label{sec:lowions}

\begin{figure*}[htbp!]
    \epsscale{1.0}
    \plotone{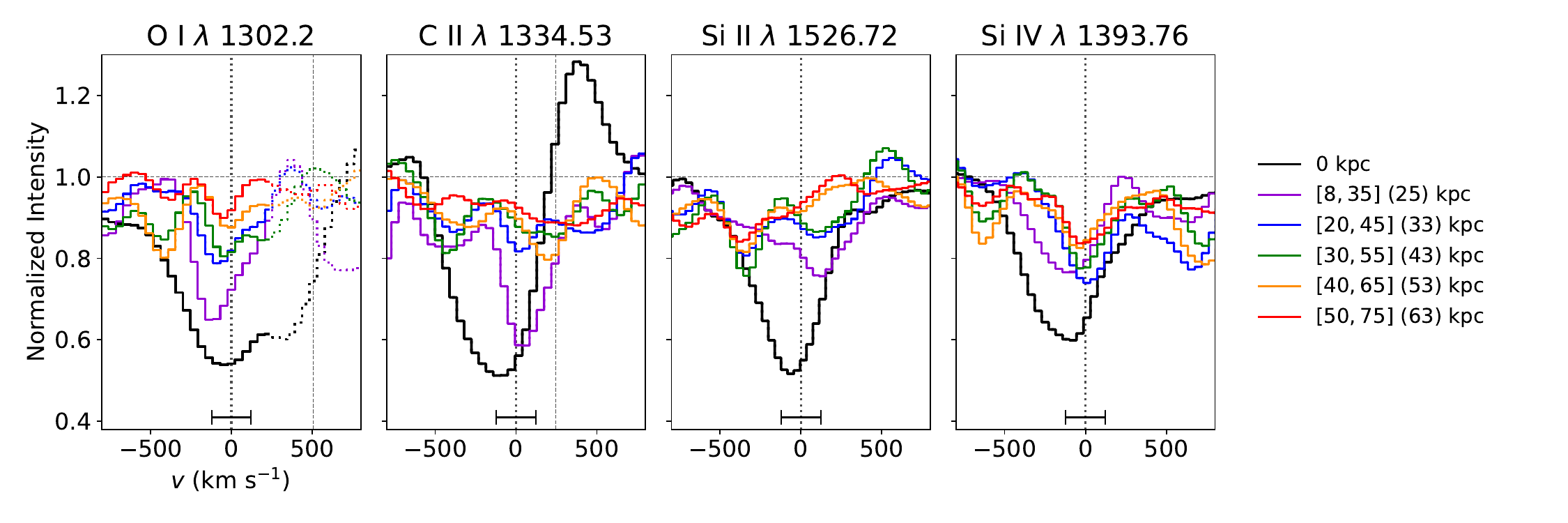}
     \caption{Normalized intensity profiles for several strong far-UV transitions, measured DTB (thick black) and at \dtran between 8 and 75 kpc (median \dtran shown in parentheses). The legend specifies the map between line color and range of \dtran. The plotted colors move from blue to red across the rainbow with increasing \dtran e.g., $\dtran\simeq 25$ kpc is blue and $\dtran\simeq 63$ kpc is red. Black vertical dotted lines show $v_\mathrm{sys}$ and dashed lines show nearby transitions where appropriate (\siiiwm and \ion{C}{2}* $\lambda 1335$). Regions of the spectrum contaminated by such nearby transitions are shown with a dotted line style. The black error bar shows the effective velocity resolution. This figure is in contrast to Figure \ref{fig:lowionsamev} which shows different transitions within the same bin.}

    \label{fig:linebins}
        
\end{figure*}

\begin{figure*}[htbp!]
    \epsscale{0.8}
    \plotone{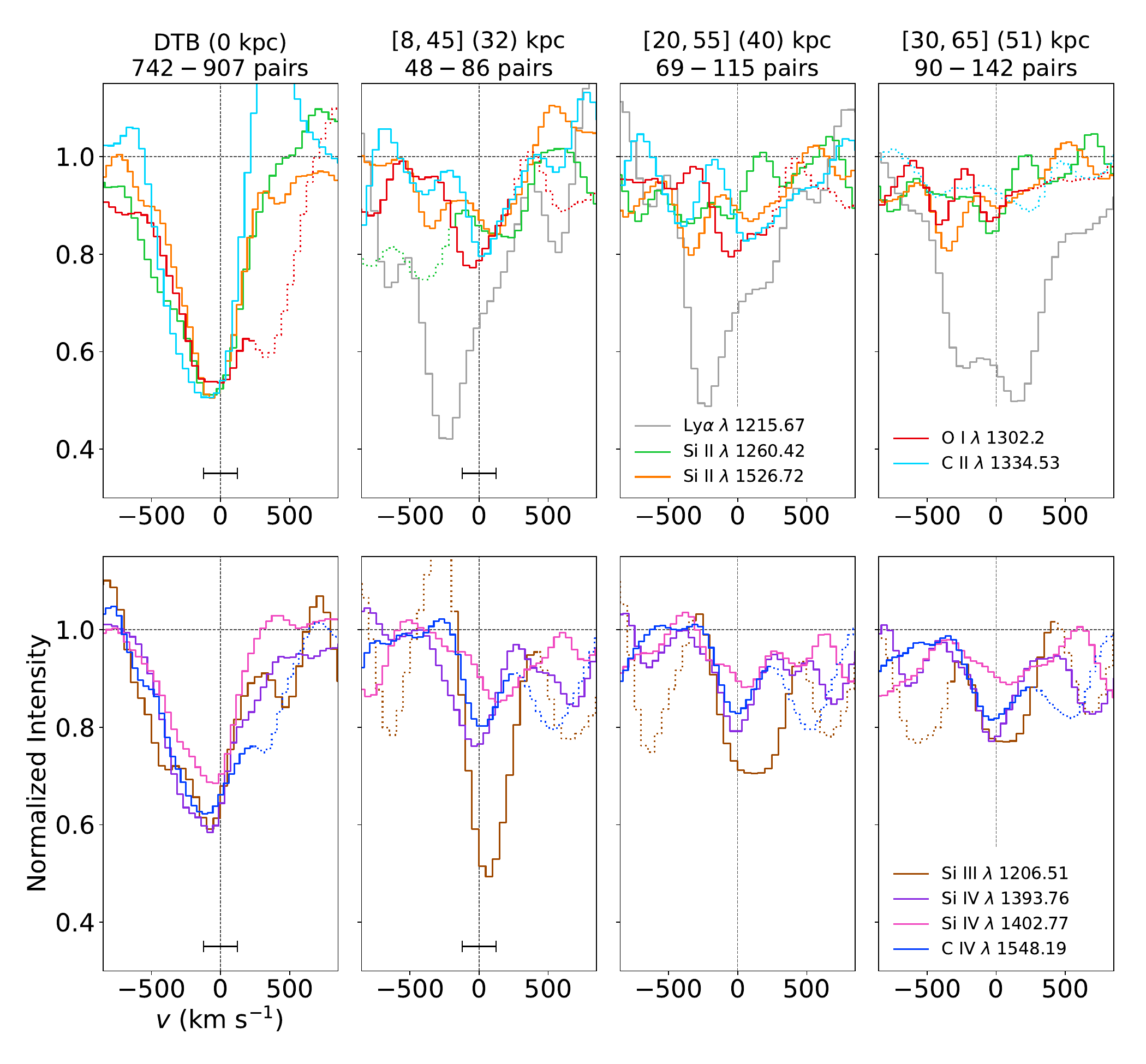}

     \caption{Normalized far-UV absorption features on a unified velocity scale for several different transitions DTB ($\dtran = 0$ kpc) and at three nonzero projected distances (separated by column). The range of \dtran and number of galaxy pairs in that range are listed in the title in a similar format to previous figures. The ions being compared are shown in the legend. The vertical dashed line shows the systemic velocity of the foreground galaxy stack and the black error bar at the bottom shows the effective spectral resolution. Spectral regions contaminated by neighboring features are shown in a dotted line style. This plot differs from Figure \ref{fig:linebins} in that several transitions are being compared at a fixed \dtran, as opposed to comparing the effect of \dtran for a fixed transition.}

    \label{fig:lowionsamev}
        
\end{figure*}

In addition to \lya and \civ, we attempted to measure additional far-UV low ionization interstellar absorption transitions in the stacked spectra. 
Figure \ref{fig:linebins} shows several far-UV transitions where the absorption profiles vary significantly with \dtran and provides a visualization of the changing \ew and $\sigma$ measured in Figures \ref{fig:ew_b} and \ref{fig:sigma_b} respectively. 
As expected, DTB ($\dtran=0$ kpc) absorption is considerably stronger for all transitions when compared with background galaxy sightlines ($\dtran>10$ kpc). Among the $\dtran>10$ kpc spectra, the trend continues:\ absorption strength for all transitions decreases with \dtran. In particular, \oi and \siiv show the largest change in both absorption strength and velocity extent over the other transitions. 

\begin{deluxetable*}{llcccccc}[htbp!] 
\tabletypesize{\scriptsize}
\tablecaption{DTB Equivalent Width Measurements and Model Fits$^a$ \label{tab:ewfits}}
\tablehead{
\colhead{Line} & \colhead{$\lambda$} & \colhead{$W_{\lambda}^b$} & \colhead{$W_{\lambda,\,\mathrm{corr}}^c$} & \colhead{$\gamma_\mathrm{out}$} & \colhead{$R_\mathrm{eff}$} & \colhead{$v_1$} & \colhead{$f_{c,\,{\rm max,\, out}}$}
\\  
  \colhead{} &\colhead{($\mathrm{\AA}$)} & \colhead{($\mathrm{\AA}$)} & \colhead{($\mathrm{\AA}$)} & \colhead{} & \colhead{(kpc)} & \colhead{(km s$^{-1}$)} & \colhead{}
  }
  \decimals
  \startdata
\ion{Si}{3} & 1206.51 & $1.09 \pm 0.06$\phm{d} & \phm{d}$2.19 \pm 0.12$ & $0.19 \pm 0.01$ & $>250$ & $711 \pm 102$ & $0.43 \pm 0.06$ \\
Ly$\alpha$ & 1215.67 & \nodata & \nodata & $0.33 \pm 0.02$ & $>250$ & $734 \pm 150$ & 0.67$^e$ \\
\ion{Si}{2} & 1260.42 & $1.38 \pm 0.03$\phm{d} & \phm{d}$2.75 \pm 0.06$ & $0.25 \pm 0.03$ & $>250$ & $709 \pm \phm{d}32$ & $0.52 \pm 0.02$ \\
\ion{O}{1} & 1302.2 & $1.00 \pm 0.02^d$ & \phm{d}$2.01 \pm 0.05$ & $0.27 \pm 0.03$ & $\phm{d}92\pm 22$ & $616 \pm \phm{d}25$ & $0.43 \pm 0.02$ \\
\ion{Si}{2} & 1304.4 & $1.16 \pm 0.02^d$ & \phm{d}$2.31 \pm 0.05$ & $0.41 \pm 0.06$ & $123\pm 37$ & $704 \pm \phm{d}28$ & $0.44 \pm 0.02$ \\
\ion{C}{2} & 1334.53 & $1.16 \pm 0.02$\phm{d} & \phm{d}$2.31 \pm 0.05$ & $0.20 \pm 0.01$ & $191\pm 78$ & $632 \pm \phm{d}23$ & $0.46 \pm 0.02$ \\
\ion{Si}{4} & 1393.76 & $1.23 \pm 0.02$\phm{d} & \phm{d}$2.47 \pm 0.04$ & $0.33 \pm 0.03$ & $>250$ & $685 \pm \phm{d}21$ & $0.44 \pm 0.01$ \\
\ion{Si}{4} & 1402.77 & $0.78 \pm 0.02$\phm{d} & \phm{d}$1.55 \pm 0.04$ & $0.24 \pm 0.04$ & $>250$ & $664 \pm \phm{d}20$ & $0.28 \pm 0.01$ \\
\ion{Si}{2} & 1526.72 & $1.39 \pm 0.02$\phm{d} & \phm{d}$2.77 \pm 0.04$ & $0.31 \pm 0.02$ & $114\pm 62$ & $655 \pm \phm{d}19$ & $0.48 \pm 0.01$ \\
\ion{C}{4} & 1549.0 & $1.50 \pm 0.02$\phm{d} & \phm{d}$3.00 \pm 0.05$ & $0.27 \pm 0.01$ & $>250$ & $822 \pm \phm{d}26$ & $0.40 \pm 0.01$ \\
\ion{Fe}{2} & 1608.45 & $0.70 \pm 0.02$\phm{d} & \phm{d}$1.40 \pm 0.04$ & $0.32 \pm 0.07$ & $102\pm 34$ & $596 \pm \phm{d}17$ & $0.25 \pm 0.01$ \\
\ion{Al}{2} & 1670.81 & $1.02 \pm 0.03$\phm{d} & \phm{d}$2.04 \pm 0.05$ & $0.44 \pm 0.09$ & $100\pm 52$ & $616 \pm \phm{d}20$ & $0.35 \pm 0.01$ 
\enddata
\tablecomments{\\
$^a$ The full set of foreground galaxies in this sample have median stellar mass $\log(\mstar/\msun)=9.6\pm 0.8$ (Figure \ref{fig:sedprops}).\\
$^b$ $W_{\lambda}$ are measured equivalent widths from DTB spectra ($\dtran=0$ kpc). $W_\lambda(\text{\civ})$ is computed by integrating over the full doublet.\\
$^c$ $W_{\lambda,\,\mathrm{corr}}$ are corrected equivalent widths assuming one is looking through a full halo endowed with a symmetric velocity distribution (i.e.\ $W_{\lambda,\,\mathrm{corr}}=2\, W_{\lambda}$). The DTB $W_{\lambda,\,\mathrm{corr}}$ are used in Figure \ref{fig:ew_b}.\\
$^d$ At the spectral resolution of the sample, the DTB \oiw and \siiiwm lines are blended. We measure $W_\lambda(1304.4)$ by scaling the \siiiwl profile to the red wing of the \siiiwm absorption feature; $W_\lambda(1302.2)$ is computed by subtracting off the scaled \siiiwl profile from the blended (\oi $+$ \siii) profile and measuring the residual absorption.\\
$^e$ $f_{c,\,{\rm max}}(\lya)$ is the sum of the two best fit $f_{c,\,{\rm max}}$ components in the 2D model. See Section \ref{sec:model} and Table \ref{tab:mcmc} for details.
}
\end{deluxetable*}

As a group, the high ions (\siiii, \civ, and \siiv) have shallower declines in $W_\lambda$ with \dtran 
compared to the low ions. Whereas the low ions (excluding \siiiwl) become undetectable within $\dtran \leq 250$ kpc, the high ions are present over the full range of \dtran sampled.
\cii, \siii, \oi, and \lya are all consistent with one another in their \ew dependence (especially at $\dtran \lesssim 100$ kpc) while weaker transitions like \alii and \siiiwh tend to fall off the fastest with projected distance. The exception to this trend is \siiiwl which tracks \siiiwh steadily until $\dtran\simeq 50$ kpc but maintains a larger \ew beyond that impact parameter.

We compare the \siiiwl and \siiiwh \ew ratios to see how \siii optical depth varies with \dtran. Down the barrel, $W_\lambda(1260)/W_\lambda(1526)=1$ (Table \ref{tab:ewfits}) implying both \siiiwl and \siiiwh are strongly saturated.
However, the DTB spectra in the top left panel of Figure \ref{fig:lowionsamev} show a divergence of velocities between \siiiwl (green) and \siiiwh (orange) on the blue side of the line suggesting that high velocity outflowing gas may be less strongly saturated than gas at $\dv\gtrsim -200$ \kms. Comparing $W_\lambda(1260)$ and $W_\lambda(1526)$ with \dtran (see Figures \ref{fig:ew_b} and \ref{fig:lowionsamev}), we find \siii remains saturated through $\dtran \simeq 50$ kpc, at which point $W_\lambda(1526)$ begins to fall off more rapidly than $W_\lambda(1260)$. At $\dtran\simeq 75$ kpc, $W_\lambda(1260)/W_\lambda(1526)\approx 2$ ($\tau(1526) \simeq 1.7$, assuming the flat part of the curve of growth). We note that $W_\lambda(1260)/W_\lambda(1526)=6.25$ in the optically thin limit so beyond $\dtran\simeq 100$ kpc, \siiiwh is unsaturated and we measure an optically thin \siii column density of $\log (N(\text{\siii})/\mathrm{cm}^{-2})\simeq12.9$. We lack sufficient $S/N$ to measure both transitions beyond $\dtran\simeq 125$ kpc.

Figures \ref{fig:linebins} and \ref{fig:lowionsamev} provide two complementary views on how each ion evolves with \dtran and how the profiles between transitions compare. While all lines show a decrease in absorption strength and velocity dispersion with increasing \dtran, the higher ionization lines like \civ and \siiv have a slightly broader velocity range at all \dtran than observed for \oi and \cii. Unlike the other transitions, \siiv has an extended red wing that extends to $\vlos\simeq 700$ \kms\ for $\dtran \lesssim 50$ kpc and similar to \civ, $W_\lambda(1393)/W_\lambda(1402)$ increases with \dtran as $N(\text{\siiv})$ decreases. The dichotomy between lower and higher ionization species is in agreement with previous QSO studies \citep[see e.g.][]{rauch_small-scale_2001,rudie_column_2019} which have shown that higher ionization species probe a wider range of both \nhi\ and velocity.  

Like the DTB spectra, Figure \ref{fig:lowionsamev} shows that all of the low ions and all of the high ions share similar respective profiles. The velocity extent of the metal lines agree to within $\sim 100$ \kms\ in the $\dtran\simeq32$ kpc bin and the amplitudes are consistent to within the bootstrap uncertainties. The decreasing absorption strength with \dtran makes this type of comparison more difficult at larger projected distances due to the reduced $S/N$ in each line, so we restrict these plots to $\dtran \lesssim 50$ kpc. Of the low ionization lines,  \oi and \cii show the best agreement with one another for $\dtran\lesssim 50$ kpc.

\subsection{2-D Absorption Maps }
\label{sec:2dmaps}

\begin{figure*}
        \centering
        \includegraphics[width=\linewidth]{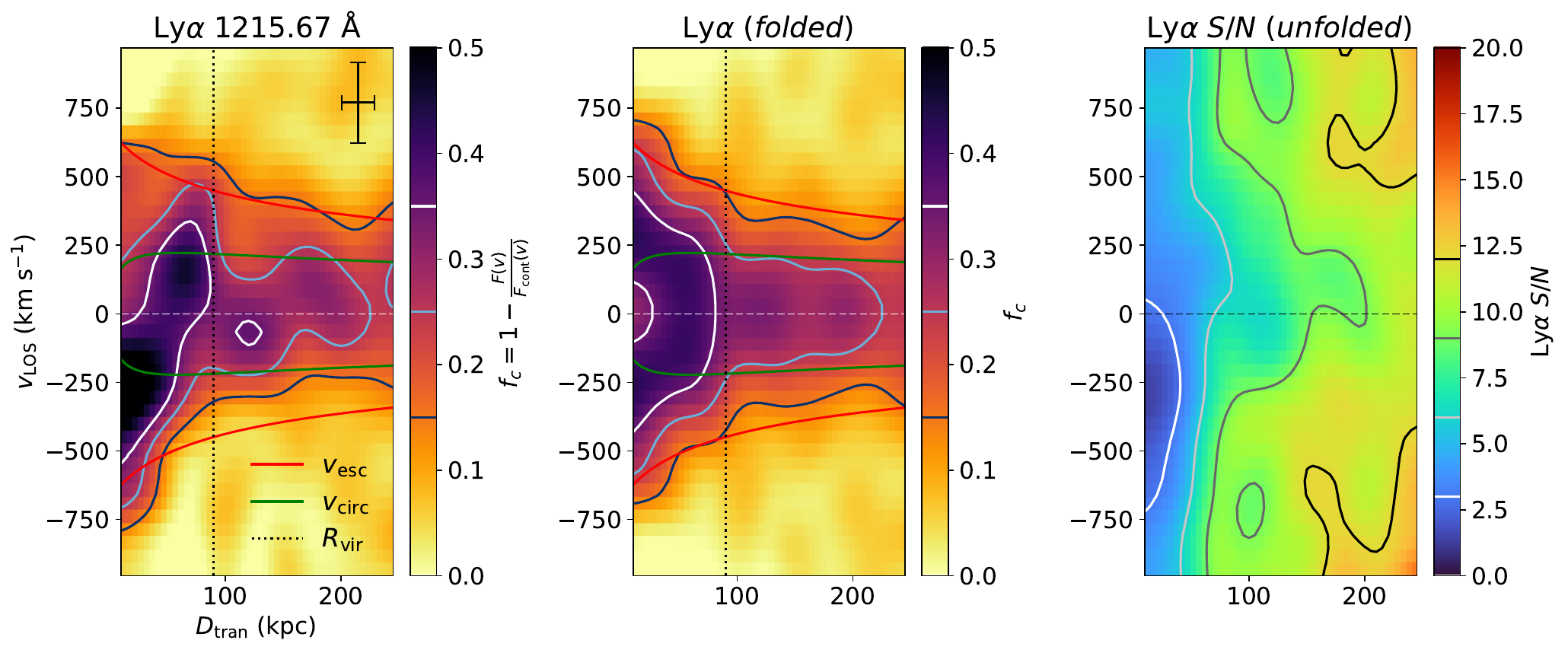}
        
        \caption{\textit{Left:} Maps of \lya covering fraction ($f_c$; Equation \ref{eqn:tau}) as a function of \vlos and \dtran. The darker regions correspond to deeper absorption troughs and thus higher covering fractions. \textit{Center:} $f_c(\lya)$ folded about $\vlos=0$ \kms. \textit{Right:} The \lya $S/N$ map, associated with the unfolded \lya map (computed over the full range of \vlos). The escape velocity, circular velocity, and virial radius for a $M_\mathrm{halo}=10^{12}\ \msun$ NFW halo are denoted in red, green, and black, respectively. Contours are plotted where applicable on each respective colorbar. \lya contours are separated by $\Delta f_c=0.1$ and $S/N$ contours proceed in increments of three i.e., $\Delta(S/N) = 3$. The black error bar in the top right of the left-hand panel shows the average bin size and effective LOS velocity resolution.}
        \label{fig:heatmaps}
\end{figure*}

In addition to one-dimensional spectra, we constructed a two-dimensional map of covering fraction
\begin{equation}
    f_c=1-\frac{F(v)}{F_\mathrm{cont}(v)},
    \label{eqn:tau}
\end{equation}
as a function of \vlos and \dtran; this is shown in Figure~\ref{fig:heatmaps}. In Equation \ref{eqn:tau}, $F(v)$ is the flux density measured for a stacked spectrum and $F_\mathrm{cont}(v)$ is the level of the continuum. When working with normalized spectra, $F_\mathrm{cont}(v)$ would naturally be one. \added{Recall that the covering fraction
$f_c$ is best interpreted as an excess optical depth (at a particular velocity and \dtran) since we do not measure the \textit{total} optical depth along the line of sight. For ease of visualization, we ``fold'' the \lya map about $\vlos=0$ \kms\ by taking the average of the red and blue halves of the velocity range. In other words, $f_c(\dtran,\vlos)=\langle f_c(\dtran,\pm\vlos)\rangle$; the folded map is shown in the middle panel of Figure \ref{fig:heatmaps}. }

We know from QSO sightlines through the CGM for a fraction of the galaxies in this sample that \hone\ and metals at these projected distances are clumpy and comprise several discrete velocity components \citep{rudie_gaseous_2012}. Although absorption is seen at velocities $|\Delta v| < 700$ \kms, the individual velocity components do not span the full range of velocities meaning large swaths of velocity space may show little to no absorption. Where present however, most complexes have $\nhi\gtrsim 10^{14.5}$ cm$^{-2}$ such that \lya is saturated. \edit1{At low resolution, as in Figure \ref{fig:heatmaps}, the \lya absorption profile shows the variation in velocity and $f_c$ of the clumps along the line of sight.}

On that point, the two-dimensional maps in Figure \ref{fig:heatmaps} illustrate the spatial and velocity extent of \lya with the corresponding $S/N$ shown in the righthand panel. Given that these are compilations of one-dimensional stacks (selected examples of which are shown in Figure \ref{fig:lyaciv}), they support the assertions made in Section \ref{sec:lyaciv}, i.e.\ \lya starts out with extended absorption at small $\dtran$ which becomes shallower (both in $f_c$ and $v_\mathrm{LOS}$) as galactocentric radius increases. The two-dimensional representation highlights the location of the velocity centroid and how that evolves with \dtran. Interestingly, \lya absorption is the strongest within $\dtran\lesssim 50$ kpc at $\vlos\simeq -250$ \kms,
indicative of a blue-dominated velocity asymmetry. Moving to larger \dtran, the velocity centroid shifts to $\vlos\simeq 100$ \kms\ by $\dtran \simeq 80$ kpc and \lya absorption appears skewed by $\dv\simeq +100$ \kms\ for $\dtran \lesssim 200$ kpc. Both of these asymmetries are discussed in Appendix~\ref{app:filling}.

\section{A semianalytic model}
\label{sec:model}

\added{In the vein of \citet{steidel_structure_2010} and \citet{chen_keck_2020}, we constructed a simple analytic model to reconcile DTB observations and absorption from composite spectra with $\dtran > 0$ kpc. This model is empirically motivated and attempts to broadly capture the trends we measure in the foreground and background spectra:\ a general decrease in LOS velocity spread and covering fraction with \dtran. We begin by introducing the model parameters, then establish models for the metal ions that reproduce the measured \ew vs.\ \dtran distribution observed in Figure \ref{fig:ew_b} (Section \ref{sec:1dmod}). In Section \ref{sec:2dmod}, we configure the model to output 2-D maps of covering fraction and use it to better understand the main features in measured the 2-D \lya $f_c$ distribution (Figure \ref{fig:heatmaps}).

\subsection{Modeling 1-D Composites}
\label{sec:1dmod}

Given its simplicity and its prior success in matching the observed change in \ew with \dtran, we apply the \citet{steidel_structure_2010} outflow model to the current dataset. We provide an overview of the model in this work and direct the reader to that study for a complete description and motivation for each parameter. As discussed in Section \ref{sec:2dmaps}, for a saturated transition observed at low resolution (i.e.\ $R\simeq 1000$ for our sample), the absorption line profile depends on $f_c$ and the velocity ($v$) of gas along the line of sight. The \citet{steidel_structure_2010} model uses a simple parameterization of these two components ($f_c$, $v$) to generate line profiles whose \ew are consistent with the \ew vs.\ \dtran distribution. We tailor this model to fit our \ew measurements (Figure \ref{fig:ew_b}) below.} 

The model velocity profile assumes an NFW halo \citep{navarro_structure_1996,navarro_universal_1997} where outflowing gas is initially accelerated to a maximum velocity $v_1$ within a galactocentric radius $r\simeq1$ kpc and subsequently decelerates with increasing \dtran solely under the influence of gravity. The outflowing radial velocity field is defined such that  

\begin{equation}
    \label{eqn:vout}
    v_\mathrm{out}(r)=\sqrt{v_1^2+A\left(-\ln{\frac{R_s+1}{R_s}}+\frac{1}{r}\ln{\frac{R_s+r}{R_s}}\right)},
\end{equation}
where the NFW scale radius $R_s=27$ kpc ($R_\mathrm{vir}\simeq 90$ kpc, given a concentration parameter $c=3.3$) and 

\begin{equation}
    \label{eqn:NFWA}
    A=\frac{8\pi G \rho_0 R_s^3}{1\,\mathrm{kpc}}\simeq 1.2 \times 10^7 \, \mathrm{km}^2\, \mathrm{s}^{-2}
\end{equation}
for a NFW halo at $z\simeq2.3$ with mass $M_\mathrm{halo}=10^{12}$ \msun. \added{The NFW halo parameters are from \citet{chen_keck_2020} and were chosen to match the mean KBSS halo.} 

\added{Analogous to \citet{steidel_structure_2010}, the gas opacity depends on a radially varying covering fraction $f_c(r)$ such that $f_{c,\,\mathrm{out}}(r)=f_{c,\,\mathrm{max,\,out}} r^{-\gamma_\mathrm{out}}$ where $f_{c,\, {\rm max,\,out}}$ is the maximum covering fraction used to normalize the distribution, $\gamma_{\rm out}$ is the power law index, and $r$ is once again the galactocentric radius. The radial power-law form is a natural choice and, based on our experiments, resulted in reasonable \ew predictions over the range of \dtran sampled. Following \citet{steidel_structure_2010}, we define $f_{c,\,\mathrm{max,\,out}}$ as the deepest part of the absorption trough, relative to the continuum, and measure $f_{c,\,\mathrm{max,\,out}}$ for each ion using the composite DTB spectrum (see e.g.\ Figure \ref{fig:fullstack}).} 

Since the velocities we measure relative to the foreground galaxy systemic redshift are projected along our LOS, we set up two grids:\ one in terms of projected distance ($\dtran$) and the other over a range of line of sight distances ($\ell$) such that each combination of of $\dtran$ and $\ell$ form a right triangle satisfying $r^2=\ell^2+\dtran^2$. The line of sight velocity for a given component is therefore
\begin{equation}
    \label{eqn:vlos}
    v_\mathrm{LOS}(\dtran, \ell)=\frac{\ell}{r} v(r)=\frac{\ell}{\sqrt{\dtran^2+\ell^2}}\,v(r).
\end{equation}

For each $(\dtran,\ell)$ grid point, we measured velocities and covering fractions
and projected each into the $v_\mathrm{LOS}-\dtran$ plane using Equation \ref{eqn:vlos}. \added{For the low ions whose \ew fall off precipitously within $\dtran \leq 250$ kpc, we fit a maximum outflow radius ($R_\mathrm{eff}$) beyond which there is no absorption. In the case of the high ions (which extend beyond $\dtran = 250$ kpc), we set $R_\mathrm{eff}=500$ kpc (twice the maximum \dtran of the pair sample) or to the radius where $f_c$ was below the noise level of the observed composite spectra to ensure $f_c$ represented absorption associated with the galaxy and not the IGM.

The predicted $W_\lambda(\dtran)$ are shown in Figure \ref{fig:ew_b} and their associated parameters are in Table \ref{tab:ewfits}. The uncertainties on the model parameters are derived from bootstrap resampling of the composite spectra (discussed in Sections \ref{sec:stacking} and \ref{sec:metalabs}). The model captures the shallow dependence of \ew on \dtran reasonably well and the initial gas velocities ($v_1$) are consistent with the maximum outflow velocities measured in the DTB spectra. We concede that $R_\mathrm{eff}$ is clearly an unphysical simplification as real outflows are unlikely to have such well-defined ``edges'', but including this parameter allows for an accurate reproduction of the rapid decline in low ion \ew with decreasing column densities.

\subsection{Modeling the 2-D Map}
\label{sec:2dmod}

Given the successful fits to \ew in Figure \ref{fig:ew_b}, we attempted to fit the 2-D \lya absorption maps in a manner similar to \citet{chen_keck_2020}. Despite having slightly different parameterizations for each component and different output formats, we note that the \citet{steidel_structure_2010} and \citet{chen_keck_2020} models fundamentally operate in the same way:\ both construct grids of $f_c$ and \vlos to make absorption line profiles as functions of \dtran; the \citet{steidel_structure_2010} model focuses on fitting \ew while the \citet{chen_keck_2020} model fits the \lya kinematics. The similarities between the current and previous two models allowed us to extend the current model to output \lya absorption maps. Importantly, our sample is restricted to $\dtran\leq 250$ kpc, while the \citet{chen_keck_2020} sample extends to $\dtran\simeq 4$ Mpc.}

We fit the \lya absorption map using the outflow component defined in Section \ref{sec:1dmod}, and to add additional opacity near $\vlos=0$ \kms, we added a random velocity component that takes on a Gaussian velocity distribution with constant velocity width ($\sigma_v$) with radius\footnote{As discussed in \citet{chen_keck_2020} and based on experimentation with the current sample, a one-component ``outflow only'' model is insufficient to reproduce the observed trends in the 2-D \lya absorption map.}. In contrast to \citet{chen_keck_2020}, we neglect the Hubble expansion due to the close-in range of \dtran sampled. This represents the simplest parameterization of the non-outflowing gas motions we observe. Again, we let the covering fraction vary radially such that $f_{c,\,\mathrm{r}}(r)=f_{c,\,\mathrm{max,\,r}} r^{-\gamma_\mathrm{r}}$ where each symbol takes on a similar meaning to those defined above. For simplicity, the random velocity component operates independently of the outflowing component. Since \lya extends beyond $\dtran=250$ kpc, we again set $R_\mathrm{eff}=500$ kpc to ensure only absorption associated with the galaxy is considered.

\subsubsection{2-D Model Results}

\begin{figure*}
    \epsscale{1.2}
    \plotone{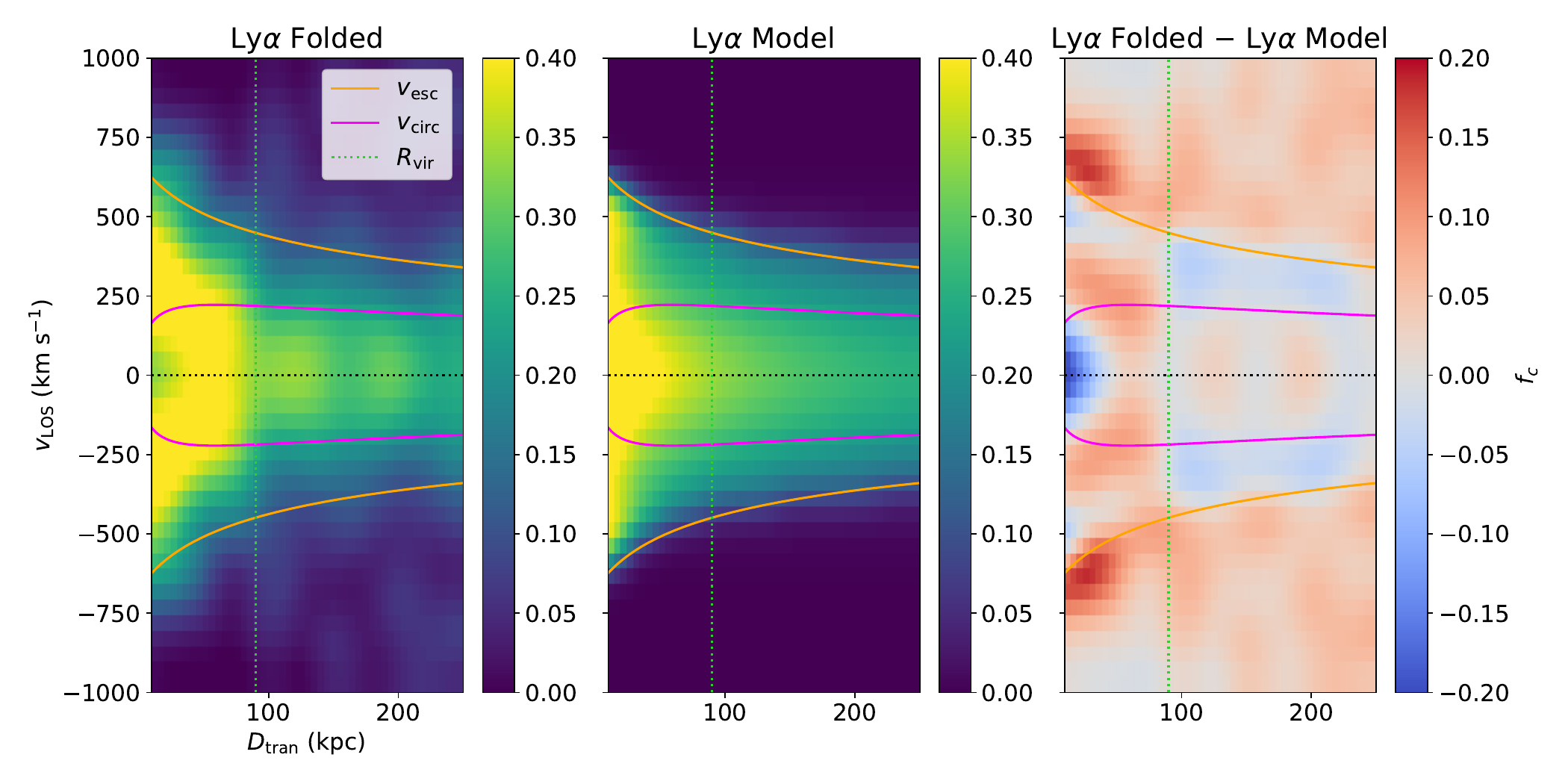}
    \caption{\textit{Left:} Measured \lya \replaced{apparent optical depth}{covering fraction} folded about $\vlos=0$ \kms; \textit{Middle:} \added{\lya best-fit (outflow and random velocity gas) model;} \textit{Right:} \lya absorption residuals (data $-$ model). The observed and model heatmaps have the same colorbar. The escape and circular velocities for a $M_\mathrm{halo}=10^{12}\ \msun$ NFW halo are shown in orange and magenta respectively in conjunction with the approximate virial radius (lime green). The best-fit model parameters are listed in Table \ref{tab:mcmc}.}
    \label{fig:outin_mod}
\end{figure*}

The observed \lya map has $\Delta v_\mathrm{LOS}=50$ \kms\ and $\Delta \dtran=5$ kpc sampling
Given the simplicity of the model and its inability to capture small variations in covering fraction, we smoothed the observed \lya maps with a 2 pix $\times$ 2 pix Gaussian kernel ($\sigma_{\dtran} = 10$ kpc and $\sigma_{v_{\mathrm{LOS}}}=100$ \kms) so the broad trends were captured and the fitting procedure was facilitated. 
Since the model is symmetric around $\vlos=0$ \kms, we use the folded map to compensate for the asymmetry in the measured \lya absorption (especially at $\dtran\lesssim 50$ kpc). We argue in Appendix~\ref{app:filling} that this asymmetry results from foreground \lya emission contributing to the background absorption feature, so the folded map bears a closer resemblance to the unaffected signal.\footnote{The ``hole'' in the folded \lya map at $(\dtran, \vlos) = (0,0)$ most likely results from foreground \lya emission.} We employ the \texttt{emcee} \citep{foreman-mackey_emcee_2013} Markov chain Monte Carlo (MCMC) sampler to find the best fit to the six ($v_1$, $\sigma_v$, $f_{c,\,\mathrm{max,\,out}}$, $f_{c,\,\mathrm{max,\,r}}$, $\gamma_\mathrm{out}$, and $\gamma_r$) model parameters given the smoothed \lya covering fraction and uncertainty maps.\footnote{Since the observed \lya map is constructed using overlapping bins, in fitting we boost the measured bootstrap uncertainties by $\simeq 7\%$ (the largest difference in uncertainty between the adopted binning scheme and a set of independent bins) to account for bin-to-bin covariance.} We require the fit $\sigma_v$ to have a velocity consistent with the random velocity dispersion of a $10^{12}~\msun$ halo.
The best fit model is shown in the middle panel of Figure \ref{fig:outin_mod} and the parameters used to generate it are enumerated in Table \ref{tab:mcmc}.

\begin{deluxetable}{lll} 
\tablecaption{Best Fit Model Parameters \label{tab:mcmc}}
\tablehead{
\colhead{Component} & \colhead{Parameter} & \colhead{Value} 
  }
  \decimals
  \startdata
Outflow & $v_1$ (km s$^{-1}$) & $781\,^{+73}_{-77}$\\
& $f_{c,\,\mathrm{max,\,out}}$ & $0.26\,^{+0.25}_{-0.19}$ \\
& $\gamma_\mathrm{out}$ & $0.40\,^{+0.21}_{-0.17}$\\
Random Velocity & $\sigma_v$ (km s$^{-1}$) & $219\,^{+32}_{-34}$\\
& $f_{c,\,\mathrm{max,\,r}}$ & $0.41\,^{+0.32}_{-0.38}$\\
& $\gamma_r$ & $0.32\,^{+0.36}_{-0.22}$\\
\enddata
\end{deluxetable}

In comparing the best fit models between \citet{chen_keck_2020} and our study we first recall that the range of \dtran sampled in \citet{chen_keck_2020} span galactocentric radii larger by an order of magnitude than our current sample. With this in mind, we note a few distinctions. The initial outflow velocity ($v_1$) is larger by 30\% in the current sample ($781\,^{+73}_{-77}$ \kms\  vs.\ $603\,^{+5}_{-11}$ \kms) and the velocity of gas near the systemic velocity ($\sigma_v$) is a factor of $\simeq 2.6$ greater in magnitude ($219\,^{+32}_{-34}$ \kms\ vs.\ $-84\,^{+6}_{-6}$ \kms). The finer sampling and higher velocity resolution of the current sample means low-$f_c$, high-\vlos gas is more readily detectable than in the \citet{chen_keck_2020} sample. Intuitively, the outflowing component in the model establishes a high velocity (relatively low optical depth) envelope while the non-outflowing velocity component adds additional opacity at lower velocities centered around $v=0$ \kms. Since the gas with the highest galactocentric velocity in both \citet{chen_keck_2020} and this sample is found at $\dtran\lesssim 60$ kpc, those are the impact parameters that fix the model parameters of the outflowing component. In contrast, an \hone overdensity with $|v|<200$ \kms\ was observed throughout the full 4 Mpc range in \citet{chen_keck_2020}, gradually decreasing in optical depth with \dtran. 
When restricted to $\dtran<250$ kpc, the velocity field appears considerably more active than at larger galactocentric radii so the constant velocity component (centered at $\vlos=0$ \kms) will naturally take on a correspondingly higher value to match that velocity spread. It is for this reason that although functionally similar to the ``inflow'' component of the \citet{chen_keck_2020} model, we associate this component with gas with random velocities or high optical depth non-radial flows since it is unlikely that the excess optical depth at these galactocentric distances and velocities results strictly from ``inflowing'' material. 

\deleted{We observe that the outflow and constant velocity gas absorption coefficients ($\tau_\mathrm{0,\, out}$, $\tau_{0,\, r}$) in the current sample are larger than the \citet{chen_keck_2020} measurements by a factor of $\sim 6$ and $\sim 25$, respectively (cf.\ $\alpha_\mathrm{0,\, in}$ and $\alpha_\mathrm{0,\, out}$ in \citealt{chen_keck_2020}). The difference in integration box sizes between the two samples may affect this quantity most significantly. Moreover, the increased spectral resolution in the current study results in more pronounced absorption troughs and therefore higher excess optical depth measurements especially near the systemic velocity. This explains why the random velocity component (centered around $v_\mathrm{sys}=0$ \kms) has the largest contrast between samples. }

The random velocity gas has a covering fraction similar to that of the outflowing component, that is, $f_{c,\,\mathrm{max,\,r}}\sim f_{c,\,\mathrm{max,\,out}}$ contrary to what was previously seen
in \citet{chen_keck_2020}. 
One can consider this element of the model as absorption due to gas with velocities $v_\mathrm{LOS} \lesssim 300$ \kms, similar to but not the same as an ISM component to a down-the-barrel absorption line profile in a galaxy with outflowing material. For our sample, the ratio $f_{c,\,\mathrm{max,\,r}}/f_{c,\,\mathrm{max,\,out}} \simeq 1$ suggests (perhaps unsurprisingly) most of the gas in the CGM ($\dtran<250$ kpc) is relatively slow moving ($v_\mathrm{LOS} \lesssim 300$ \kms) but there exists an appreciable component of outflowing material.

More quantitatively, Figure \ref{fig:outin_mod} shows the circular and escape velocities for a $M_\mathrm{halo}=10^{12}$ \msun\ NFW halo. These curves treat $r=\dtran$ and do not account for velocity projection effects so the real radii may be larger and LOS velocities may be smaller. As expected, the circular velocity approximately matches the random velocity component since both are constrained by the gravitational potential well of the average host galaxy. We observe the escape velocity at the virial radius $v_\mathrm{esc}(R_\mathrm{vir})\approx 400$ \kms\ is close to the maximum velocity associated with the outflowing component. While most of the absorbing gas has velocities less than the escape velocity (since those are the velocities where this type of map is most sensitive), a nontrivial fraction has the possibility of escaping the halo. As discussed in Section \ref{sec:lowions}, \lya absorption is not the best probe of high velocity gas since higher ionization lines sample a larger cross-section of the CGM, but the fact that a nominal amount of \hone\ exceeds $v_\mathrm{esc}$ supports this hypothesis.

As discussed in \citet{chen_keck_2020} \added{and above}, the strictly radial dependence of velocity and covering fraction is clearly an oversimplification of a complex environment. With the improved resolution in both \vlos and \dtran, this simplification becomes even more suspect. We measure a clumpy distribution of \lya absorption due to variance in the constituent galaxy stacks; two examples of regions with higher than average opacity can be seen in the left panel of Figure \ref{fig:outin_mod} at $\dtran\simeq125$ kpc and $\dtran\simeq175$ kpc.
To model these semianalytically, either non-radial covering fraction and velocity profiles would need to be implemented or a third component would need to be added for this express purpose, but since these features are most likely systematic noise, further complications to the model may be unjustified. \added{That being said, the qualitative picture of outflowing material superposed with ambient CGM gas bound to the galaxy is supported by both the \lya map and model.}

Interestingly, the observed \lya map (Figure \ref{fig:heatmaps}) has a step in the velocity field at approximately the virial radius for the average foreground galaxy halo ($R_\mathrm{vir}\simeq 80$ kpc) where the velocity spread decreases from $\sim 750$ \kms\ to $\sim 500$ \kms; we denote the approximate virial radius as a green dotted line in the left panel of Figure \ref{fig:outin_mod}. Despite its simplicity, the random velocity component of the model shows a marked downturn in absorption strength at approximately the same projected distance. We conjecture that this could be a site where outflowing and inflowing gas are both present and would therefore result in a smaller average velocity when measured spectroscopically. The data are unable to show the degree (or lack thereof) to which the two components interact or mix but considering that the model treats the outflowing and random velocity components as separate sources of excess absorption, even noninteracting gas streams may produce this ``pinch'' in the velocity field. In a situation where the two components interact, the effect on the line of sight velocities could be more pronounced.

\begin{deluxetable*}{lccccccc}[htbp!]
\tabletypesize{\scriptsize} %
\tablecaption{Absorption Strength (\ew) vs. Foreground Galaxy Properties\label{tab:sedew}}
\tablehead{
\colhead{Composite$^a$} & \colhead{} & \colhead{$W_\lambda(\lya)$} & \colhead{} & \multicolumn{4}{c}{$W_\lambda(\text{\ionl{C}{4}{1549}})^b$}\\
\cmidrule(lr){2-4} \cmidrule(lr){5-8}
\colhead{} & \colhead{$[8, 70]~(51)^c$} & \colhead{$[70, 100]~(86)$} & \colhead{$[100,250]~(188)$} & \colhead{DTB$^d$} &\colhead{$[8, 70]~(51)$} & \colhead{$[70,100]~(86)$} & \colhead{$[100,250]~(188)$}\\
\colhead{} & \colhead{(\AA)} & \colhead{(\AA)} & \colhead{(\AA)} & \colhead{(\AA)} & \colhead{(\AA)} & \colhead{(\AA)} & \colhead{(\AA)}}

\decimals
\startdata
$\log(\mstar/\msun)$ $= 9.0 \pm 0.5$ & $0.85 \pm 0.20$ & $1.03 \pm 0.14$ & $0.63 \pm 0.07$ & $1.77 \pm 0.03$ & $0.47 \pm 0.10$ & $0.71 \pm 0.09$ & $0.23 \pm 0.05$ \\
$\log(\mstar/\msun)$ $= 9.6 \pm 0.3$ & $2.33 \pm 0.25$ & $1.70 \pm 0.21$ & $1.01 \pm 0.07$ & $3.35 \pm 0.02$ & $1.04 \pm 0.12$ & $0.99 \pm 0.13$ & $0.25 \pm 0.04$ \\
$\log(\mstar/\msun)$ $= 10.2 \pm 0.4$ & $1.74 \pm 0.22$ & $2.21 \pm 0.16$ & $0.80 \pm 0.08$ & $3.69 \pm 0.02$ & $1.65 \pm 0.11$ & $0.99 \pm 0.11$ & $0.44 \pm 0.04$ \\
\hline
$\log[{\operatorname{SFR}}/(M_\odot\, \mathrm{ yr}^{-1})]$ $= 0.6 \pm 0.3$ & $1.82 \pm 0.19$ & $1.31 \pm 0.14$ & $0.74 \pm 0.06$ & $2.35 \pm 0.02$ & $0.75 \pm 0.09$ & $0.86 \pm 0.08$ & $0.43 \pm 0.03$ \\
$\log[{\operatorname{SFR}}/(M_\odot\, \mathrm{ yr}^{-1})]$ $= 1.2 \pm 0.3$ & $1.36 \pm 0.15$ & $1.15 \pm 0.14$ & $0.77 \pm 0.05$ & $3.75 \pm 0.01$ & $1.13 \pm 0.09$ & $0.98 \pm 0.10$ & $0.31 \pm 0.03$ \\
\hline
$\log({\operatorname{sSFR}}/\mathrm{Gyr}^{-1})$ $= -0.2 \pm 0.4$ & $1.62 \pm 0.19$ & $1.27 \pm 0.15$ & $0.76 \pm 0.06$ & $3.24 \pm 0.02$ & $0.90 \pm 0.08$ & $0.59 \pm 0.08$ & $0.33 \pm 0.03$ \\
$\log({\operatorname{sSFR}}/\mathrm{Gyr}^{-1})$ $= 0.6 \pm 0.5$ & $1.38 \pm 0.17$ & $1.00 \pm 0.15$ & $0.79 \pm 0.05$ & $3.05 \pm 0.01$ & $0.67 \pm 0.10$ & $1.03 \pm 0.09$ & $0.18 \pm 0.03$ \\
\hline
$E(B-V)$ $= 0.07 \pm 0.04$ & $2.03 \pm 0.15$ & $1.23 \pm 0.14$ & $0.74 \pm 0.05$ & $2.95 \pm 0.01$ & $0.84 \pm 0.07$ & $0.89 \pm 0.09$ & $0.19 \pm 0.03$ \\
$E(B-V)$ $= 0.12 \pm 0.05$ & $1.18 \pm 0.18$ & $1.44 \pm 0.14$ & $0.89 \pm 0.06$ & $3.63 \pm 0.02$ & $1.44 \pm 0.09$ & $1.08 \pm 0.08$ & $0.28 \pm 0.03$ \\
\hline
Full Sample & $1.42 \pm 0.09$ & $0.90 \pm 0.09$ & $0.76 \pm 0.04$ & $3.00 \pm 0.05$ & $0.79 \pm 0.06$ & $0.67 \pm 0.06$ & $0.26 \pm 0.02$ \\
\enddata
\tablecomments{\\
$^a$ The pair sample is split into tertiles in \mstar; the remaining subsamples are formed by splitting the pair sample in half at the median value of the stellar population parameter. Each subsample's median value and interquartile range are shown; see Figure \ref{fig:sedprops} for the precise ranges of each subsample. The corresponding spectra from which these values are derived are shown in Figures \ref{fig:sed4x4} and \ref{fig:sedciv3}. Values from the full pair sample are shown in the last row for comparison.\\
$^b$ $W_\lambda(\text{\civ})$ is computed by integrating over the full doublet.\\
$^c$ In order:\ the minimum, maximum, and median \dtran (in kpc) for each bin.\\
$^d$ The DTB $W_\lambda(\text{\civ})$ are corrected as discussed in Section \ref{sec:metalabs} (cf. $W_{\lambda,\,\mathrm{corr}}$ in Table \ref{tab:ewfits}).}
\end{deluxetable*}

\begin{deluxetable*}{lccccccc}[htbp!]
\tablecaption{Absorption Second Moments ($\sigma$) vs. Foreground Galaxy Properties\label{tab:sedsig}}
\tablehead{
\colhead{Composite} & \colhead{} & \colhead{$\sigma(\lya)$} & \colhead{} & \multicolumn{4}{c}{$\sigma(\text{\ionl{C}{4}{1549}})$}\\
\cmidrule(lr){2-4} \cmidrule(lr){5-8}
\colhead{} & \colhead{$[8, 70]~(51)$} & \colhead{$[70, 100]~(86)$} & \colhead{$[100,250]~(188)$} & \colhead{DTB} &\colhead{$[8, 70]~(51)$} & \colhead{$[70,100]~(86)$} & \colhead{$[100,250]~(188)$}\\
\colhead{} & \colhead{(\kms)} & \colhead{(\kms)} & \colhead{(\kms)} & \colhead{(\kms)} & \colhead{(\kms)} & \colhead{(\kms)} & \colhead{(\kms)}}
\decimals
\startdata
$\log(\mstar/\msun)$ $=9.0\pm0.5$ & \phn $209$ $\pm$ $73$ & \phn $232$ $\pm$ $64$ & \phn $210$ $\pm$ $49$ & \phn $473$ $\pm$ $35$ & \phn \phn $89$ $\pm$ $89$ & \phn $343$ $\pm$ $51$ & \phn $457$ $\pm$ $72$ \\
$\log(\mstar/\msun)$ $= 9.6 \pm 0.3$ & \phn $299$ $\pm$ $28$ & \phn $281$ $\pm$ $36$ & \phn $215$ $\pm$ $28$ & \phn $544$ $\pm$ \phn $8$ & \phn $282$ $\pm$ $54$ & \phn $352$ $\pm$ $48$ & \phn $475$ $\pm$ $40$ \\
$\log(\mstar/\msun)$ $= 10.2\pm0.4 $  & \phn $284$ $\pm$ $32$ & \phn $359$ $\pm$ $14$ & \phn $231$ $\pm$ $35$ & \phn $592$ $\pm$ \phn $5$ & \phn $360$ $\pm$ $22$ & \phn $274$ $\pm$ $52$ & \phn $246$ $\pm$ $47$ \\
\hline
$\log[{\operatorname{SFR}}/(M_\odot\, \mathrm{ yr}^{-1})]$ $= 0.6 \pm 0.3$ & \phn $274$ $\pm$ $32$ & \phn $266$ $\pm$ $48$ & \phn $234$ $\pm$ $29$ & \phn $454$ $\pm$ $16$ & \phn $142$ $\pm$ $71$ & \phn $370$ $\pm$ $28$ & \phn $492$ $\pm$ $18$ \\
$\log[{\operatorname{SFR}}/(M_\odot\, \mathrm{ yr}^{-1})]$ $= 1.2 \pm 0.3$ & \phn $259$ $\pm$ $40$ & \phn $282$ $\pm$ $42$ & \phn $166$ $\pm$ $36$ & \phn $558$ $\pm$ \phn $5$ & \phn $353$ $\pm$ $26$ & \phn $301$ $\pm$ $36$ & \phn $214$ $\pm$ $52$ \\
\hline
$\log({\operatorname{sSFR}}/\mathrm{Gyr}^{-1})$ $= -0.2 \pm 0.4$ & \phn $300$ $\pm$ $29$ & \phn $304$ $\pm$ $34$ & \phn $240$ $\pm$ $27$ & \phn $516$ $\pm$ \phn $8$ & \phn $250$ $\pm$ $44$ & \phn $263$ $\pm$ $83$ & \phn $420$ $\pm$ $27$ \\
$\log({\operatorname{sSFR}}/\mathrm{Gyr}^{-1})$ $= +0.6 \pm 0.5$ & \phn $233$ $\pm$ $48$ & \phn $252$ $\pm$ $54$ & \phn $172$ $\pm$ $34$ & \phn $551$ $\pm$ \phn $8$ & \phn $165$ $\pm$ $79$ & \phn $298$ $\pm$ $34$ & \phn $255$ $\pm$ $83$ \\
\hline
$E(B-V)$ $= 0.07 \pm 0.04$ & \phn $319$ $\pm$ $17$ & \phn $229$ $\pm$ $51$ & \phn $190$ $\pm$ $33$ & \phn $501$ $\pm$ \phn $7$ & \phn $136$ $\pm$ $54$ & \phn $428$ $\pm$ $27$ & \phn $338$ $\pm$ $71$ \\
$E(B-V)$ $= 0.12 \pm 0.05$ & \phn $192$ $\pm$ $53$ & \phn $325$ $\pm$ $24$ & \phn $234$ $\pm$ $20$ & \phn $563$ $\pm$ \phn $6$ & \phn $401$ $\pm$ $18$ & \phn $248$ $\pm$ $75$ & \phn $365$ $\pm$ $35$ \\
\hline
Full Sample & \phn $261$ $\pm$ $47$ & \phn $224$ $\pm$ $21$ & \phn $211$ $\pm$ $18$ & \phn $514$ $\pm$ $17$ & \phn $232$ $\pm$ $46$ & \phn $288$ $\pm$ $87$ & \phn $378$ $\pm$ $26$ \\
\enddata
\tablecomments{Similar to Table \ref{tab:sedew}, now with second moment measurements. The DTB $\sigma(\text{\civ})$ are measured over the full doublet and corrected as discussed in Section \ref{sec:metalabs}.}
\end{deluxetable*}

\section{Dependence of CGM Absorption on Galaxy Properties}
\label{sec:bin}
 
\added{The current sample of (\zfg, \zbg) galaxy pairs is larger in number, spans a larger range of foreground galaxy properties, and has significantly higher effective spectral resolving power than previous galaxy (\zfg, \zbg) pair samples at similar redshifts. In this section, we investigate whether sub-samples (in bins of \dtran) can reveal significant differences in the absorption \ew  and kinematics ($\sigma$) when divided according to inferred properties of the foreground galaxies.} 

Using SED measurements of the foreground galaxies introduced in Section~\ref{sec:paircons}, we constructed subsamples based on \mstar, SFR, sSFR, and $E(B-V)$. As discussed in Appendix \ref{app:stackmethod}, the stacked spectra we construct are always ``pair limited,'' i.e., the $S/N$ of the spectra and the corresponding ease of detecting absorption features are both proportional to the number of sightlines being averaged. For this reason, the \dtran sampling achievable with the SED-based sub-samples is considerably coarser that what was possible using the full sample.

After some experimentation with various \dtran bins, we settled on three independent intervals of \dtran: $\dtran = 8 - 70$ kpc, $70 -100$ kpc, and $100 -250$ kpc, which have median \dtran of 51, 86, and 188 kpc, respectively; the two inner bins contain a total of 208 and 202 galaxy pairs, respectively. We focus on \lya and \civ absorption, the only lines that are strong enough to allow for splitting the samples into two or three sub-samples according to galaxy properties while maintaining statistical power.  

The distributions of galaxy properties inferred from SED fits for this sample are shown in Figure \ref{fig:sedprops}. 
While the $\dtran=$ 100 -- 250 kpc bin has $\simeq 80\%$ of the total foreground galaxies and is thus representative of the full sample, the number of galaxies in the $\dtran=$ 8 -- 70 kpc and $\dtran=$ 70 -- 100 kpc bins is smaller by almost an order of magnitude with respect to the $\dtran=$ 100-- 250 kpc bin. That being said, the overall distributions of galaxy properties sampled within each \dtran bin remain similar to one another, and representative of the full sample of foreground galaxies.

\begin{figure*}[htbp!]
    \centering
    \includegraphics[width=\linewidth]{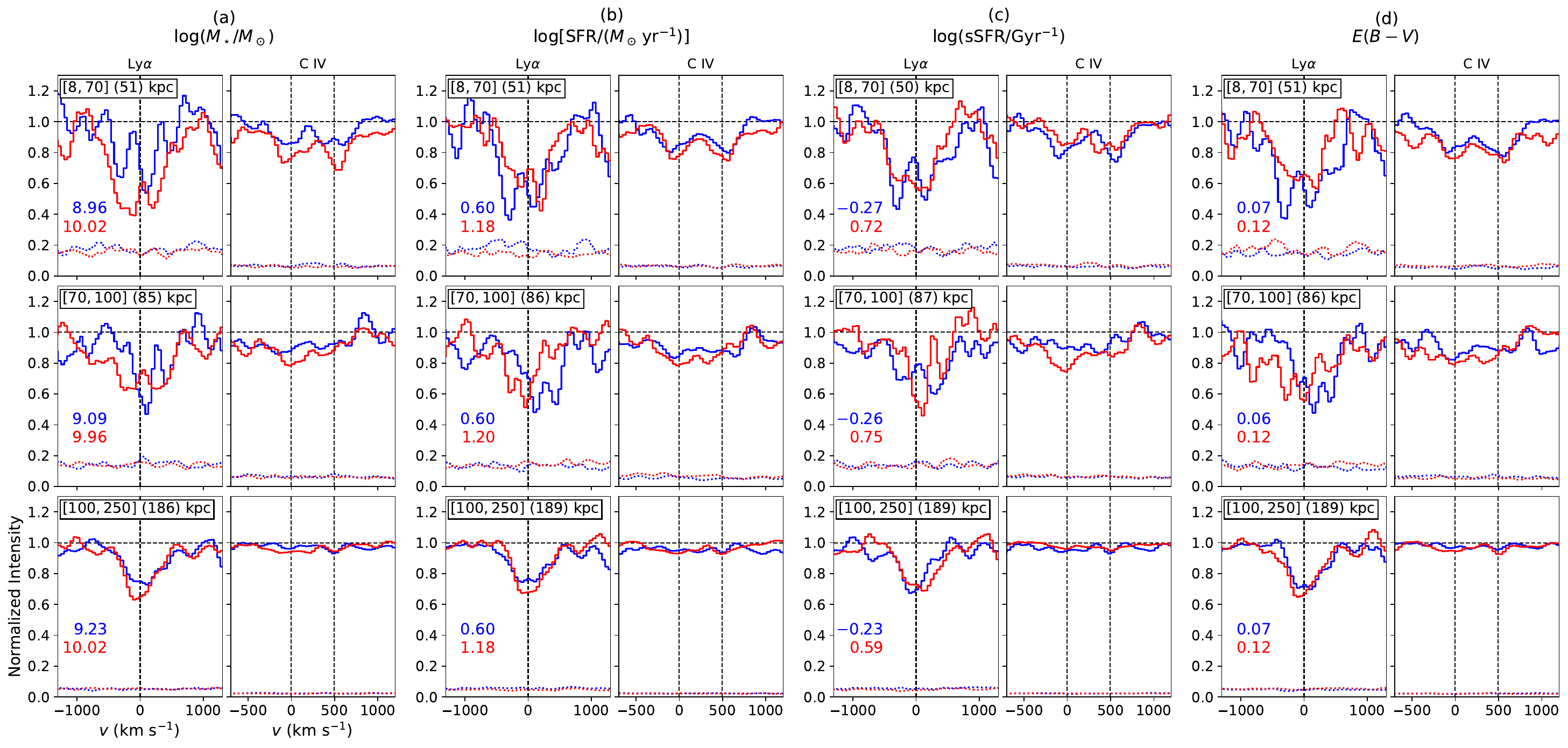}
    \caption{Composite spectra in the vicinity of \lya (left column of each panel) and \civ (right column of each panel) for sub-samples divided less than (blue) and greater than (red) the median value of each of four SED-derived galaxy properties:\ (a) \mstar,  (b) SFR,  (c) sSFR, and (d) $E(B-V)$. Within each panel, the top spectra compare the results for \dtran$=8-70$ kpc, the middle spectra for \dtran$=70-100$ kpc, and the bottom spectra for \dtran$=100-250$ kpc. The median value of the stellar population parameter is printed in each, color-coded in the same way as the plotted spectra, for the ``low'' and ``high'' sub-samples. Error spectra are shown as color-coded dotted lines; the continuum level and systemic velocity of the foreground galaxy are indicated with black dashed lines. Figure \ref{fig:sedprops} shows the full distributions of the parameters.}
    \label{fig:sed4x4}
\end{figure*}

Initially, we split the four SED-derived distributions in half at their median value, and constructed ``low'' and ``high'' sub-samples; see the leftmost column of Table \ref{tab:sedew}. Each subsample based on foreground galaxy properties spans approximately 1.5 dex in \mstar, 1 dex in SFR and sSFR, and 0.1 magnitudes in $E(B-V)$. For each subsample, we followed the same procedure (see Section \ref{sec:paircons}) as the full sample to construct stacked background galaxy spectra that contain foreground objects satisfying a particular galaxy property constraint (e.g.\ $\log[{\operatorname{SFR}}/(M_\odot\, \mathrm{ yr}^{-1})]\leq 0.6$). These stacks were combined using a trimmed mean, renormalized, and line profile measurements were computed for \lya and \civ. \ew 
and $\sigma$ for each bin and absorption line are given in Tables \ref{tab:sedew}
and \ref{tab:sedsig}, respectively. Spectral cutouts of \lya and \civ for each galaxy property are shown in Figure \ref{fig:sed4x4}.
The top row of plots corresponds to spectra from the $\dtran = 8 - 70$ kpc bin, the middle has $\dtran = 70-100$ kpc, and the bottom shows $\dtran = 100 - 250$ kpc composites. 
The results for \civ are shown in Figure~\ref{fig:sedciv3}, discussed below.  

\subsection{\ion{C}{4}}
\label{subsec:civ}

\begin{figure*}[htbp!]
    \plotone{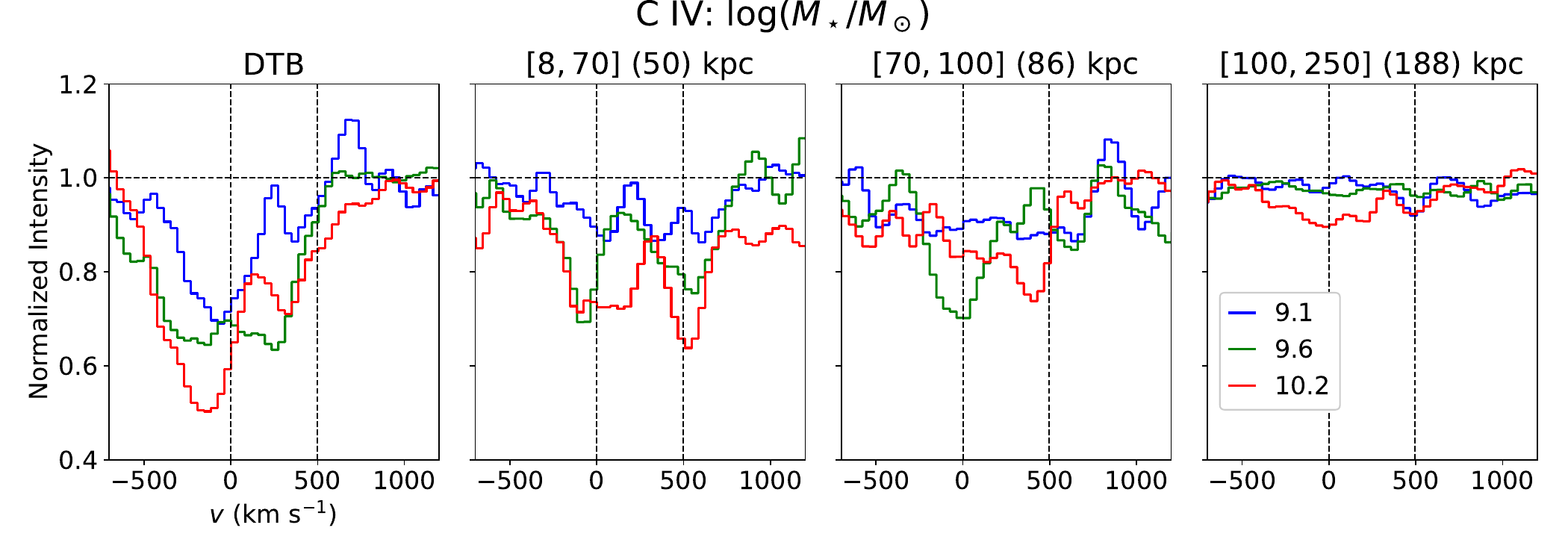}
    \caption{\civ absorption profiles for three independent sub-samples of \mstar, in four bins of \dtran: from left to right, DTB ($\dtran=0$ kpc) spectra, the $\dtran = 8 - 70$ kpc bin, $\dtran = 70 - 100$ kpc bin, and \dtran $=$ 100 -- 250 kpc bin are shown. Blue, green, and red spectra correspond to the first, second, and third tertiles of \mstar. The legend shows the median $\log(\mstar/\msun)$ for each sub-sample.}
    \label{fig:sedciv3}
\end{figure*}

\begin{figure*}[htbp!]
    \epsscale{0.9}
    \plotone{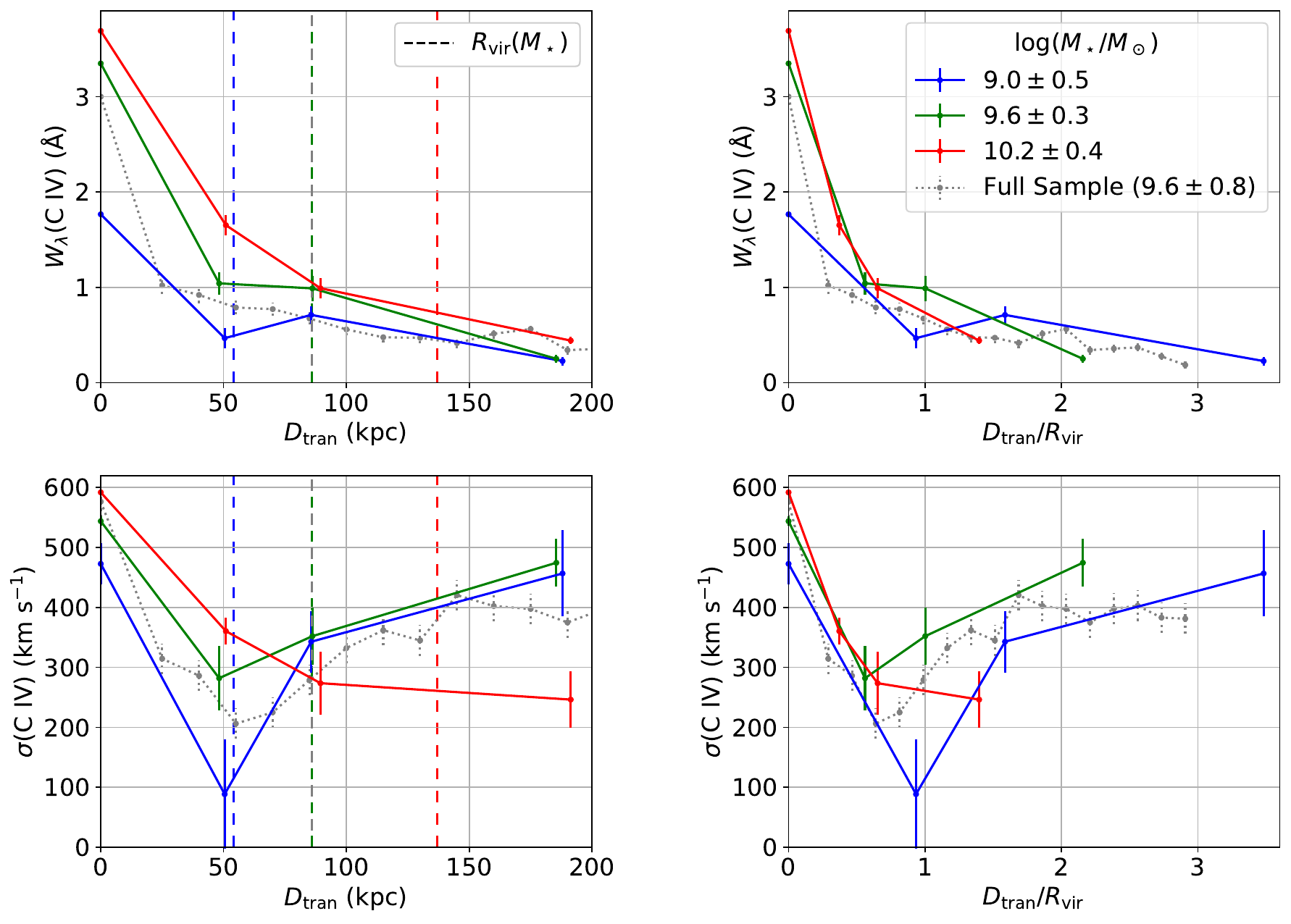}
    \caption{{\it Top:} $W_\lambda(\text{\civ})$ vs.\ \dtran (left; see Table~\ref{tab:sedew}) and projected distance normalized by $R_{\rm vir}$ (right) for the three \mstar\ tertiles (blue, green, and red curves show increasing \mstar, as indicated in the legend).  {\it Bottom:} The line-of-sight velocity second moment $\sigma$(\civ) vs. \dtran (left; see Table~\ref{tab:sedsig}) and vs.\ $\dtran/R_{\rm vir}$ (right). The plotted values of $\sigma$ have been corrected for the instrumental contribution as described in Section \ref{sec:sigma_inst}. The vertical dashed lines in the two lefthand panels indicate the estimated $R_{\rm vir}$ for the corresponding \mstar\ bin, as discussed in the text. The ``Full Sample'' data, which provide finer sampling of \dtran, are the same as those plotted in Figures~\ref{fig:ew_b} and \ref{fig:sigma_b}.}
    \label{fig:civ_rvir}
\end{figure*}

\added{The most significant correlation between a galaxy property and the corresponding CGM absorption is found for \mstar\ versus \civ absorption, where the higher-\mstar\ sub-sample has $W_{\lambda}({\rm CIV}) \simeq 2\times$ greater than the low-\mstar\ bin at all \dtran shown in Figure~\ref{fig:sed4x4}. Given the difference in \civ strength for the two \mstar\ subsamples, we split the sample into three bins of \mstar\footnote{Splitting the sample into a larger number of \mstar\ bins results in stacks with an insufficient number of pairs to successfully detect \civ.}; the results for \civ in tertiles of \mstar\ are shown in Figure \ref{fig:sedciv3}, where we have included the DTB spectra for each \mstar\ sub-sample. The first three rows of Tables~\ref{tab:sedew} and~\ref{tab:sedsig} summarize the numerical results. 

The upper left panel of Figure~\ref{fig:civ_rvir} summarizes the $W_{\lambda}({\rm CIV})$ measured (as a function of \dtran) from each mass-binned spectrum. Figure~\ref{fig:sedciv3} shows the \civ absorption profiles for three \mstar\ bins, with median \mstar\ values log($\mstar/M_{\odot}) = 9.0$, 9.6, and 10.2 in four different bins of \dtran: $\langle \dtran \rangle = 0$, $\langle \dtran \rangle = 51$, $\langle \dtran \rangle = 86$, and $\langle \dtran \rangle = 188$ kpc. As the figure shows (see also Table~\ref{tab:sedew}), the trend of monotonically increasing \civ absorption strength with \mstar\ exists in all four \dtran bins, illustrated graphically in the upper lefthand panel of Figure~\ref{fig:civ_rvir}. 
The highest mass \mstar\ bin has $W_{\lambda}({\rm CIV})$ exceeding that of the lowest-mass bin by factors ranging from $1.9-3.5$, with the largest difference found for 
the $\langle \dtran \rangle =51$ kpc bin, where the spectra for all three \mstar\ bins are detected with very high significance. 

It has become common practice to assume that the galactocentric distance to which one expects to observe CGM metals associated with a given galaxy may be related to the virial radius $R_{\rm vir}$ of the corresponding dark matter halos.  Figure~\ref{fig:civ_rvir}a includes vertical dashed, color-coded lines indicating the approximate $R_{\rm vir}$ corresponding each of the three \mstar\ bins.  The values adopted are $R_{\rm vir} = 53$, 86, and 137 kpc for the three \mstar\ bins, based on dark matter halos with $\langle {\rm log}(M_{\rm halo}/M_{\odot}) \rangle =11.9$ at the median stellar mass ${\rm log}(\mstar/M_{\odot}) = 9.6$ for the KBSS sample \citep{trainor_halo_2012}. We estimated $R_{\rm vir}$ for the other two \mstar\ bins by assuming that the ratio $\mstar/M_{\rm h}$ is the same as that of the full KBSS sample\footnote{The median halo masses for the low-\mstar\ and high \mstar\ bins under this assumption would be $\log(M_{\rm halo}/M_{\odot}) = 11.3$ and 12.5, respectively.}), so that $\langle R_{\rm vir} \rangle \propto \mstar^{1/3}$. The upper right panel of Figure~\ref{fig:civ_rvir} shows the same data as the upper left panel, but with \dtran\ scaled by the estimated $R_{\rm vir}$ for each \mstar\ sub-sample.
The similar appearance of $W_{\lambda}({\rm CIV})$ vs. $\dtran/R_{\rm vir}$ 
for the three \mstar\ sub-samples certainly suggests that $R_{\rm vir}$ may play a significant role in determining the extent of absorbing gas -- with the largest changes in the total \ion{C}{4} absorption strength occurring within $R_{\rm vir}$, while all three \mstar\ bins reach a similar ``plateau'' in $W_{\lambda}({\rm CIV})$ between $\dtran \sim 1-3~R_{\rm vir}$.

The average $W_{\lambda}({\rm CIV})$ among the \mstar\  bins is affected by both the depth of the absorption features -- which depends on the covering fraction and/or the mean $N$(\civ) as a function of LOS velocity --  and on the extent in velocity space of the absorption feature. The latter can be characterized by the second moment $\sigma$(\civ), which is shown
in the bottom panels of Figure~\ref{fig:civ_rvir}. The lower left panel shows that the dispersion in velocity of the mean \civ absorption has a similar monotonic behavior of increasing $\sigma$(\civ) with \mstar, particularly in the inner two bins of \dtran, and is largely responsible for the most significant difference in absorption strength vs.\ \mstar, in the $\langle \dtran \rangle = 51$ kpc bin.  The lowest \mstar bin reaches a clear minimum $\sigma$(\civ) at $\langle \dtran \rangle = 51$ kpc, where it is consistent with being unresolved ($\sigma= 89\pm 89$ \kms) -- and corresponds closely to the estimated $R_{\rm vir}$ for $\langle {\rm log}(\mstar/M_{\odot})\rangle = 9.0$ galaxies. 

The values of $\sigma$(\civ) for the \mstar\ sub-samples become statistically indistinguishable in the $\langle \dtran \rangle = 86$ kpc bin; by $\langle \dtran \rangle = 188$ kpc the two lower mass bins have nominal $\sigma(\text{\civ})\simeq 500$ \kms, which suggests that, though net \civ absorption is detected, it is not concentrated at the systemic redshift of the foreground galaxy, but is instead nearly uniformly distributed within the velocity window used for the measurement.  In contrast, the highest \mstar\ bin has a significantly {\it lower} $\sigma$(\civ) at large $\dtran$, suggesting that the \civ absorbing gas remains gravitationally influenced by the presence of the foreground galaxy, with a line-of-sight velocity dispersion consistent with the circular velocity of the host halos ${\rm log}(M_{\rm h}/M_{\odot}) \simeq 12.5$.

Aside from the \mstar\ dependence of \civ\ absorption, there is some evidence that 
the high SFR sub-sample has significantly higher $W_{\lambda}$(\civ) and $\sigma$(\civ) than that of the lower-SFR sub-sample, at least for the DTB and $\langle \dtran \rangle = 50$ kpc bins (Tables~\ref{tab:sedew} and \ref{tab:sedsig}). A similar trend may be present for the $E(B-V)$ sub-samples, in the sense that more reddened galaxies exhibit stronger \civ\ absorption with higher $\sigma$(\civ); however, both of these trends might be explained by the correlation of higher SFR and $E(B-V)$ with \mstar\ within the foreground galaxy ensemble. }

\subsection{\lya}

\begin{figure*}[htbp!]
    \epsscale{0.9}
    \plotone{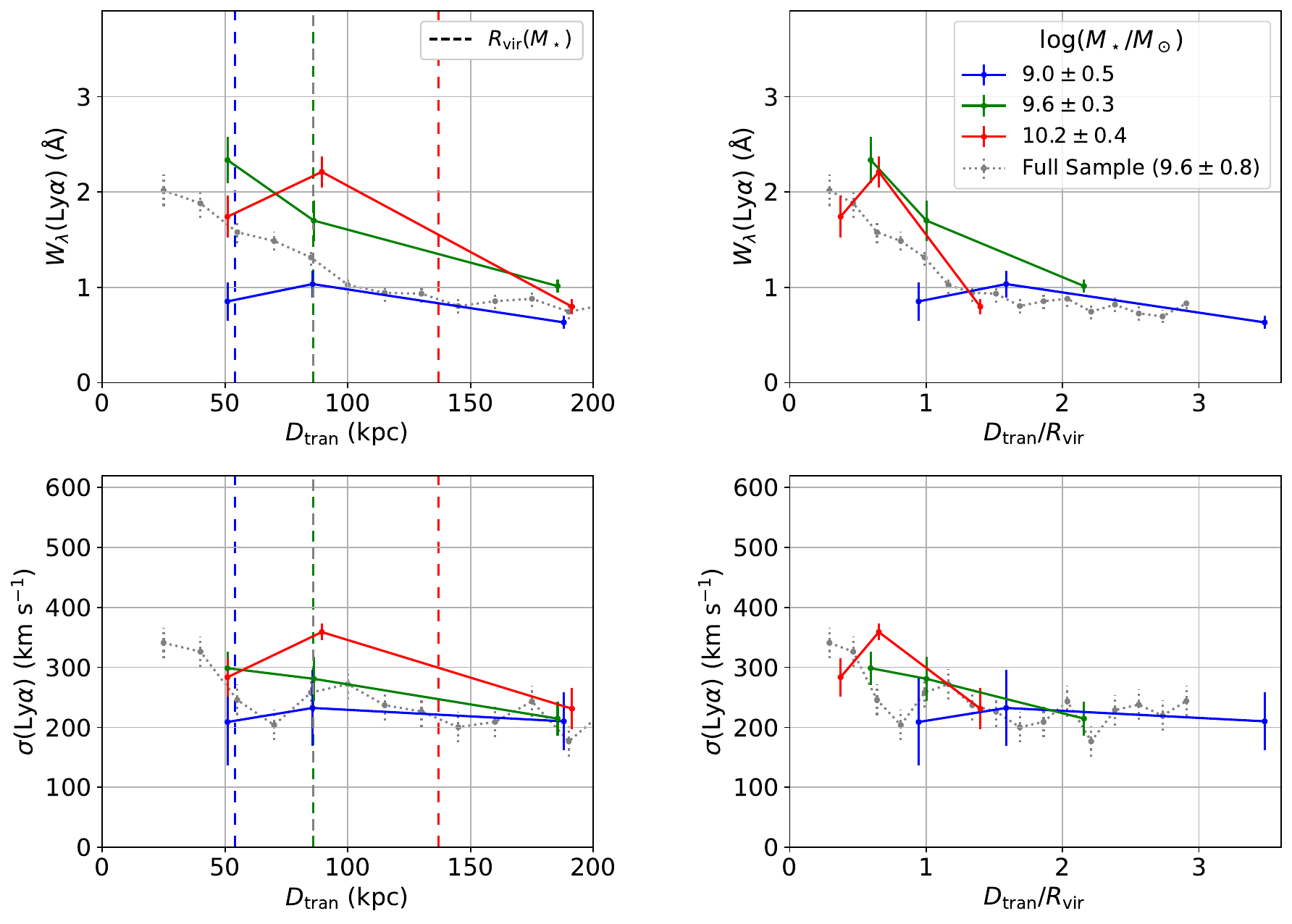}
    \caption{As for Figure~\ref{fig:civ_rvir}, but for \lya\ absorption (see Tables~\ref{tab:sedew} and \ref{tab:sedsig}).
    The ``Full Sample'' data, which provide finer sampling of \dtran, are the same as those plotted in Figures~\ref{fig:ew_b} and \ref{fig:sigma_b}.}
    \label{fig:lya_rvir}
\end{figure*}

\added{While \lya absorption strength $W_{\lambda}(\lya)$ and velocity dispersion $\sigma(\lya)$ generally increase with \mstar\footnote{The DTB \lya\ absorption is confused by strong \lya\ emission in the centers of the foreground galaxies, so we do not attempt a $\dtran = 0$ measurement; see Figure~\ref{fig:lyaciv}.}, the trend is not monotonic at all \dtran, and the differences between \mstar\ sub-samples are significant at $> 2\sigma$ level only 
within the $\langle \dtran \rangle = 86$ kpc bin, where both $W_{\lambda}(\lya)$ and $\sigma(\lya)$ are positively correlated with \mstar\ (Figure~\ref{fig:lya_rvir}; Tables~\ref{tab:sedew} and \ref{tab:sedsig}). Unlike what was found for \civ\ (Figure~\ref{fig:civ_rvir}), there is no significant difference in $\sigma(\lya)$ in the $\langle \dtran \rangle = 188$ kpc bin of impact parameter, with values of $\sigma(\lya)$ comparable to that of the highest \mstar\ sub-sample for $\sigma$(\civ) at the same \dtran\ -- likely due to the fact that \lya\ remains well-detected ($\gtrsim 10\sigma$) for all 3 \mstar\ sub-samples out to $\gtrsim 200$ kpc. The large number of galaxy pairs available at large \dtran, together with a ``smoother'' local continuum level against which to observe the excess \lya\ absorption, helps to suppress the systematic noise in the composite spectra. We discuss the behavior of \lya\ kinematics and absorption strength vs. \dtran for the full sample in Section~\ref{sec:model}.

The \lya profiles in the two independent sub-samples of SFR, sSFR, and $E(B-V)$ are largely consistent with one another (Figure~\ref{fig:sed4x4}). The profiles become smoother at larger \dtran because larger number sightlines are being averaged, but there is no discernible trend across all three bins. 
Although the sub-samples divided by galaxy properties do not show large variations in \lya absorption, we again note that the apparent \lya\ absorption profiles (both in velocity and apparent optical depth) in the composite spectra of background galaxies for $\dtran \lesssim 100$ kpc can be significantly affected by extended \lya\ emission from the foreground galaxies. This effect, which depends on the intensity and kinematics of the \lya\ emission halos of the foreground galaxies relative to the apparent magnitude of the background galaxies against which the absorption is being measured, may be responsible for some of the stochastic behavior of \lya\ absorption profiles. We discuss this issue further in Appendix \ref{app:filling}, and also in Section~\ref{sec:model}. }

\section{Discussion}
\label{sec:disc}

\added{The current study further constrains the distribution of baryons in the \ztwo\ CGM. Fundamentally, we notice a decrease in \ew and $\sigma$ with increasing \dtran for all transitions studied. The CGM ionization state changes with \dtran:\ $W_\lambda(\text{\cii})/W_\lambda(\text{\civ})$, $W_\lambda(\text{\siii})/W_\lambda(\text{\siiii})$, and $W_\lambda(\text{\siii})/W_\lambda(\text{\siiv})$ all decrease with \dtran consistent with a reduction of cool, dense gas along the line of sight. The decreasing $W_\lambda$ with approximately constant $\sigma$ in the low ions can be explained by a decreasing gas column density, either in the metals themselves or in the \hone that self-shields them.

Kinematically, $\sigma(\lya)$ and $\sigma(\text{\civ})$ decrease with \dtran and both reach a minimum at approximately the virial radius of the average galaxy in our sample. Taken together with the blueshifted absorption features in the DTB spectra, it follows that the gas motions at $\dtran \lesssim R_{\rm vir}$ result from optically thick outflows, the fastest of which are localized to within $\dtran \lesssim 30$ kpc of the average foreground galaxy. \lya and \civ within $\dtran \lesssim 50$ kpc have absorption at velocities in excess of the escape velocity suggesting that some gas is unbound from the galaxy. This technique is only sensitive to \textit{excess} absorption associated with foreground galaxies so it is not possible to quantify precisely how much gas escapes the system; gas with $\vlos > v_{\rm esc}$ would typically appear as contamination at nearby wavelengths in the composite spectra. We interpret the minimum in the $\sigma$ vs.\ \dtran distribution as a caustic between outflows that dominated the absorption within $\dtran \lesssim 100$ kpc and gas with velocities less than $v_{\rm circ}$. This represents the extent of the galaxy's influence on its CGM. The absorption at $\dtran \simeq R_{\rm vir}$ may result from inflowing gas, which tends to have $\sigma < v_\mathrm{circ}$ \citep[e.g.][]{rubin_direct_2012}, or the bound ambient CGM. We attribute the increase in $\sigma(\text{\civ})$ at $\dtran \gtrsim 100$ kpc to an increasing number of nearby (but unrelated) \civ systems that serve to broaden the composite \civ absorption profile. These systems are increasingly unlikely to be causally connected with the CGM of the foreground galaxy as \dtran increases, but with the current stacking strategy, it is not possible to directly isolate those associated with the foreground system.

In addition to the full sample, the mass-separated \lya and \civ measurements provide additional clarity into the dynamical signatures of the \ztwo\ CGM. The strong dependence of \civ absorption strength on stellar mass suggests that sightlines through higher mass halos are sampling larger CGM volumes. The general agreement between mass subsamples when \dtran is normalized by the virial radius further supports this assertion and suggests that $R_\mathrm{vir}$ -- and by extension dark matter halo mass -- is a fundamental quantity underpinning the gas distribution of the CGM. Moreover, $R_\mathrm{vir}$ sets a relative scale over which a galaxy can have a significant effect on its CGM. The overall agreement of $\sigma$ vs.\ $\dtran/R_{\rm vir}$ may suggest that \civ is similarly distributed throughout halos of varying sizes. Despite the coarse sampling, the slight dependence of $\sigma$ on \mstar\ (especially at $\dtran \lesssim R_{\rm vir}$) supports a picture of feedback where more massive galaxies drive faster outflows, and the outflowing gas associated with more massive galaxies in turn appears to permeate further into the CGM.}

\subsection{Comparison to Prior Observations}
\label{sec:priorobs}

The enhanced excess \lya absorption extending out to the maximum \dtran of the sample ($\dtran \simeq 250$ kpc) is consistent with \citet{rudie_gaseous_2012} and \citet{turner_metal-line_2014} using KBSS QSO lines of sight and \citet{steidel_structure_2010} and \citet{chen_keck_2020} with KBSS galaxy pairs. \edit1{The measured \lya equivalent widths
and the inferred radial dependence of the covering fraction ($f_c(r)\propto r^{-0.33}$) in particular closely match the dependencies seen in \citet{chen_keck_2020} and \citet{steidel_structure_2010}, respectively}. The \citet{chen_keck_2020} $W_\lambda(\lya)$ measurements are shown in Figure \ref{fig:ew_b} (brown curve) for comparison and both $W_\lambda(\lya)$ agree well with one another. Note that in \citet{steidel_structure_2010}, $W_\lambda(\lya)$ at $100\lesssim \langle\dtran \rangle/\mathrm{kpc}\lesssim 250$ were computed from HIRES spectra of QSO-galaxy pairs while \citet{chen_keck_2020} used galaxy-galaxy pairs for all $\dtran \lesssim 4$ Mpc. In addition, we find a reduced in $W_\lambda(\lya)$ and $\sigma(\lya)$ between $75 \lesssim \dtran/\mathrm{kpc} \lesssim 125$ in accordance with the \citet{chen_keck_2020} results. Given the large overlap in galaxy pairs between \citet{steidel_structure_2010}, \citet{chen_keck_2020}, and this sample, it is not surprising that the equivalent widths show large agreement.

Outside of the KBSS sample, the anticorrelation between \ew and \dtran for \lya and metals in the CGM has been well established for galaxies across a range of masses and redshifts. \citet{kacprzak_evidence_2020} recently compiled \ew vs.\ \dtran measurements for \lya and several far-UV transitions from studies \citep{prochaska_probing_2011,werk_cos-halos_2013,bordoloi_cos-dwarfs_2014,liang_mining_2014,borthakur_connection_2015,burchett_deep_2016,johnson_extent_2017,pointon_relationship_2019} of $9\leq \log(\mstar/\msun)\leq 11$ galaxies at $z\lesssim 0.3$. Each transition shows a decrease in absorption strength with \dtran with the metals showing the most rapid falloff. Analogous to \siii in our sample, at $z\lesssim 1.4$, there is a similar trend between $W_\lambda(\text{\ion{Mg}{2}})$ and $\sigma(\text{\ion{Mg}{2}})$ with \dtran \citep[e.g.][]{martin_kinematics_2019,hamanowicz_muse-alma_2020,chen_circumgalactic_2024} in line with the previously listed studies at $z\lesssim 0.3$. \added{While the general trends are similar between low and high redshift star forming galaxies, \lya and the metal ions have steeper falloffs with \dtran at low-$z$ compared to the current $z\sim 2$ sample. The galaxies studied above also tend to have smaller \ew and overall quieter velocity fields than seen at $z\sim 2$. Indeed, the low-$z$ COS-Burst ``starburst'' galaxies \citep{Heckman17} may offer the most direct comparison in terms of \ew magnitudes and slope (with \dtran) to the current sample.} It is worth noting that the previously listed studies all use QSOs as their background source which lead to more stochastic gas distributions in comparison to foreground/background galaxy pairs which probe gas on $\sim 4$ kpc scales (\citealt{steidel_structure_2010}; the typical half-light diameter of a galaxy in our sample).

At $z\sim 2$, \citet{steidel_structure_2010} modeled gas covering fraction $f_c$ from both \hone\ and metals as power laws of the form $f_c\propto r^{-\gamma}$ and found $0.2\lesssim \gamma\lesssim 0.6$ at $\dtran \lesssim 250$ kpc; \added{applying a similar model to our dataset, we find similarly shallow power law indices:\ $0.2\lesssim \gamma\lesssim 0.5$ suggesting a comparable covering fraction evolution with \dtran. When projected along our line of sight, we observe a faster \ew falloff in the low ions when compared to the high ions.} 
At face value, this suggests that the column density of the low ions decreases faster than the high ions over the range of projected distances or that the low ions sample a narrower range of CGM velocities and absorbing columns; a result also put forth by \citet{rudie_column_2019}.

While there is good agreement in \vlos between \citet{chen_keck_2020} and this study (see Section \ref{sec:2dmaps}) especially in the high velocity gas at $\dtran< 100$ kpc, $\sigma(\lya)$ and $\sigma(\text{\civ})$ are larger by a factor of $\simeq 2$ than the \citet{liang_mining_2014} velocity measurements of gas surrounding $z<0.176$ galaxies within $\dtran < 250$ kpc. Moreover, \citet{liang_mining_2014} find gas moving with $\vlos\gtrsim 200$ \kms\ almost exclusively arises from $\dtran > 250$ kpc sightlines, contrary to our results. The authors speculated that the broad velocity wings may be associated with large-scale filaments; in our case, most of the $\vlos\gtrsim250$ \kms\ excess \lya absorption likely arises from star formation induced outflows.

Our $W_\lambda(\text{\civ})$ vs.\ \dtran dependence is consistent with the results of \citet{adelberger_connection_2005}, \citet{steidel_structure_2010}, and \citet{turner_metal-line_2014} where the samples have overlap in \dtran.  \citet{adelberger_connection_2005} reported \civ with a velocity spread $\simeq 260$ \kms\ for $\dtran \lesssim 80$ kpc using galaxy-galaxy pairs and \citet{steidel_structure_2010} extended these measurements to $\dtran \simeq 100$ kpc. The $W_\lambda(\text{\civ})$ measurements from both studies are in agreement with our sample (which recall is a broad superset of theirs); the additional galaxy-galaxy pairs at $\dtran > 100$ kpc in our sample enable an analysis that was not possible previously. At $\langle \zfg\rangle\sim2.6$, \citet{mendez-hernandez_metal_2022} measured $W_\lambda(\text{\civ})$ vs.\ \dtran around 238 star-forming galaxies at $\dtran < 173$ kpc. Both this sample and the \citet{mendez-hernandez_metal_2022} measurements are consistent in their slopes and approximate $W_\lambda(\text{\civ})$ values although the \citet{mendez-hernandez_metal_2022} \dtran sampling and spectral resolution are considerably coarser. Surprisingly, the $z\simeq3.3$ \lya emitters studied in \citet{muzahid_musequbes_2021} do not show a $W_\lambda(\text{\civ})$ vs.\ \dtran dependence. A possible explanation for this discrepancy is that $D_\mathrm{tran,\, med}=165$ kpc for the \citet{muzahid_musequbes_2021} sample and the largest variability in $W_\lambda(\text{\civ})$ tends to be at $\dtran \lesssim 100$ kpc. Indeed our sample shows only a factor of $\simeq 2$ change in $W_\lambda(\text{\civ})$ between $\dtran\simeq 100$ kpc and $\dtran\simeq 250$ kpc. 

\begin{figure}
    \centering
    \includegraphics[width=0.8\linewidth]{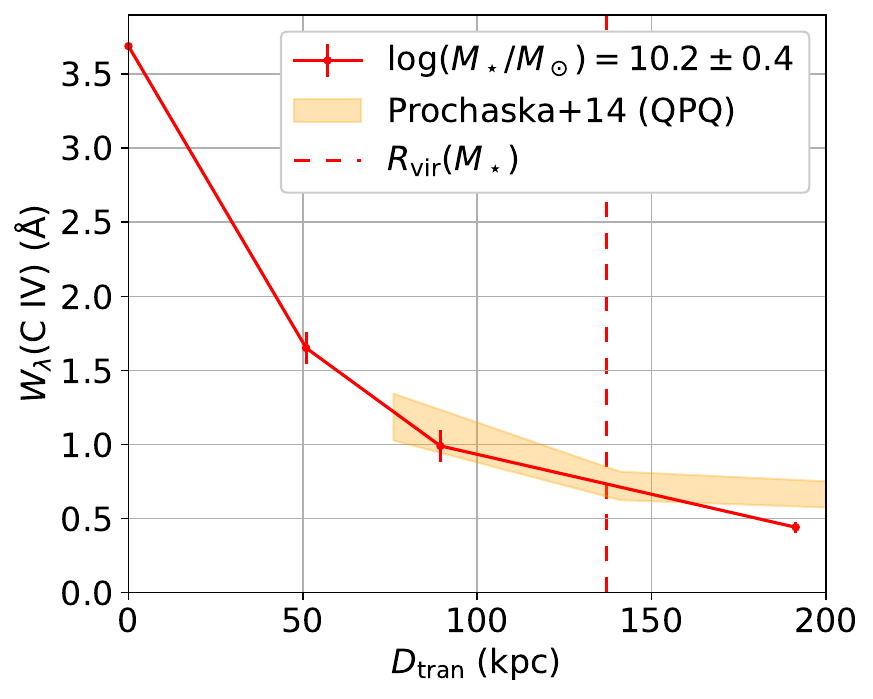}
    \caption{$W_\lambda(\text{\civ})$ vs.\ \dtran for the high mass tertile ($\log(\mstar/\mstar)\approx 10.2$) of galaxy pairs split by \mstar\ in red (similar color coding to Figure \ref{fig:civ_rvir}) and in orange, the average $W_\lambda(1548)$ computed from individual sightlines binned by \dtran from the Quasars Probing Quasars (QPQ) survey \citep{prochaska_quasars_2014}. The bounds on the orange band represent the dispersion of the individual measurements (see Table 3 in \citealt{prochaska_quasars_2014}). To ensure a fair comparison between samples, we boost the $W_\lambda(1548)$ values by 50\% to match the $W_\lambda(\text{\civ})$ one would measure integrating over a full (optically thin) doublet, as is done in this study.
    The virial radius for a $\log(M_\mathrm{halo}/\mstar)\approx 12.5$ halo (which both sets of galaxies are expected to inhabit) is plotted as a dashed vertical line.}
    \label{fig:prochaska}
\end{figure}

\added{We estimated that the galaxies in the high mass \civ tertile reside in dark matter halos with masses $M_\mathrm{halo}\approx10^{12.5}~\msun$ (see Section \ref{subsec:civ} and \citealt{trainor_halo_2012}) and as such, offer an interesting comparison to the sample of $z\simeq 2.3$ QSO-QSO pairs in \citet{prochaska_quasars_2014} which probe the CGM around QSOs with comparable halo masses. We find the two sets of $W_\lambda(\text{\civ})$ measurements agree well and the dependence of $W_\lambda(\text{\civ})$ on \dtran is consistent (Figure \ref{fig:prochaska}). Evidently, the average \civ halos in the CGM of star-forming galaxies and QSOs with halo masses $M_\mathrm{halo}\approx10^{12.5}~\msun$ are quite similar. In other words, the agreement of $W_\lambda(\text{\civ})$ between samples suggests the CGM metal distribution at $\dtran \gtrsim 70$ kpc ($\sim 0.5 R_{\rm vir}$) for halos at these masses appears largely unaffected by the presence of a central QSO.}

We observe that $W_\lambda(1548)/W_\lambda(1550)$ transitions from $\simeq 1$ to $\simeq 2$ around $\dtran\simeq 60$ kpc (Figure \ref{fig:lyaciv}) and the velocity centroid over the full \civ doublet becomes consistently blueshifted at a similar projected distance (Figure \ref{fig:dv}) suggesting \civ becomes optically thin around that point. At $z \leq 0.1$, \citet{bordoloi_cos-dwarfs_2014} found $W_\lambda(\text{\civ})$ falls off as a power law with \dtran and remained detectable out to $\dtran \approx 100$ kpc. At $\dtran \simeq 100$ kpc, $W_\lambda(1548) \lesssim 0.1$ \AA\ and $\log\left[N(\text{\civ})\right]\lesssim 13.5$; within a factor of $\simeq 2$ of our observed $W_\lambda(\text{\civ})$ and consistent with optically thin \civ. 
Using QSO-galaxy pairs at $z\sim 2$, \citet{turner_metal-line_2014} also computed $W_\lambda(\text{\civ})$ and $N(\text{\civ})$. In their first and second \dtran bins (40 -- 130 kpc and 130 -- 180 kpc), like \citet{bordoloi_cos-dwarfs_2014}, they measured $\log\left[N(\text{\civ})\right]\simeq 13.0-13.5$ and $W_\lambda(\text{\civ})\simeq 0.01-0.1$ \AA\ again suggesting optically thin \civ at a similar \dtran to this study.

The most significant trend when splitting the pair sample up by SED-derived galaxy properties is between \civ absorption strength and \mstar, followed tentatively by SFR. The trend of increased $W_\lambda(\text{\civ})$ in higher mass galaxies is observed at $z\sim 0$ \citep{burchett_deep_2016}, $z\sim 1$ \citep{kumar_nature_2024} and at $z\sim 2$ \citep{mendez-hernandez_metal_2022}. \citet{bordoloi_cos-dwarfs_2014} also detected a mild correlation between \civ absorption strength and sSFR in $z<0.1$, $\log(\mstar/\msun)\leq 10$ galaxies but we do not observe this at higher redshift. \citet{bordoloi_cgm_2018} and \citet{wilde_cgm2_2021} report a correlation between \mstar\ and \lya absorption strength at $z<1$ but evidence for this is not seen in the \citet{mendez-hernandez_metal_2022} $z\sim 2$ sample. These findings can be reconciled since our sample extends to lower stellar masses (by approximately an order of magnitude) and the most significant change we observe in \lya absorption occurs between the two lowest mass bins in our sample ($\log(\mstar/\msun) \simeq 9$ and $\log(\mstar/\msun) \simeq 9.6$). It is interesting to note that $W_\lambda(\text{\ion{Mg}{2}})$ depends most strongly on \mstar\ and SFR for galaxies at $z\sim 0.5$ \citep{bordoloi_radial_2011,rubin_galaxies_2018,chen_circumgalactic_2024}, the same stellar population parameters that correlate with \civ, but we lack sufficient $S/N$ in the split pair samples to make a more thorough comparison. 

\added{With the advent of \textit{JWST}, observations of the CGM surrounding higher redshift galaxies have been facilitated. Using a sample of 29 star-forming galaxies at $2.3<z<6.3$ (all but four have $z > 4$) discovered within 300 kpc of a $z\sim 6.33$ QSO, \citet{bordoloi+24} find $W_\lambda(\text{\ion{Mg}{2}})$ decreases rapidly with \dtran and, like our sample, absorption is detected well beyond the virial radii of their host galaxies. In contrast, the majority of their rest-frame $W_\lambda(\text{\ion{Mg}{2}})$ are below our detection threshold in the $z\sim 2$ composites. The strongest absorption systems are preferentially detected in massive galaxies and those with high SFR. A small fraction of the \ion{Mg}{2} complexes have velocity spreads in excess of the galactic escape velocity, and interestingly, the median \dv is redshifted by 135 \kms. Unlike the net redshifted \lya we observe at $\dtran\sim 100$ kpc, this measurement may simply result from combining a relatively small number of inherently stochastic QSO sightlines.
In aggregate however, the trends observed in the \citet{bordoloi+24} sample are emblematic of our results seen at $z\sim 2$ and the preponderance of absorption beyond $R_{\rm vir}$ may suggest metals in these systems are not bound to the galaxies in which they were produced.}

\section{Summary}
\label{sec:conc}

We have measured the mean spatial and kinematic distribution of absorbing gas around $\langle\zfg\rangle=2.03 \pm 0.36$ star forming galaxies at projected distances $8\leq \dtran/\mathrm{kpc} \leq 250$ using 2738 KBSS foreground/background galaxy pairs (see Figure \ref{fig:indobjs} for some examples and Figure \ref{fig:hist} for sample statistics). The foreground galaxies studied have a median stellar mass $\log(\mstar/\msun)=9.6$ and span the range $8\lesssim {\rm log(\mstar/M_{\odot})} \lesssim 11$ (Figure \ref{fig:sedprops}). For each galaxy foreground/background pair, we shifted background KCWI and LRIS galaxy spectra to the redshift of their associated foreground galaxies and constructed stacks in the rest-frame of the foreground galaxy, binned by projected distance \dtran (Figure \ref{fig:fullstack}). We measured the properties of the foreground absorbing gas using rest-UV interstellar absorption features in the composite spectra. Our principal conclusions are:\

\begin{enumerate}
    
    \item We detected significant \lya absorption throughout the full range of \dtran sampled (Figures \ref{fig:ew_b} and \ref{fig:lyaciv}a), consistent with previous results using background QSOs \citep{adelberger_galaxies_2003,adelberger_connection_2005,rudie_gaseous_2012,turner_metal-line_2014} and background galaxy pairs \citep{steidel_structure_2010,chen_keck_2020}. The \lya covering fraction falls off as 
    \edit1{$f_c\propto r^{-0.33}$}, in agreement with the previously listed $z\sim 2$ studies, but noticeably shallower than most results at lower redshift (Table \ref{tab:ewfits}). 

    \item After \lya, \civ is the strongest absorption feature detected in stacks. \edit1{\civ closely tracks \lya ($f_c(\text{\civ})\propto r^{-0.27}$) for all measured \dtran (Figure \ref{fig:ew_b}), and both have similar line-of-sight kinematics at $\dtran \lesssim 100$ kpc. In addition to the shallower \ew vs.\ \dtran dependence, the $z \sim 2$ velocity fields are considerably more active than those observed at low-$z$. The \civ line-of-sight velocity second moment $\sigma(\text{\civ})$ reaches a minimum at $\dtran\simeq 60$ kpc ($\simeq0.7\,R_\mathrm{vir}$) signaling the approximate region over which the galaxy has the largest influence on the properties of its CGM (Figures \ref{fig:sigma_b} and \ref{fig:civ_rvir}).}

    \item Intermediate ionization species (\siiii and \siiv) are detected out to $\dtran \simeq 200$ kpc,  while the absorption strength of low ionization species (e.g., \siii, \cii, and \oi)  decline more steeply with increasing \dtran (Table \ref{tab:ewfits}),  with weaker species falling below our detection limit between $\dtran \simeq 60$ kpc (e.g., \siiiwh) and $\dtran \simeq 100$ kpc (e.g., \oi; Figure \ref{fig:ew_b}). In general, higher ionization metallic species trace gas with a wider range of %
    LOS velocity dispersions when compared to the low ions (Figures \ref{fig:sigma_b} and \ref{fig:linebins}). 
    Using $W_\lambda(1260)/W_\lambda(1526)$, we estimated the saturation of the \siii transitions with \dtran and found that \siii is strongly saturated for $\dtran \lesssim 50$ kpc, and becomes optically thin at $\dtran \gtrsim 70$ kpc. We showed how absorption strength varies on a per-transition basis (Figure \ref{fig:linebins}) and demonstrated that the low and high ions with similar absorption profiles in DTB spectra of the foreground galaxies DTB (e.g., \siii and \cii) tend also to exhibit similar velocity profiles at $\dtran \simeq 30-50$ kpc when significantly detected (Figure \ref{fig:lowionsamev}).

   \item \edit1{We constructed 2-D maps of \lya absorption strength parametrized as the covering fraction (absorption depth) as functions of \dtran and \vlos (Figure \ref{fig:heatmaps}). \lya absorption spans the full velocity range $\left|\vlos\right|\lesssim v_\mathrm{esc}$ of the average foreground galaxy, with some absorption detected in excess of $v_\mathrm{esc}$ at all \dtran, suggesting that a significant fraction of the CGM may be unbound. As in \citet{chen_keck_2020}, we observed a distinct ``edge'' in the $f_c(\lya)$ map at $\dtran\simeq 80$ kpc ($\simeq R_{\rm vir}$ for the average galaxy in the sample) that corresponds to a local minimum in the line of sight velocity dispersion $\sigma(\lya)$, beyond which there is a general flattening of $\sigma(\lya)$. We argued that $\dtran \sim 80$ kpc represents a transition from primarily outflowing gas at $\dtran\lesssim 80$ kpc to inflowing and/or ambient CGM gas beyond $\dtran\gtrsim 80$ kpc. }

    \item \edit1{We split the pair sample into sub-samples based on foreground galaxy \mstar, SFR, sSFR, and reddening parametrized by $E(B-V)$. The strongest observed trend is between \civ absorption strength and \mstar\ (Figures \ref{fig:sed4x4}, \ref{fig:sedciv3}, and \ref{fig:civ_rvir}):\ $W_\lambda(\text{\civ})$ increases by a factor of $\simeq 3.5$ at a fixed \dtran between galaxies in the lowest ($\log(\mstar/\msun)\simeq 9.0$) and highest  ($\log(\mstar/\msun)\simeq 10.2$) \mstar\ bins. Over the same range in \mstar, $W_\lambda(\lya)$ differs by a factor of $\simeq 2$ at $\dtran \simeq 70$ kpc, but the difference becomes insignificant beyond $\dtran \sim 150$ kpc (Figure \ref{fig:lya_rvir}). While a weaker trend of $W_\lambda(\text{\civ})$ with SFR is also observed, it may be a by-product of the well-known \mstar--SFR correlation.  The self-similarity of $W_\lambda(\text{\civ})$ and $\sigma(\text{\civ})$ across the \mstar\ sub-samples when \dtran is normalized by $R_{\rm vir}$ suggests that dark matter halo mass sets the scale for both the spatial extent and kinematics of the CGM gas distribution at $z \simeq 2$. We show that the foreground galaxy sub-sample with estimated halo mass similar to that of QSOs at $z \sim 2-3$ ($\log(M_{\rm halo}/M_{\odot}) \simeq 12.5$) exhibits comparable CGM \ion{C}{4} absorption at $\dtran \gtrsim 70$ kpc where the comparison is possible. Evidently, the presence of an active QSO does not appear to significantly alter the CGM absorption (at least for $\dtran \gtrsim 70$ kpc) of the host galaxy compared to those without active QSOs (Figure~\ref{fig:prochaska}), which may have implications for our understanding of AGN feedback for relatively massive galaxies during the peak of the QSO era.}

\end{enumerate}

\edit1{KCWI has significantly improved the effective spectral resolution and spatial coverage of the CGM for $\dtran \le 250$ kpc of typical star-forming galaxies at $z\sim2$. This sample, which augments the earlier KBSS studies using galaxy-galaxy pairs \citep{adelberger_connection_2005,steidel_structure_2010,chen_keck_2020}, represents a data set that is complementary to those using QSO sightlines at $z \sim 2-3$. With the current sample, we have consistently detected \lya and \civ over $\dtran=8-250$ kpc and improved the \dtran sampling of several weaker far-UV resonance transitions for $\dtran \lesssim 100$ kpc. For the first time for galaxies at $z \simgt 2$ we have shown that significant correlations exist between the CGM properties (absorption strength and LOS velocity dispersion) and their host galaxies extending from $\dtran \sim 0$ (DTB) to projected distances well beyond the virial radii of the host halos.} %

\edit1{Galaxy-galaxy foreground/background pairs have the advantage of sampling larger swaths of the CGM and converging more efficiently to the mean gas distribution compared to surveys using background QSOs, at the expense of $S/N$ and spectral resolving power for individual sightlines. At small projected distances ($\dtran \lesssim 50$ kpc), we have shown that \lya\ photons produced in the foreground galaxy and scattered {\it into} our line of sight toward a background galaxy affects both the equivalent width and the apparent line-of-sight velocity distribution of \lya\ absorption measured in the spectrum of the background galaxy, confirming speculation along similar lines in \citet{chen_keck_2020}.}
\edit1{IFUs offer a means of ``correcting'' this contamination given their ability to chart both emission and absorption halos around galaxies contiguously over fields the size of the average $z\sim 2$ CGM. In the future, deep integrations targeting the emission halos around individual galaxies together with background galaxies (or QSOs) sampling absorption in the same galaxy CGM at similar \dtran will allow interesting constraints on the structure and kinematics of the clumpy CGM that must be responsible for both. }

\bigskip 

N.\ Z.\ P.\ thanks Yuanze Ding, Zhihui Li, Evan Nu\~{n}ez, and Zhuyun Zhuang for thought-provoking discussions, sumptuous meals, and general camaraderie during the writing of this paper. N.\ Z.\ P.\ additionally thanks Cameron Hummels for his insights into modeling the results presented in this study. The authors thank Rosalie McGurk, Greg Doppmann, Mateusz Matuszewski, and Don Neill for constructive conversations about KCWI. 
The authors are indebted to Kurt Adelberger, Milan Bogosavljevi\'{c}, Dawn Erb, Matthew Hunt, David Law, Olivera Rakic, Naveen Reddy, Gwen Rudie, Alice Shapley, Allison Strom, Rachel Theios, Ryan Trainor, and Monica Turner for their efforts towards the KBSS survey over the past $\sim$two decades. \added{The authors thank the referee for thoughtful feedback that improved the final version of this paper.}
N.\ Z.\ P.\ and C.\ C.\ S.\ were supported in part by NSF grant AST-2009278. This research has made use of the Astrophysics Data System, funded by NASA under Cooperative Agreement 80NSSC21M00561. Finally, the authors wish to recognize and acknowledge the very significant cultural role and reverence that the summit of Maunakea has always had within the Native Hawaiian community. We are most fortunate to have the opportunity to conduct observations from this mountain.

\software{\texttt{matplotlib} \citep{hunter_matplotlib_2007}, \texttt{numpy} \citep{harris_array_2020}, \texttt{scipy} \citep{virtanen_scipy_2020}, \texttt{Astropy} \citep{the_astropy_collaboration_astropy_2013, the_astropy_collaboration_astropy_2018,the_astropy_collaboration_astropy_2022}, \texttt{YT} \citep{turk_yt_2011}, \texttt{TRIDENT} \citep{hummels_trident_2017}, \texttt{tqdm} \citep{da_costa-luis_tqdm_2024}, \texttt{scikit-image} \citep{van_der_walt_scikit-image_2014}, multiprocess \citep{mckerns_building_2012}, corner \citep{foreman-mackey_cornerpy_2016}, \textsc{IRAF}\footnote{NOIRLab IRAF is distributed by the Community Science and Data Center at NSF NOIRLab, which is managed by the Association of Universities for Research in Astronomy (AURA) under a cooperative agreement with the U.S. National Science Foundation.} \citep{tody_iraf_1986,tody_iraf_1993,national_optical_astronomy_observatories_iraf_1999,fitzpatrick_modernizing_2024}, \texttt{seaborn} \citep{waskom_seaborn_2021}, \texttt{plotly} \citep{plotly}, \texttt{L.A.COSMIC} \citep{vandokkum_cosmicray_2001}, \textsc{IPython} \citep{perez_ipython_2007}, QFitsView \citep{qfitsview}}

\facilities{Keck:II (KCWI), Keck:I (LRIS), Keck:I (MOSFIRE), Keck:I (HIRES)} 

\appendix

\begin{figure}
    \epsscale{1}
    \plotone{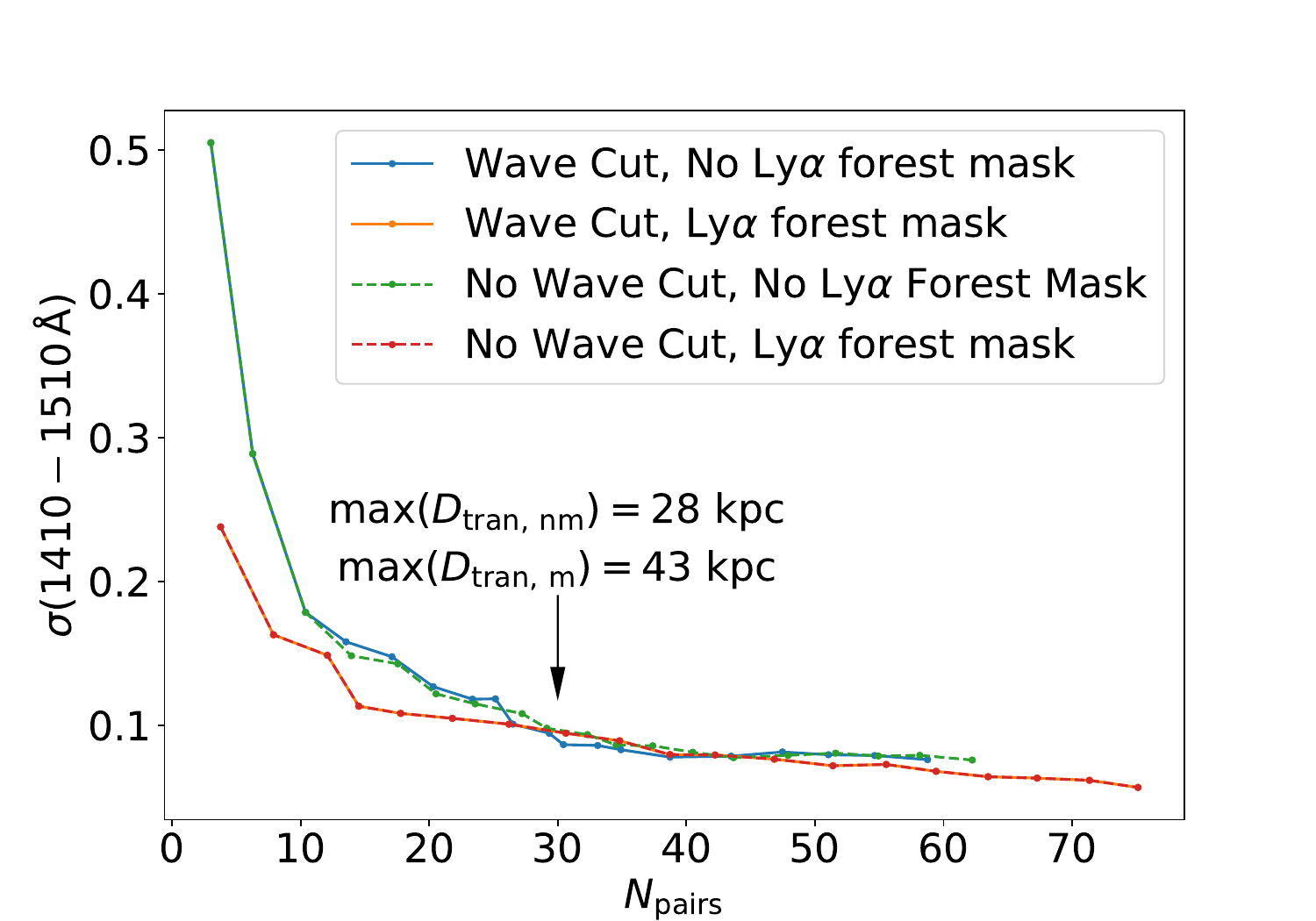}
    \caption{Continuum noise ($\sigma$) for a given number of pairs ($N_\mathrm{pairs}$) depending on wavelength trimming method. The two options include manual wavelength cuts or a redshift dependent rejection that excludes objects with foreground absorption that falls within the \lya forest of the background spectrum.}
    \label{fig:trim}
\end{figure}

\section{Wavelength Trimming}
\label{app:wavetrim}

In deciding which galaxies contribute to the galaxy-galaxy pair sample, both \citet{steidel_structure_2010} and \citet{chen_keck_2020} imposed an upper bound on the redshift difference between the foreground and background pairs. The former uses $\Delta z \leq 1$ while the latter imposes $\frac{\Delta z}{1+\zfg} < 0.3$; for the median foreground and background galaxy redshifts in our sample, these criteria are very similar. We observe that 
\begin{equation}
    \lambda_\mathrm{bg}(1+\zbg) = \lambda_\mathrm{fg}(1+\zfg),
\end{equation}
or equivalently, 
\begin{equation}
    \frac{\Delta z}{1+\zfg}=\frac{\lambda_\mathrm{fg}}{\lambda_\mathrm{bg}}-1
\end{equation}
where $\Delta z = \zbg - \zfg$ and $\lambda_\mathrm{fg}$ and $\lambda_\mathrm{bg}$ are the rest wavelengths of a particular transition in the foreground and background frames, respectively. To prevent foreground \civ absorption from being contaminated by unrelated \lya forest absorption, we chose $\lambda_\mathrm{fg}=1549\,\mathrm{\AA}$ and $\lambda_\mathrm{bg}=1216\,\mathrm{\AA}$ so only pairs with $\frac{\Delta z}{1+\zfg}< 0.27$ are included. Naturally, if one observed significant noise from the background \lya forest contributing to foreground absorption, this requirement could be strengthened at the cost of many potential foreground/background galaxy pairs; for example, requiring $\frac{\Delta z}{1+\zfg}< 0.15$ would prevent \lya forest absorption from contributing to the foreground \siiv profiles.

In our sample, we experimented with several such redshift cuts and found that the continuum noise level is indeed reduced, especially when the number of pairs in a bin is small (see the red dashed curve in Figure \ref{fig:trim}). In addition to a blanket cut, we also tried wavelength trimming for each individual spectrum, choosing the highest $S/N$ regions in the bandpass to contribute to the final stacks. This typically involved masking the short wavelength end of the spectrum in the \lya forest and (in the case of LRIS) the wavelengths near the dichroic cutoff near $5600$ \AA. The fact that the noise levels in the stacks remain similar with or without a manual wavelength mask suggests that such trimming is not necessary (see the solid vs.\ dashed curves in Figure \ref{fig:trim}). 

While both noise mitigation techniques converge once the number of pairs in a bin exceeds 30, the range of \dtran differs. Figure \ref{fig:trim} shows the maximum \dtran in a bin of 30 pairs drawn from a pair list sorted by \dtran. In other words, when applying a redshift cut, the 30 pairs with the smallest separations probe $\dtran\lesssim 43$ kpc; however, if such a cut is not imposed and each spectrum is manually trimmed of noisy regions, the first 30 pairs sorted by impact parameter probe gas within 28 kpc of an average galaxy halo. Since the noise levels in stacks with $N\gtrsim 30$ pairs are comparable and all of the stacks in our sample contain at least thirty pairs, we chose to avoid using the blanket redshift cut (dashed green curve in Figure \ref{fig:trim}), effectively maximizing the number of galaxy pairs contributing to a stack.

\section{HIRES}
\label{app:hires}

\begin{figure*}

     \plottwo{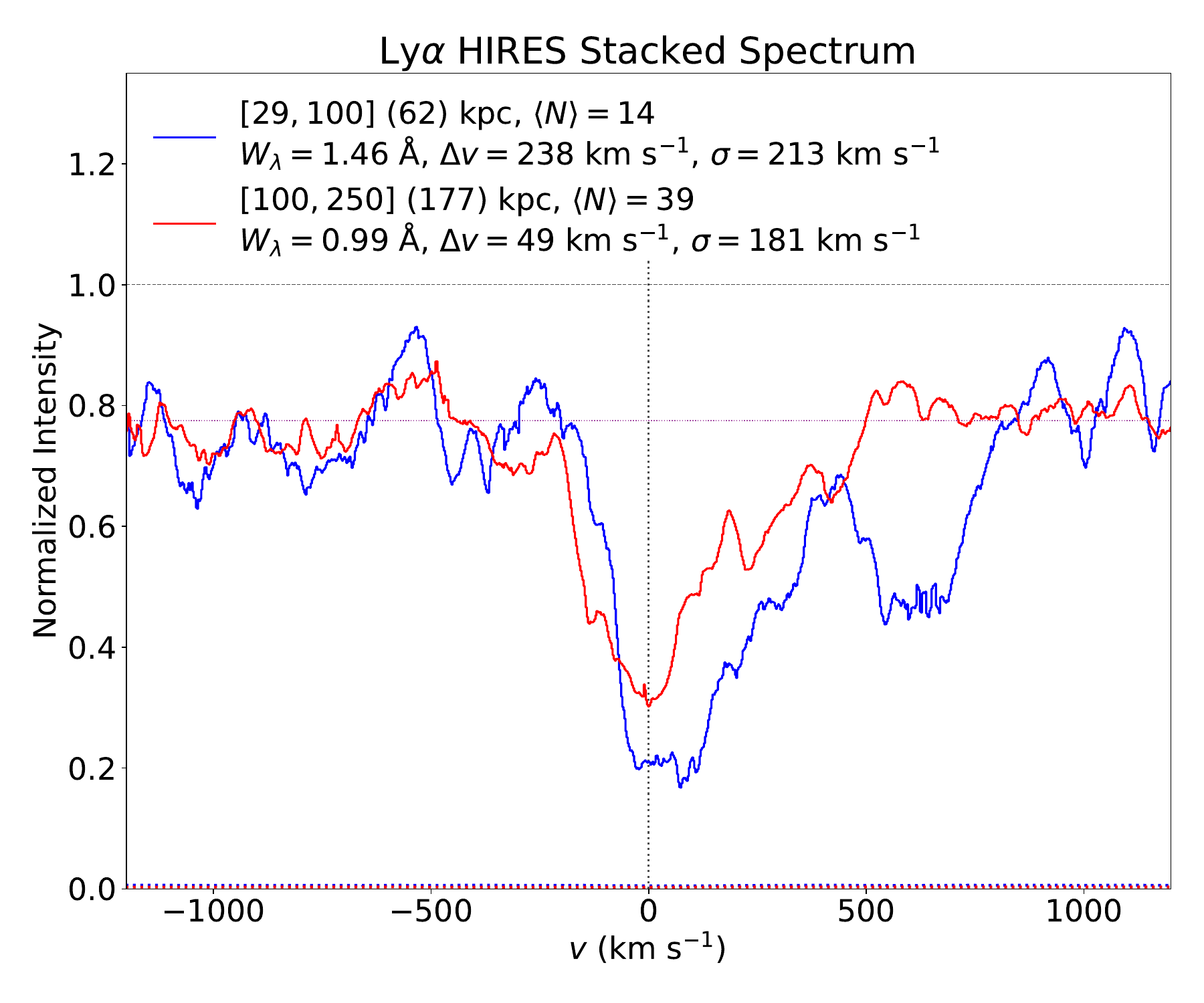}{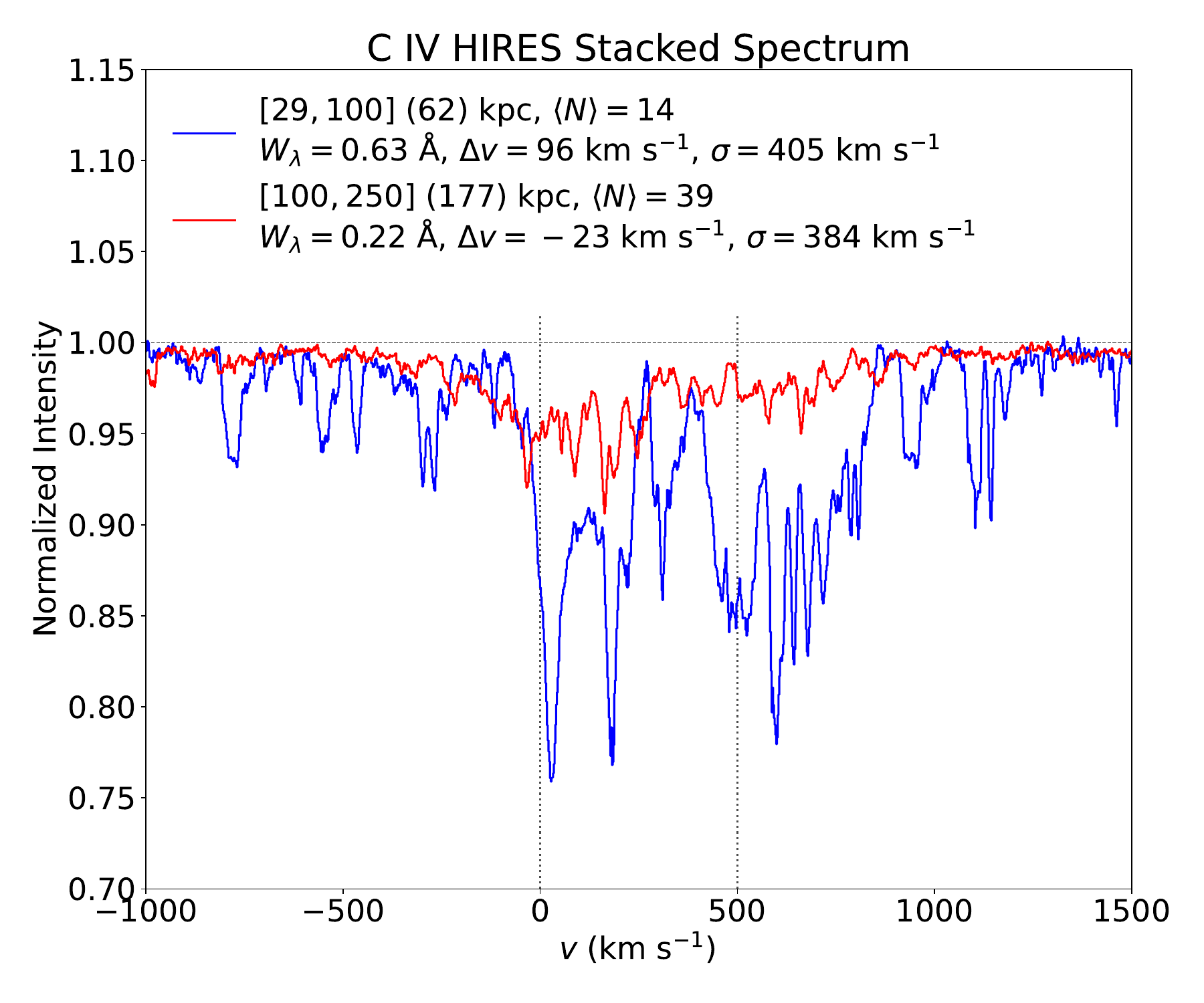}
    \caption{Stacked HIRES absorption spectra shifted into the rest frame of the foreground systems centered around the \lya and \civ transitions. The range of \dtran and the number of pairs contributing are enumerated in the plot legend. For each bin, $W_\lambda$, $\Delta v$, and $\sigma$ are also shown. The horizontal dotted purple line denotes the suppressed continuum used for the $W_\lambda(\lya)$ measurement. 
    }
    \label{fig:hires}
\end{figure*}

In order to provide some insight into the cause of asymmetries observed for some of the galaxy-galaxy pair samples,  we assembled a sample of QSO-galaxy pairs using existing HIRES spectra of the central QSOs in the KBSS fields (see \citealt{rudie_gaseous_2012,turner_metal-line_2014}).
Briefly, the KBSS QSOs have $\langle z \rangle \simeq 2.7$, and their HIRES spectra have $R\simeq 45,\!000$ with $S/N\simeq 50-100$~pix$^{-1}$ over the observed wavelength range $3100-6000$~\AA.
Figure~\ref{fig:hires} shows stacks constructed using the same method as for the galaxy-galaxy pair samples, for 53 foreground galaxies with $\dtran \le 250$ kpc. The results are shown in two broad bins of \dtran, for the \lya and \civ transitions. The bins were chosen to be similar to those used for the galaxy-galaxy pair samples discussed in Section~\ref{sec:bin}.) We note that the QSO-galaxy pair measurements pertain to foreground galaxies in the same KBSS survey regions and at similar redshifts, but are completely independent of the galaxy-galaxy pair measurements since the QSO sightlines sample different \dtran even when probing the same foreground galaxy. 

Figure~\ref{fig:hires} shows that, even when the background spectra have very high $S/N$ and spectral resolution, the mean profiles from 
combining QSO-galaxy pairs can be strongly affected by sample variance -- 
stochastic variation of peculiar velocities and component line strength contributed by different QSO-galaxy pairs -- as well as by unrelated absorption lines from other redshifts that fall happen to fall within the same wavelength intervals. Applying outlier rejection in forming the stacks reduces the effects of unrelated absorption features and/or unusual absorption profiles in one or a few individual QSO-galaxy pairs; however, with small pair numbers 
the average profiles can appear quite asymmetric in velocity (e.g., the blue curves in both panels of Figure~\ref{fig:hires}). 

For absorption systems with a high incidence rate such as \lya (lefthand panel of Figure~\ref{fig:hires}), an additional effect of combining many systems shifted to the redshift of a foreground galaxy is that the net absorption is superposed on a suppressed local continuum, reflecting the average line blanketing in the \lya forest at the mean redshift of the foreground galaxy sample -- in this case, $\langle z \rangle \sim 2.3$. For stacks with a relatively small number of contributing QSO-galaxy pairs, sample variance in the local \lya forest can cause ``noise'' in the net absorption profile due to fluctuations in the local continuum. The apparent continuum fluctuations become smaller for samples with larger numbers of QSO-galaxy pairs, as do the net velocity asymmetries (i.e., centroid shift relative to $v=0$; compare the blue and red composites in the lefthand panel of Figure~\ref{fig:hires}.)

For metal line transitions such as the \ion{C}{4} doublet, when limited to redshifts that place \ion{C}{4} at wavelengths longward of \lya in the spectrum of the background QSO (as in the righthand panel of Figure~\ref{fig:hires}), the continuum remains well-defined. However, small samples may still produce asymmetric \ion{C}{4} profiles that do not resemble a simple velocity-broadened \ion{C}{4} doublet with equivalent width ratio $1 \le W_{\lambda}(1548)/W_{\lambda}(1550) \le 2$, and the velocity centroid of the absorption may also appear shifted with respect to the foreground galaxy systemic redshift for a similar reason. When a larger number of QSO/galaxy pairs is averaged (red spectrum in the righthand panel of Fig.~\ref{fig:hires}) the net profile is much smoother, and has an apparent doublet ratio close to that expected for unsaturated absorption. We observe similar behavior, albeit using galaxy spectra with $\sim 20-30$ times lower resolving power and $S/N$, in the galaxy-galaxy pair samples covering similar ranges of \dtran.

For \lya and \civ, the values of $W_\lambda$, \dv, and $\sigma$ for each \dtran bin are shown in the legend of Figure~\ref{fig:hires}. 
We find general agreement in $W_\lambda(\lya)$, \dvla, and absorption profile morphology between the HIRES $\dtran=100-250$ kpc stack of \lya and the galaxy pair stacks over the same range of \dtran plotted in Figure \ref{fig:sed4x4}. The computed $\sigma(\lya)$ for the HIRES stacks are $50-100$ \kms\ lower than their full sample or SED-binned counterparts, but we note that $\sigma$ depends sensitively on continuum normalization. Similar to the lower resolution stacks, we measure a redshifted velocity centroid at small \dtran ($\dv=238$ \kms\ at $\dtran=62$ kpc) that approaches the systemic velocity with increasing \dtran ($\dv=49$ \kms\ at $\dtran=177$~kpc).
The \civ profiles all show broad ($\sigma \sim 400$ \kms), nearly symmetric absorption in velocity space with a $\simeq 100$ \kms\ shift towards a blueshifted centroid between the inner (blue) and outer (red) bins. This global trend is consistent with a decreasing optical depth where the individual \civ systems transition from an equivalent width ratio $W_{\lambda}(1548)/W_{\lambda}(1550)$ of one to two.

\section{Stacking Method}
\label{app:stackmethod}

\begin{figure}

    \plotone{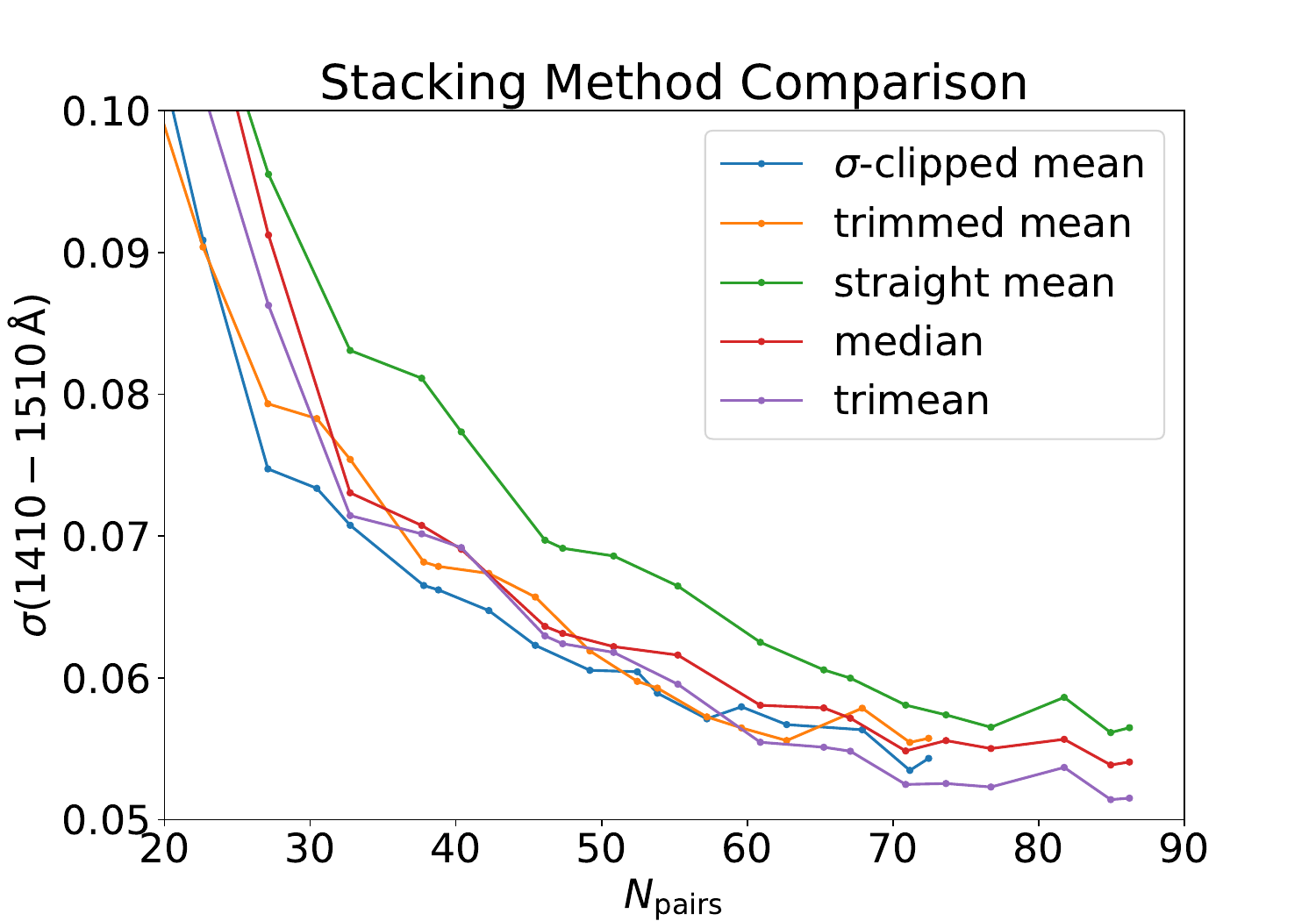}
    \caption{Similar to Figure \ref{fig:trim}, the continuum noise level over the 1410--1510 \AA\ wavelength interval as a function of the number of pairs in a bin is plotted for several methods of combining spectra within that bin.}
    \label{fig:stackmethodcomp}
\end{figure}

Before deciding on a trimmed mean (averaging the 5th -- 95th percentile spectral pixels) for constructing stacks, we experimented with several other stacking options for the final composite spectra. In theory, with high resolution normalized spectra, any absorption manifests as a narrow feature centered at the rest wavelength in the frame of the foreground galaxy. Foreground galaxies without significant absorption will have normalized intensities close to unity. The distribution of intensities at a given rest wavelength is expected to have a mode near unity and a skew towards smaller normalized intensity caused by real absorption. For any distribution, the mean absorption profile will converge faster than the median for a fixed number of galaxy pairs; on the other hand, combining shifted spectra using the pixel-by-pixel mean can result in noise caused by unrelated absorption at other redshifts. This source is clearly exacerbated by averaging over only a few galaxy pairs where background \lya forest absorption systems can produce large noise spikes at random foreground wavelengths.

In dealing with relatively low resolution ($R\simeq 670$) LRIS spectra, \citet{chen_keck_2020} found that the median of all foreground galaxies in a stack agrees well with a clipped mean using the same spectra, but opted for the median. Now incorporating moderate resolution ($\langle R\rangle\simeq 1500$ at \lya and $\langle R\rangle\simeq 1900$ at \civ) KCWI spectra, we revisit this issue. Figure \ref{fig:stackmethodcomp} shows the foreground continuum noise level for a stack as a function of the number of pairs included in the bin. The list of pairs begins at the minimum \dtran and is sorted by \dtran so more pairs probe larger \dtran. The stack is either the mean (green), sigma-clipped mean (blue), trimmed mean (orange), median (red), or the trimean (purple; Equation \ref{eqn:trimean}) of the pixels at each wavelength point in that bin. The trimmed mean clips pixels that are either below the 5th percentile or above the 95th percentile at a given wavelength. In the small sample size limit, this algorithm rejects the minimum and maximum values; for larger samples, more pixels may be rejected (effectively 4 -- 6 objects for most of our \dtran bins). The trimean is defined as  
\begin{equation}
\label{eqn:trimean}
    \operatorname{TM}\equiv \frac{1}{2}\left(Q_2 + \frac{Q_1 + Q_3}{2}\right)
\end{equation}
where $Q_i$ are the $i$th quartiles of the underlying distribution. Although a less efficient estimator than the mean, the trimean emphasizes both the central values and the extrema of a distribution. For our spectra, this measure of central tendency helps mitigate continuum noise while preserving the strength of absorption features.

We note that the straight mean has $\sim 10\%$ higher noise levels overall compared to the other statistics, which are consistent with one another (to within $\simeq 5\%$). The sigma-clipped mean has a minor advantage for $N<50$ pairs, while the trimean has the lowest noise for $N>60$ pairs. With $N>30$ pairs, each measure of central tendency (straight mean excluded) is consistent with the others, suggesting they are at that point all reflective of the underlying distribution. The straight mean begins to converge to the rest of the metrics with $N>60$ pairs.

\begin{figure*}

    \plotone{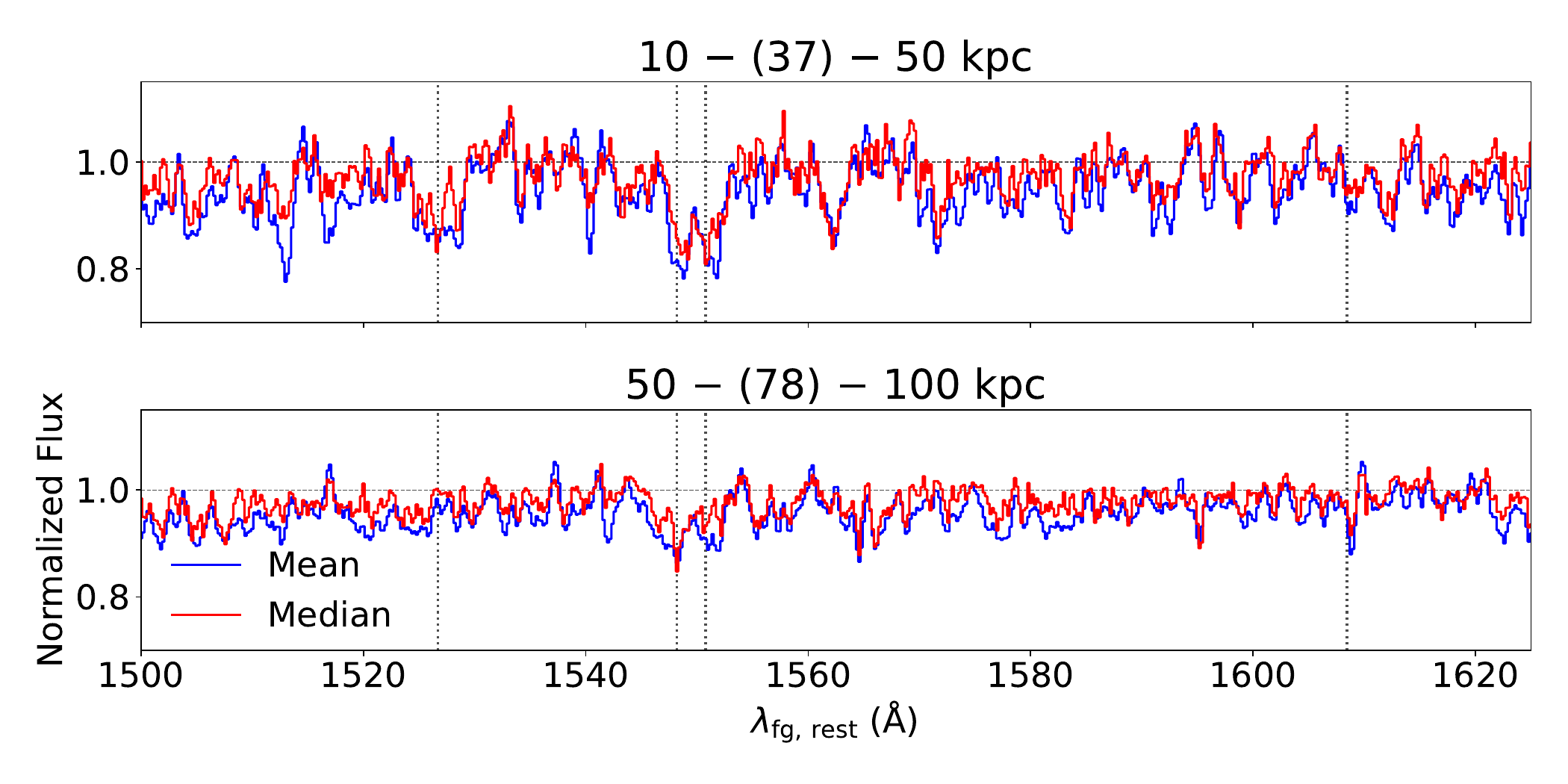}
    \caption{Background spectra shifted to the foreground rest frame and stacked using a trimmed mean (blue) and median (red) for two ranges of \dtran (\textit{top:} 10 -- 50 kpc, \textit{bottom:} 50 -- 100 kpc). The vertical dotted lines show \siiiwh, \civ, and \feiiw and each input spectrum has been normalized as described in Section \ref{sec:stacking}.}
    \label{fig:medmeanstack}
\end{figure*}

\begin{figure}

    \plotone{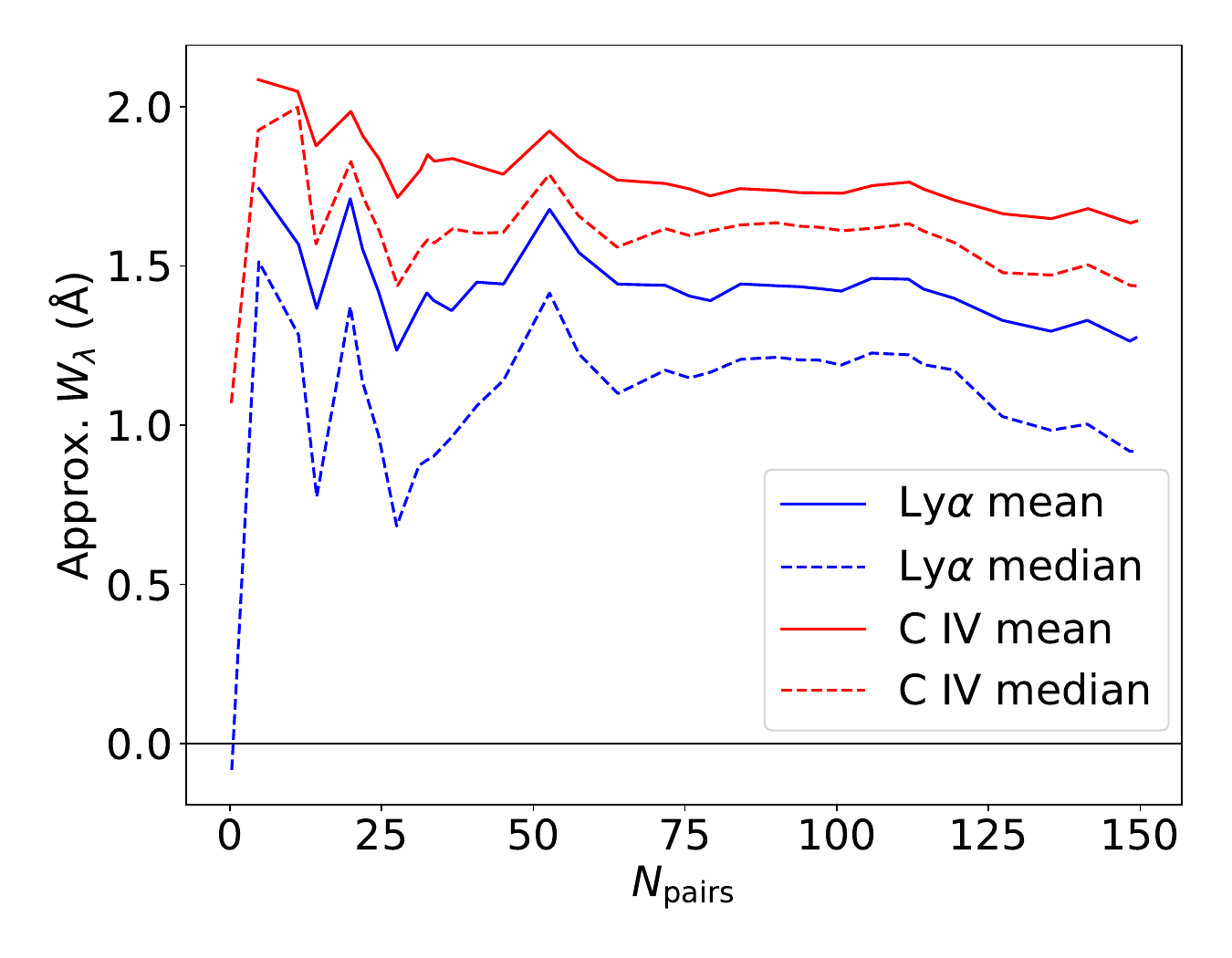}
    \caption{\lya (blue) and \civ (red) approximate equivalent width is measured from stacks with varying numbers of constituent galaxy pairs as a test of line profile convergence. The curves start at $\dtran\simeq10$ kpc and add pairs with progressively larger separations reaching up to $\dtran \simeq 52$ kpc. Each stack is combined using either a trimmed mean (solid lines) or median (dashed lines).}
    \label{fig:lyacivconv}
\end{figure}

While the continuum noise level is consistent to within $\sim10\%$ regardless of method used, the same is not true for the absorption features. Figure \ref{fig:medmeanstack} shows two \dtran bins where the foreground absorption spectra were combined using a trimmed mean (blue) and a median (red). For each absorption line, we observe higher variability near the systemic velocity when using the median over the trimmed mean. As a result, the continuum level in the vicinity of the absorption features is less stable and the equivalent widths one would measure are systematically smaller (see the solid vs.\ dashed lines in Figure \ref{fig:lyacivconv}). Indeed, Figure \ref{fig:lyacivconv} shows spectra combined with a trimmed mean have $W_\lambda\sim 50\%$ larger than those measured in median combined spectra, regardless of the line being measured. The approximate $W_\lambda$ were measured by integrating over $|v| \leq 700$ (500) \kms\ around the systemic velocity for \lya (\civ). When considering small \dtran, the contribution of foreground \lya emission to the background galaxy spectrum (Appendix \ref{app:filling}) will bias both the median and average toward smaller \ew, but if the velocity spread between absorption and emission is not the same, the median may show more variability around the systemic velocity than the mean. We therefore choose to use the trimmed mean as it provides the most distinguishable absorption signal when present.

In using the trimmed mean, we experimented with several different bounds (e.g., asymmetric, Q1 \& Q3) but found that clipping the top and bottom 2 -- 6 (depending on the number of pairs in a bin) background spectra was sufficient for achieving noise levels comparable to that of the median while still detecting strong absorption. Put differently, $\lesssim 5\%$ of the spectra in a particular bin were noisy enough by themselves to result in severe contamination when stacked with $N \gtrsim 30$ others. It is worth remembering that when normalizing each background spectrum, we only allow regions where the spline fit is consistently above the local (DRP-computed) noise level (evaluated in $\Delta\lambda\simeq 50$ \AA\ intervals) to enter into the stack (see Section \ref{sec:stacking}). Such a requirement ensures that there is a continuum with at least $S/N\gtrsim 1$ from which to measure absorption. It therefore follows that few of the spectra entering into the stack would need additional trimming.

\begin{figure}
    \epsscale{1.2}
    \plotone{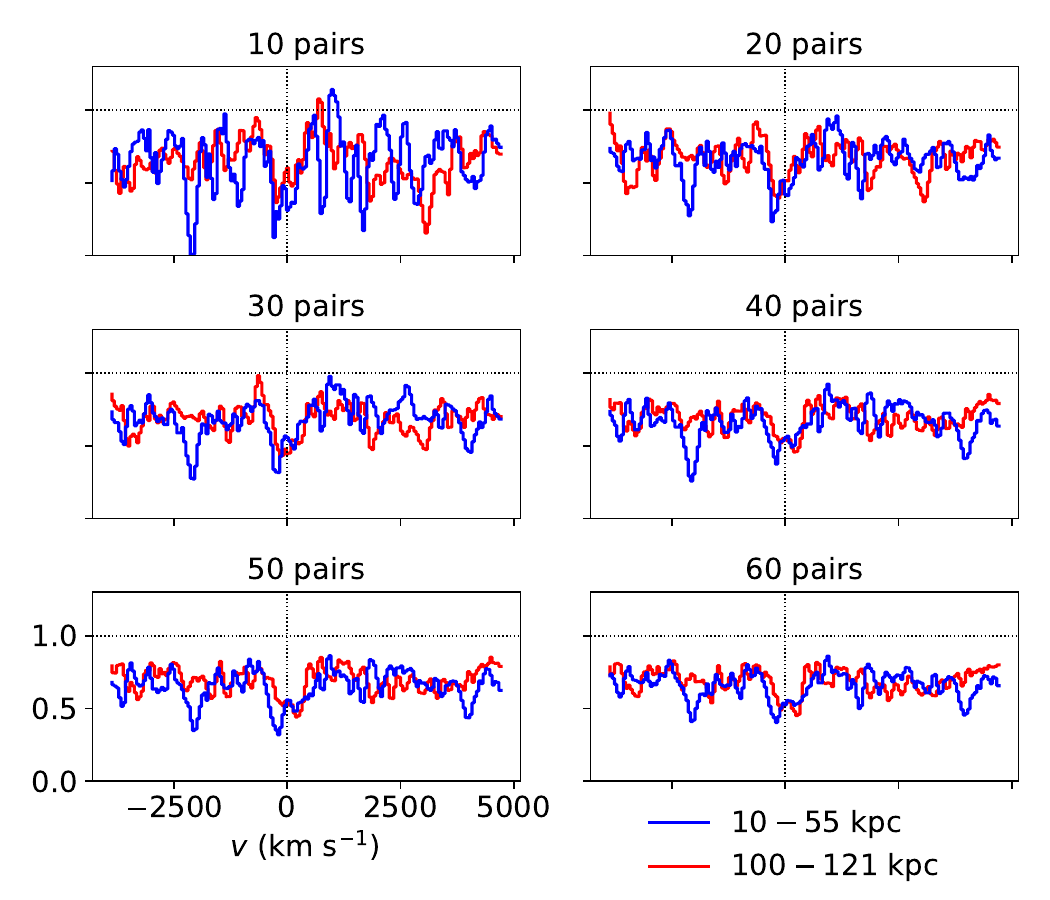}
    \caption{A spectral representation of Figures \ref{fig:stackmethodcomp} and \ref{fig:lyacivconv} showing how \lya absorption converges to a well-defined profile as more galaxy pairs are averaged. Two bin ``starting points'' are shown:\ $\dtran\simeq 10$ kpc in blue and $\dtran\simeq 100$ kpc in red; as more pairs are added, the median \dtran increases but \lya becomes easier to distinguish from the continuum. Note that the continuum is suppressed by the \lya forest when a sufficient number of pairs are averaged.}
    \label{fig:specconv}
\end{figure}

In addition to noise properties, we investigated the convergence of the \lya absorption profile for a given number of pairs. Figure \ref{fig:specconv} shows how the number of pairs averaged (using a trimmed mean) influences the net \lya profile. For transitions measured within the \lya forest of the background galaxy spectrum, a coherent absorption feature does not materialize until at least 10 pairs are averaged. Once 20 pairs are combined, \lya is discernible from the continuum and additional pairs help mitigate \lya forest noise spikes. \lya absorption strength is reduced with increasing \dtran (see Figure \ref{fig:ew_b}) so a larger number of pairs is required to produce absorption profiles with a comparable $S/N$. Although there is some sample variance, this effect can be seen by comparing the blue and red spectra in Figure \ref{fig:specconv}:\ for a given number of pairs, the larger \dtran bins (red) are marginally noisier around \lya than their small \dtran counterparts.

More quantitatively, Figure \ref{fig:lyacivconv} shows that the \ew varies considerably for stacks with fewer than 20 pairs due to stochastic background \lya forest absorption. However, once the number of pairs reaches $N \simeq 30$, \ew converges and remains roughly constant with increasing numbers of pairs. If one keeps adding pairs to the bin, the average \dtran sampled will increase and since pairs at larger \dtran tend to have weaker absorption (recall absorption \ew drops with \dtran; Figure \ref{fig:ew_b}), the measured \ew in the stacks will begin to fall off when $\gtrsim 100$ pairs are included, as seen in Figure \ref{fig:lyacivconv}. A similar trend is observed in \civ, albeit with slightly less variability when the number of pairs is small.

Putting all this together, for the trimmed mean stacks, we seek to optimize the number of pairs in each bin such that the continuum noise is minimized and the absorption feature is unambiguous, while still providing as finely sampled of a grid in galactocentric radius as possible. From this analysis and some experimentation with various bins, we find optimum stacks at $\dtran\lesssim 50$ kpc contain $N\simeq 30-80$ pairs and design the bins used in this study accordingly.

\section{The effects of L\lowercase{y}$\alpha$ emission halos of foreground galaxies}
\label{app:filling}

\begin{figure*}
\plottwo{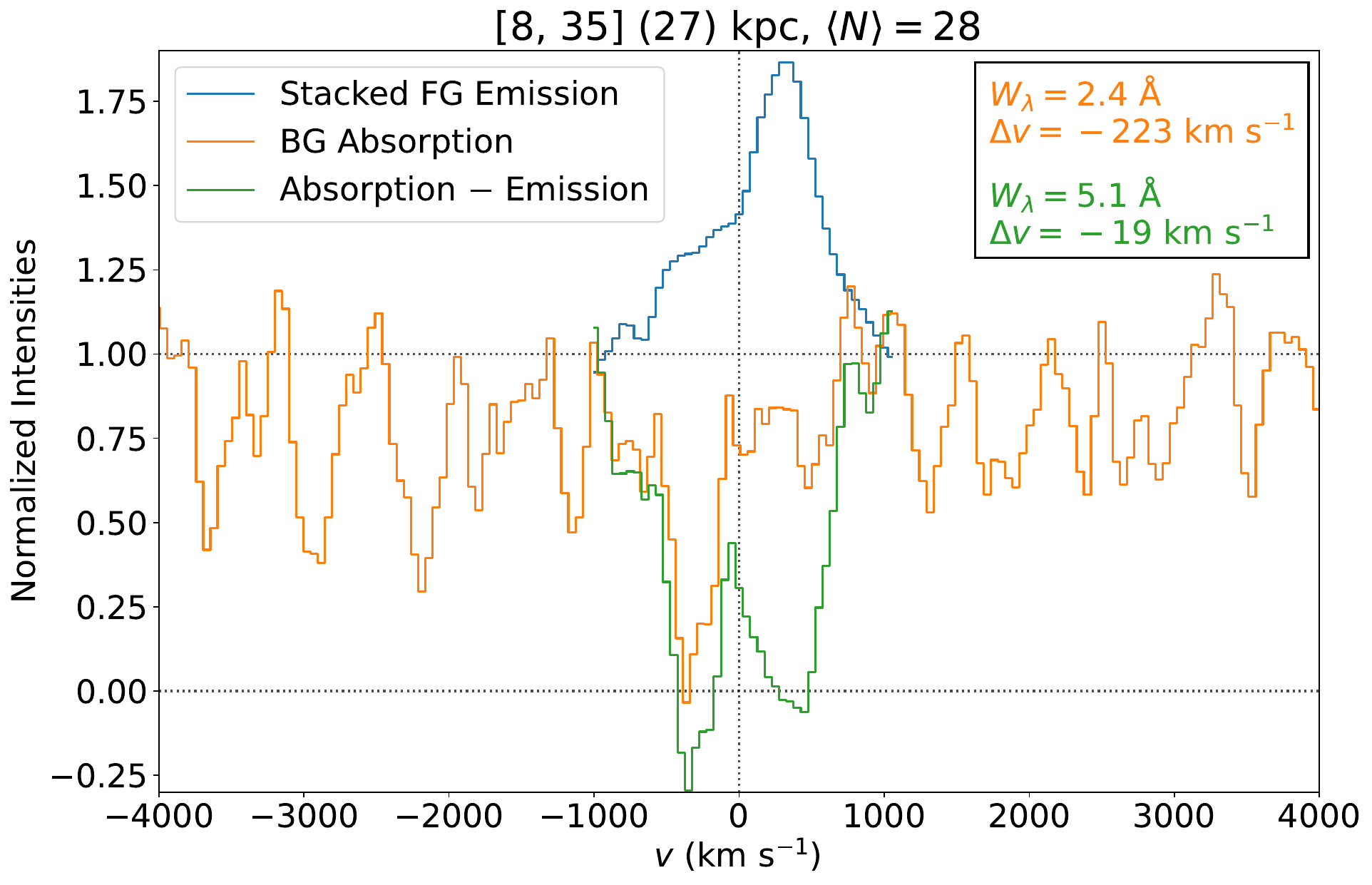}{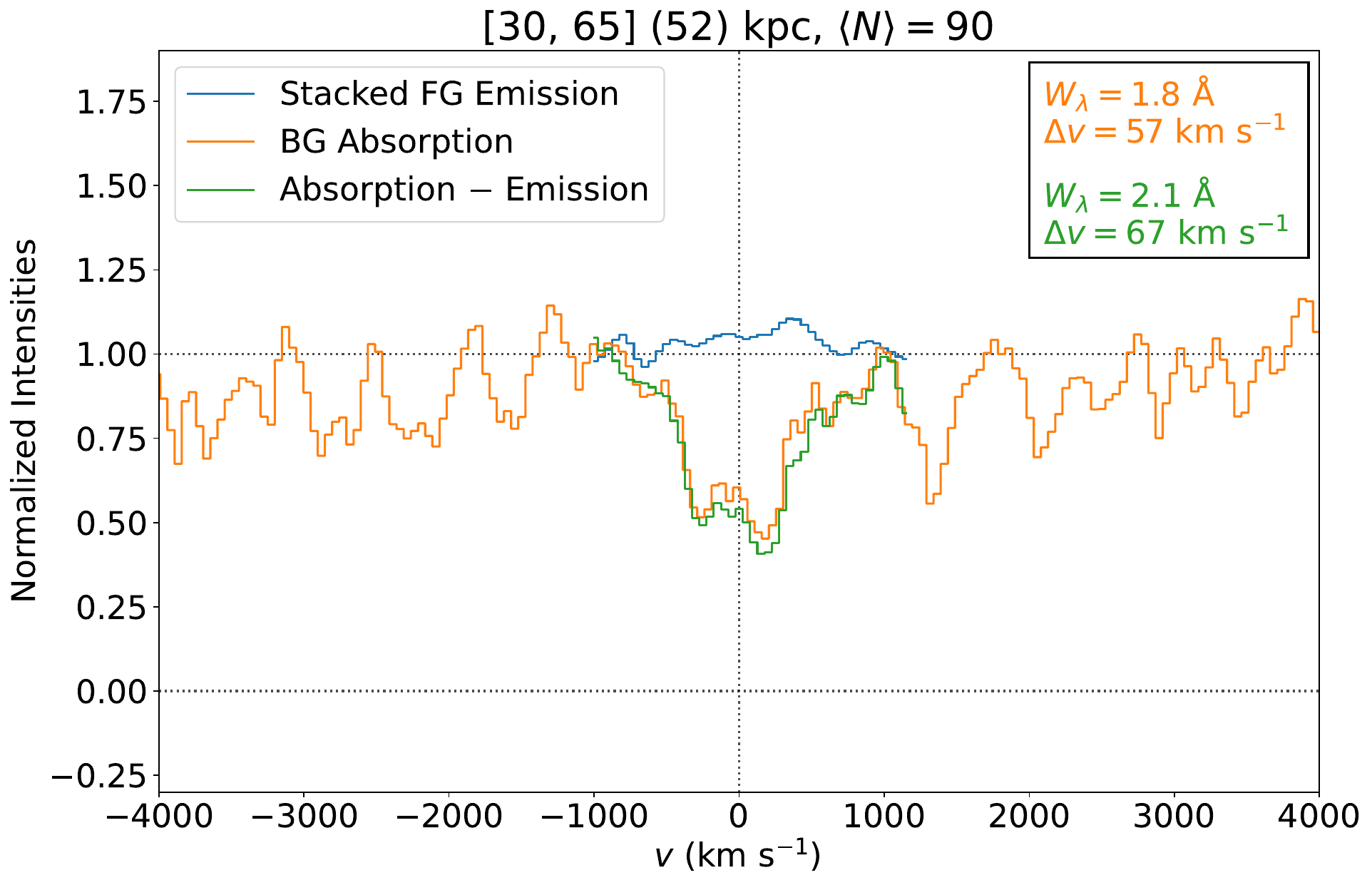}
    \caption{The contribution of \lya emission from the foreground galaxy (blue) to the \lya absorption profile measured in the background galaxy spectrum (orange) for two \dtran bins. The green spectrum is an estimate of the unaffected absorption profile, computed by subtracting the average emission feature from the absorption. $W_\lambda(\lya)$ and $\Delta v(\lya)$ from the observed and ``corrected'' spectra are shown in the top right of each plot as a color-coded list. The number of galaxies contributing to each stack is shown in each plot title.}
    \label{fig:emfill}
\end{figure*}

\begin{figure*}
\plotone{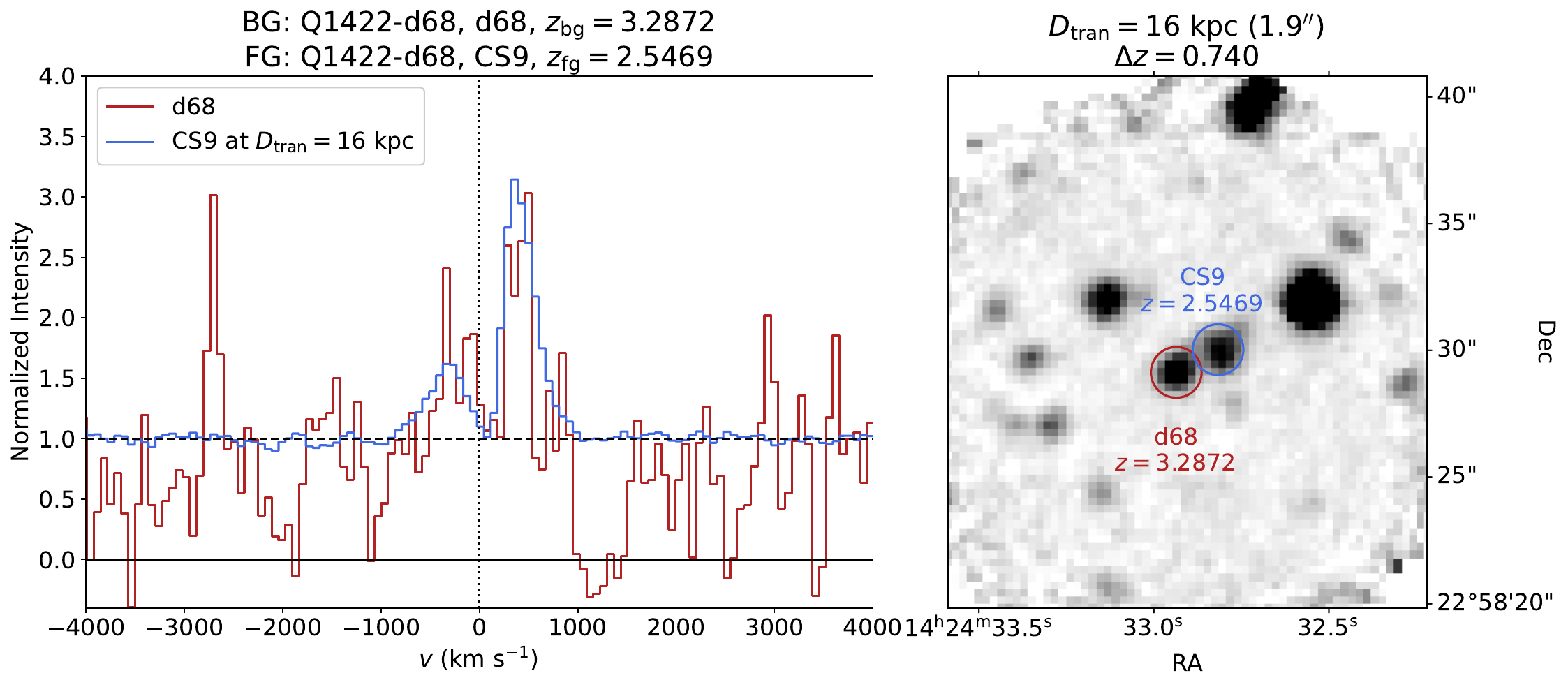}
    \caption{\textit{Left:} The spectra of Q1422-d68 (red) and Q1422-d68-CS9 at $\dtran=16$ kpc (blue) centered around \lya in the rest frame of CS9. \textit{Right:} White light image of the Q1422-d68 cube with d68 and CS9 circled (same color scheme). We observe emission from CS9 (foreground) contributing to the d68 (background) spectrum effectively masking any absorption within $\vlos\simeq1000$ \kms\ of the systemic velocity.}
    \label{fig:d68}
\end{figure*}

\citet{chen_keck_2020} noted a significant asymmetry in the average excess \lya profile, both at $\dtran \lesssim 40$ kpc and at \mbox{$50\lesssim \dtran/\mathrm{kpc}\lesssim 150$} and suggested that the asymmetry results from foreground \lya emission contaminating the background \lya absorption feature. In this work, we confirm the presence of net blueshifted \lya absorption at $\dtran \lesssim 40$ and net redshifted absorption at $50\lesssim \dtran/\mathrm{kpc}\lesssim 150$ (see e.g.\ Figure \ref{fig:heatmaps}). 

For the net redshifted absorption at $\dtran \simeq 100$ kpc, we rule out noise arising from a particular set of background spectra (i.e.\ sample variance) since this feature is larger than the effective resolution element and with $\langle N\rangle\sim 400$ galaxies being averaged together, $S/N$ is not a major concern. We also see this effect in the HIRES stacked spectra (Figure \ref{fig:hires}) which have background sources considerably brighter than the foreground galaxies they probe. Since both means of measuring foreground galaxy absorption show redshifted \lya absorption, the underlying cause may be related to a hitherto unknown bias in the set of background galaxies or another effect that remains to be explained. One possibility put forth by \citet{chen_keck_2020} involves accretion-dominated inflows at large \dtran that could result in blueshifted \lya photons being preferentially scattered in the direction of the observer. Inflowing \lya would tend to have a blue-dominated profile when scattered into our line of sight and when superposed with symmetric \lya absorption, the observed net \lya profile would appear redshifted. While the existing KBSS-KCWI spectra lack the sensitivity to characterize the \lya emission contribution to the measured \lya absorption signature at $\dtran>R_{\rm vir}$, this possibility cannot yet be ruled out.

We instead focus on understanding the (net blueshifted) \lya profile at small \dtran, where in addition to the inherent variability associated with combining \lya forest absorption systems, \lya photons in emission from the foreground galaxy modulate absorption measured from the background galaxy spectrum. We build upon the order-of-magnitude calculation presented in \citet{chen_keck_2020} using a stack of KCWI \lya halos from \citet[][2025]{chen_kbsskcwi_2021} to measure the flux in emission from the foreground galaxy that gets ``added'' to the background absorption signal. We start by constructing the radial surface brightness (SB) distribution of \lya for the mean galaxy halo in the sample. The surface brightness integrated over the full \lya profile is $\sim 2\times 10^{-17}$ erg s$^{-1}$ cm$^{-2}$ arcsec$^{-2}$ within $1\arcsec-2\arcsec$ from the center of the average halo. The halo falls off exponentially:\ at $\dtran=20$ kpc, the SB is down by a factor of $\sim 10$, and by $\dtran=50$ kpc, a factor of $\sim 100$ (see also \citealt{steidel_diffuse_2011} and \citealt{chen_kbsskcwi_2021}).

The emission spectra in Figure \ref{fig:emfill} (shown in blue) were derived by integrating the foreground galaxy surface brightness profile over the \dtran of interest weighted by the underlying galaxy pair distribution in that bin. The stack of the foreground galaxies at $\dtran\simeq27$ kpc has an average continuum flux density $f_{\lambda,\, \rm cont}\sim 5\times 10^{-19}$ erg s$^{-1}$ cm$^{-2}$ \AA$^{-1}$ and an average \lya emission flux density close to $2\times 10^{-18}$ erg s$^{-1}$ cm$^{-2}$ \AA$^{-1}$. The resultant flux density is normalized with respect to the continuum level and translated up by $+1$ for ease of visualization. 

Figure \ref{fig:emfill} shows absorption and emission profiles for $\dtran\simeq 27$ kpc (left) and $\dtran\simeq 52$ kpc (right). These \dtran represent edge cases but a similar comparison can be made at other \dtran. Whereas the \citet{chen_keck_2020} order-of-magnitude calculation compared fluxes, the spectra in Figure \ref{fig:emfill} are flux densities normalized by the stacked background galaxy continuum level ($f_{\lambda,\, \rm abs}\simeq 5\times10^{-19}$ erg s$^{-1}$ cm$^{-2}$ \AA$^{-1}$). The orange curves in both panels show normalized background composite spectra similar to those in the left panel of Figure \ref{fig:lyaciv}.

In Figure \ref{fig:emfill}, the noise properties of the emission and absorption spectra depend oppositely on \dtran. As the range of \dtran increases, the $S/N$ of the background absorption spectra increases, while at the same time, the \lya surface brightness is decreasing and the emission spectra become noisier. The two bins illustrated in Figure \ref{fig:emfill} do not meet the fiducial cutoff for the minimum number of pairs in a impact parameter bin ($N_\mathrm{pairs}\gtrsim 30$; Appendix \ref{app:stackmethod}), but represent a compromise where the changing emission and absorption profiles can be seen despite the low $S/N$. Both spectra are therefore smoothed with a one pixel ($\sigma=50$ \kms) Gaussian kernel.

At small \dtran, the \lya absorption and emission profiles resemble inverted and reflected images of one another (about the intensity and velocity axes respectively). This can be explained simply by considering a symmetric \lya absorption trough where red-dominated emission contributes to the profile resulting in a blue-dominated absorption trough ($\Delta v_{\rm meas}=-223$ \kms\ at $\dtran\simeq 27$ kpc). We note that subtracting off the expected foreground emission from the background (modulated) absorption profile (the green curve in Figure \ref{fig:emfill}) results in a quasi-symmetric \lya absorption line ($\Delta v_{\rm corr}=-19$ \kms\ at $\dtran\simeq 27$ kpc). The ratio of measured to ``corrected'' equivalent widths suggests that \lya emission contributed along the line of sight by foreground galaxies reduces the observed absorption strength by $\simeq 47\%$.
Since \lya emission falls off exponentially with \dtran, we would expect this effect to be reduced significantly by $\dtran \simeq 50$ kpc. Indeed, the observed absorption profile in the right panel of Figure \ref{fig:emfill} at $\dtran \simeq 50$ kpc is considerably more symmetric since the foreground galaxies modulate the background signal by is modulated by $\simeq 14\%$. 
While red-dominated foreground \lya emission can help explain the net blueshifted absorption spectra at $\dtran \lesssim 50$ kpc, it is therefore unlikely to contribute significantly at $\dtran\simeq 100$ kpc, and thus cannot be the sole contributor to the net redshifted \lya profiles observed at $50\lesssim \dtran/\mathrm{kpc}\lesssim 150$.

As a specific example that illustrates an extreme, we plot the Q1422-d68-CS9 galaxy pair in Figure \ref{fig:d68} which has a separation of $\dtran\simeq 16$ kpc. Going off of the full sample $W_\lambda$ vs.\ \dtran evolution (Figure \ref{fig:ew_b}), at $\dtran\simeq 16$ kpc, the average \lya absorption system has $W_\lambda(\text{\lya})\simeq 2$ \AA. Instead, the background (d68, red) spectrum at the wavelength of \lya in the foreground shows \textit{emission} with a red-dominated profile similar to that of foreground galaxy (CS9, blue). Only once the \lya emission from foreground galaxy has diminished and no longer contributes (at $|v| \simeq 1000$ \kms) do we observe absorption. For this particular system, it is likely that the unaffected absorption signal would cover $|\vlos| \lesssim 1000$ \kms, but a brighter background source (e.g.\ a QSO) would be needed to confirm this. This effect represents one of the pitfalls in using comparably luminous sources as pairs especially at small \dtran where \lya emission contributes to the background galaxy signal.

\end{document}